\documentclass{aa}
\usepackage[varg]{txfonts}
\usepackage{graphicx}
\usepackage[colorlinks=true,citecolor=blue,linkcolor=blue]{hyperref}

\begin{document}

\titlerunning{GRB 130427A, GRB 160509A, and GRB 160625B} 

\authorrunning{Liang Li, et al.}

\title{Self-Similarities and Power-laws in the Time-resolved Spectra of GRB 190114C, GRB 130427A, GRB 160509A, and GRB 160625B}

\author{
Liang~Li\inst{1}\and
R. Ruffini\inst{1,2,3,4,5,6}\and
J.~A.~Rueda\inst{1,4,5,6,9}\and
R.~Moradi\inst{1,2,7}\and
Y.~Wang\inst{1,2,7}\and
S.~S.~Xue \inst{1,2}
}

\institute{International Center for Relativistic Astrophysics Network, Piazza della Repubblica 10, I-65122 Pescara, Italy,
\email{ruffini@icra.it},
\and
ICRA, Universit\`a  di Roma ``La Sapienza'', Piazzale Aldo Moro 5, I-00185 Roma, Italy,
\and
ICRANet-Rio, Centro Brasileiro de Pesquisas F\'isicas, Rua Dr. Xavier Sigaud 150, 22290--180 Rio de Janeiro, Brazil
\and
ICRANet-Ferrara, Dipartimento di Fisica e Scienze della Terra, Universit\`a degli Studi di Ferrara, Via Saragat 1, I--44122 Ferrara, Italy
\and 
Dipartimento di Fisica e Scienze della Terra, Universit\`a degli Studi di Ferrara, Via Saragat 1, I--44122 Ferrara, Italy
\and
INAF, Viale del Parco Mellini 84, 00136 Rome, Italy
\and
INAF -- Osservatorio Astronomico d'Abruzzo,Via M. Maggini snc, I-64100, Teramo, Italy.
\and
INAF, Istituto di Astrofisica e Planetologia Spaziali, Via Fosso del Cavaliere 100, 00133 Rome, Italy.}

\date{Received date /
Accepted date }

\abstract
{A new time-resolved spectral analysis performed on GRB 190114C has allowed to identify in its prompt emission observed by {\it Fermi}-GBM three specific Episodes predicted to occur in BdHNe I. Episode 1, which includes the ``SN-rise'' with a characteristic cutoff power-law and blackbody spectra; the Episode 2, initiated by the moment of formation of the BH, temporally coincident with the onset of the GeV emission and the onset of the ultra-relativistic prompt emission (UPE) phase a characterized by cutoff power-law and blackbody spectra; Episode 3,  the ``cavity'', with its characteristic featureless spectrum recently described in a companion paper \citep{2019ApJ...883..191R}. An extreme time-resolved analysis performed on an iterative process in a sequence of ever decreasing time interval, has allowed to find self-similar structures and power-laws in the UPE of GRB 190114C; 
see e.g., the companion paper \citep{2019arXiv190404162R}. This has led to the first evidence for the identification of a discrete quantized emission in the GeV and MeV emission presented in the companion papers \citep{2018arXiv181200354R,2019arXiv190708066R}.}
{To identify and verify the BdHNe I properties in the additional sources GRB 160509A, GRB 160625B and GRB 1340427A, and compare and contrast the results with the ones of a BdHN II source GRB 180728A \citep{2019ApJ...874...39W}. We have also identified in all four sources, following the analysis GRB 130427A in the companion paper \citep{2018arXiv181200354R}, the GeV radiation during and following the UPE phase. Also in all the four sources, we describe the spectral properties of their afterglow emission, including the mass estimate of the $\nu$NS, following the results presented in the companion paper \citep{2019arXiv190511339R}.}
{In GRB 160509A and GRB 160625B, we have first identified the aforementioned three BdHN I Episodes. In the UPE phase, we have performed the time-resolved spectral analysis following the iterative process in a sequence of ever decreasing time intervals. We have also examined both the GeV radiation and the afterglow phases. The same procedure has been repeated in the case of GRB 130427A with the exception of the UPE phase, in view of a pile-up problem. The case of GRB 180728A, a BdHN II, has been used as a counterexample.}
{The results of the spectral analysis have validated the common properties in all BdHNe I: the three Episodes as well as the self-similar structures and the associated power-laws in the UPE phase. The profound similarities of the results have made a significant step forward in the taxonomy of GRBs and in evidencing a standard composition of the BdHN I. This opens the opportunity of a vaster inquire of the astrophysical nature of their components in the population synthesis approach: e.g., the BH formation in all  BdHN I occurs due to accretion of the SN ejecta in a tight binary system with a neutron star companion which  reaches its critical mass, leading to the formation of the BH. The SN-rise in all five BdHNe are compare and contrasted.} 
{The most far reaching discovery of self-similarities and power-laws here extensively confirmed, thanks also to the conclusions presented in the companion papers \citep{2018arXiv181200354R, 2019arXiv190404162R}, leads to the existence of a discrete quantized repetitive polarized emission, both in the GeV and MeV observed by {\it Fermi}-GBM and {\it Fermi}-LAT, on a timescale as short as $10^{-14}$~s. These results open new paths in the discovery of fundamental physical laws.}

\keywords{gamma-ray bursts: general -- binaries: general -- stars: neutron -- supernovae: general -- black hole physics}

\maketitle

\section{Introduction}\label{sec:intro}

As pointed out in the well documented book by Bing Zhang \citep{2018pgrb.book.....Z}, the traditional approach in the spectroscopic data analysis of BATSE on-boarded the Compton Gamma-Ray Observatory (CGRO) \citep{2000ApJS..126...19P} has typically addressed a time-integrated spectral analysis on the entire duration of $T_{90}$ and in finding commonalities in all GRBs. This approach has been continued all the way to the current {\it Fermi}-GBM observations and the observations of the BAT instrument on-board the Niels Gehrels \textit{Swift} Observatory \citep[see e.g.,][by the {\it Fermi} team]{2009Sci...323.1688A,2019GCN.23707....1H}. The time-integrated spectrum has been traditionally fitted by a smoothly-connected, broken power-law function, named the ``Band'' function \citep{1993ApJ...413..281B}. The Band function is based on four parameters whose values vary from source to source without reaching universal values. A complementary spectral analysis limited to the brightest time bin has been addressed by fitting with power-laws, smoothly broken power-laws, Comptonized and Band models \citep{2014ApJS..211...12G}.

\citet{2012A&A...548L...5I,2014A&A...565L..10R} started a time-resolved spectral analysis approach with strong sources (e.g. GRB 090618). Since 2018, this approach has been improved starting with GRB 130427A \citep{2015ApJ...798...10R} and other 16 GRBs \citep{2018ApJ...852...53R}. This approach is now been adopted in GRB 190114C \citep{2019arXiv190404162R}.

We have correspondingly defined our  priorities: 

1) to address only the brightest GRBs observed by {\it Fermi}-GBM, {\it Fermi}-LAT as well as by the Niels Gehrels {\it Swfit} Observatory, so addressing a more limited number of sources with high significance $S$ and in a much wider range of spectral energies; 

2) in view of the strongest significance $S$, to identify Episodes which present specific spectral structures and determine the duration $\Delta T$ of each Episode in the source rest-frame;

3) to perform an even more detailed time-resolved spectral analysis on ever decreasing time intervals, within the total duration $\Delta T$, which has led to identify the presence of self-similar structures and associated power-laws. We have determined new statistical significant spectral distributions and evaluated the corresponding luminosity in the cosmological rest-frame. 

\begin{figure*}
    \centering
    \includegraphics[width=0.8\hsize,height=17cm]{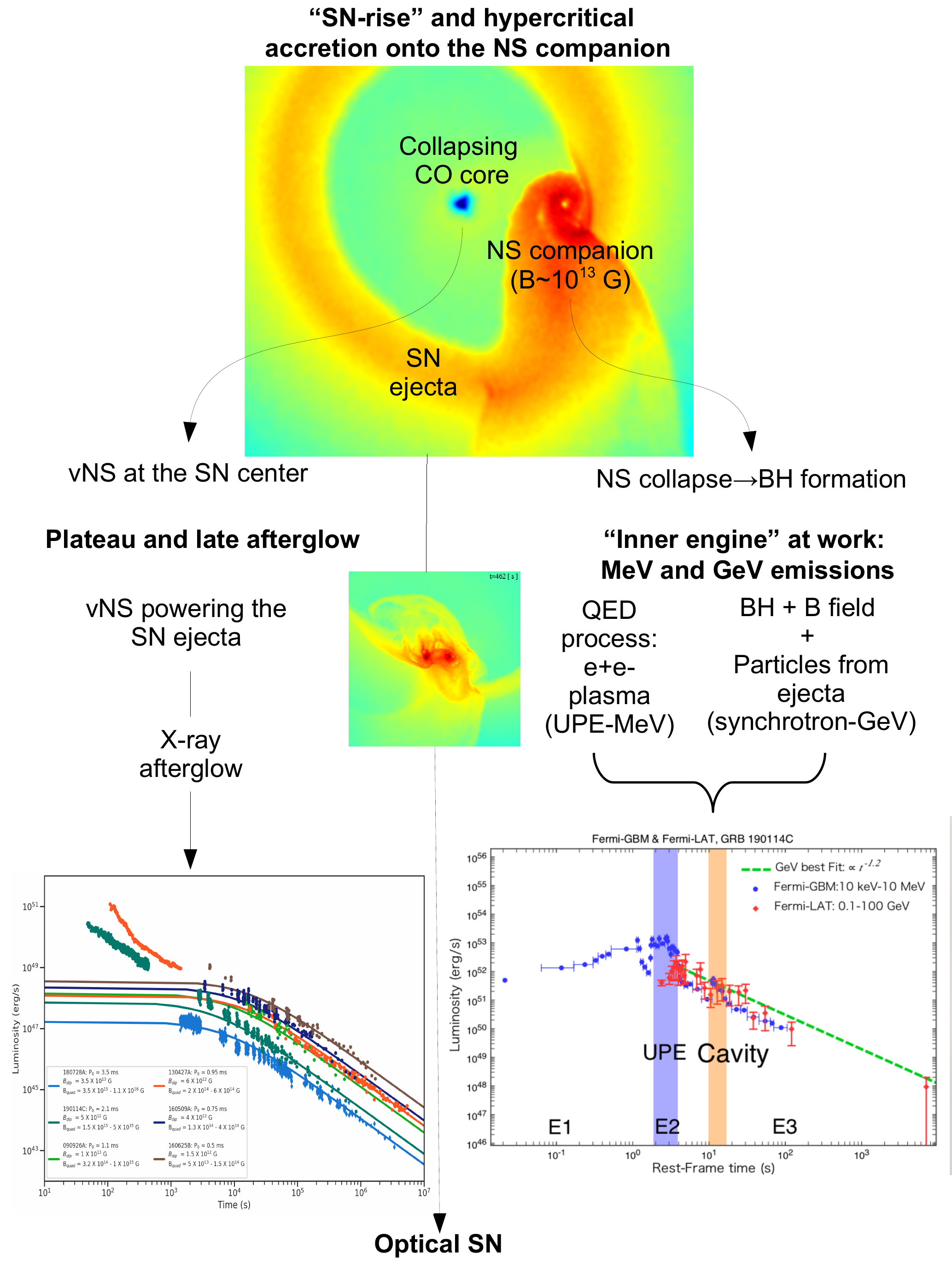}
    \caption{Episodes of a BdHN I. This scheme starts at the second SN explosion in the evolutionary path of a massive binary leading to a BdHN progenitor \citep[see e.g.][]{2015PhRvL.115w1102F}, namely a carbon-oxygen star (CO$_{\rm core}$) forming a tight (orbital period $\sim 5$~min) binary with a neutron star (NS) companion. The CO$_{\rm core}$ of mass $\lesssim 9$--$10~M_\odot$ undergoes core-collapse forming at its center a newborn NS (hereafter $\nu$NS) and, at the same time, ejecting the outermost layers in a type Ic SN explosion. The ejecta expand and their first observational appearance is what we call the ``SN-rise''. The ejecta reach the NS companion triggering a hypercritical accretion process onto it also thanks to a copious neutrino-antineutrino emission \citep{2014ApJ...793L..36F,2018ApJ...852..120B}. Numerical simulations have shown that the NS companion, by accretion, reaches the critical mass for gravitational collapse, hence forming a BH. This was first shown by two-dimensional simulations in \citet{2015ApJ...812..100B} and by three-dimensional ones, first in \citet{2016ApJ...833..107B}, and more recently in improved smoothed-particle-hydrodynamics (SPH) simulations in \citet{2019ApJ...871...14B}, from which the simulated images shown in this figure have been taken. The fundamental contribution of these simulations has been to provide a visualisation of the SN morphology that is modified from its original sphericity. A low-density region, a ``cavity'', is carved by the NS companion and, once its collapses, it is further depleted to a density as low as $\sim 10^{-14}$~g~cm$^{-3}$ by the BH formation process \citep[see][]{2019ApJ...883..191R}. The newborn Kerr BH, embedded in the magnetic field inherited from the collapsed NS, and aligned with the BH rotation axis, and surrounded by the low-density ionized plasma of the cavity, is what conform the ``inner engine'' of the GRB, see \citet{2018arXiv181200354R,2019arXiv190708066R}. The ``inner engine'' leads to MeV emission due to the $e^+e^-$ plasma created by vacuum polarization in the ultra-relativistic prompt emission (UPE) and to GeV emission by the synchrotron emission of accelerated electrons moving in the magnetic field. Details of these quantum and classic electrodynamics processes driven by the ``inner engine'' are given in companion papers; see Ruffini, Moradi, et al. (2019, submitted) and Ruffini, et al. (2019, submitted). The portion of the $e^+e^-$ plasma that enters the high density region of the ejecta produce X-ray flares observed in the early afterglow \citep{2018ApJ...852...53R}. The synchrotron emission by relativistic particles injected from the $\nu$NS into the expanding ejecta in the $\nu$NS magnetic field, explain the X-ray afterglow and its power-law luminosity \citep{2018ApJ...869..101R,2019ApJ...874...39W}. Finally, the optical emission from the ejecta due to the traditional nickel decay is observed in the optical bands few days after the GRB trigger.}
    \label{fig:BdHNepisodes}
\end{figure*}

On January 15, 2019, we indicated that GRB 190114C, discovered by {\it Fermi}-GBM on January 14, 2019 \citep{2019GCN.23707....1H}, with a redshift $z=0.424$ observed by NOT \citep{2019GCN.23695....1S}, had to be identified as a BdHN \citep{2019GCN.23715....1R}. As a BdHN, see Fig.~\ref{fig:BdHNepisodes}, within $18.8\pm 3.7$~days, a SN should be expected to appear in the same location of the GRB. After an extended campaign involving tens of observatories worldwide, the expectation of the optical SN signal was confirmed \citep{2019GCN.23983....1G,2019ApJ...874...39W}. This success and the detection of TeV radiation by MAGIC \citep{2019GCN.23701....1M} make GRB 190114C one of the best example of multi-messenger astronomy. 

For the first time, all the BdHN phases (see Fig.~\ref{fig:BdHNepisodes}) have been observed in GRB 190114C \citep[see companion paper][]{2019arXiv190404162R}. The main purpose of this article is to verity that the results obtained in GRB 190114C are not an isolated case, but on the contrary, they are verified to exist to an equal level of confidence in the other BdHN I: GRB 130427A, GRB 160509A, and GRB 160625B. Particular attention is given to the accuracy of the spectral analysis to identify the above three Episodes, as well as the much more complex iterative statistical analysis on the UPE to identify the self-similarities and the associated power-laws.

In all these sources, starting from the GBM trigger and a well determined redshift, we have progressed to identify: in Episode 1, the precursor including the first appearance of the SN (the ``SN-rise'') and the accretion of the SN ejecta onto the companion neutron star (NS). For GRB 160625B, see  Fig.~\ref{fig:joinedGRB160625B} (upper left panel); for GRB 160509A, see Fig.~\ref{fig:joinedGRB160509A} (upper left panel); for GRB 130427A, see Fig.~\ref{fig:joinedGRB130427A} (upper left panel). 
In Episode 2, the moment of formation of the BH, the simultaneous onset of the GeV emission and the onset of of the UPE phase with its characteristic cutoff power-law plus blackbody spectra observed by \textit{Fermi};  For GRB 160625B, see Fig.~\ref{fig:spectra_resolved_160625B};  for GRB 160509A, see Fig.~\ref{fig:spectra_resolved_160509A}. In Episode 3, the X-ray emission from the ``cavity'' recently modeled in the companion paper \citet{2019ApJ...883..191R}. For GRB 160625B, see  Fig.~\ref{fig:joinedGRB160625B} (upper right panel); for GRB 160509A, see Fig.~\ref{fig:joinedGRB160509A} (upper right panel); for GRB 130427A, see Fig.~\ref{fig:joinedGRB130427A} (upper right panel). As for the emission of the afterglow, for GRB 160625B, see Fig.~\ref{fig:joinedGRB160625B} (lower right panel); for GRB 160509A, see Fig.~\ref{fig:joinedGRB160509A} (lower right panel); for GRB 130427A, see Fig.~\ref{fig:joinedGRB130427A} (lower right panel), and see Fig.~9 in the companion paper  \citet{2019arXiv190511339R}.

In an unprecedented detailed spectral analysis performed on  ever decreasing time steps, we have here observed the self-similarities and power-laws in the GBM emission of the UPE phase in GRB 160509A and GRB 160625B and reported in the \ref{fig:spectra_resolved_160625B} and Fig. \ref{fig:spectra_resolved_160509A} . The details of the self similarities spectral features in the Table \ref{tab:160625B} and \ref{tab:160509A} , as well as the details of the numerical values of the fitting parameters for each GRB. We have extended the validity of of the corresponding results obtained for GRB 190114C in \citet{2019arXiv190404162R}. In the case of GRB130427A this analysis has not been possible in view of the piled up data in the UPE phase. For all sources the spectral properties of the SN rise have been obtained quote Tables \ref{table:episodes_160625B}, \ref{table:episodes_160509A}, and \ref{table:episodes_130427A}, as well as the spectral properties of the GeV emission following the UPE phase and the spectral properties of the ``cavity'' and of their afterglows.

The results here presented indicate a clear progress in ascertaining the taxonomy of a standard BdHN I the comparison and contrast with the BdHN II \citep{2019ApJ...874...39W} has also greatly contributed in clarifying the role of the formation and/or absence of formation of the BH in BdHNe. There are three main new direction  of research open: a) to submit to additional analysis each astrophysical component of each BdHN in the context of the different physical conditions characterising each Episode and harvest the physical novelties made possible by the observations of these previously unexplored regimes, not yet reproducible in earth based experimental facilities; b) to insert the BdHN observations within the larger context of population synthesis analysis, see e.g. discussion in \citet{2015PhRvL.115w1102F} and c) address the  micro-physical and physical origin of the self-similarities and the associated power-laws whose existence has been firmly confirmed in this article. In \citet[][in press]{2018arXiv181200354R}, we have addressed the nature of the ``inner engine'' of BdHNe originating the structure of self-similarity: a Kerr BH embedded in a magnetic field aligned with its rotation axis and surrounded a very low density electron-ion plasma \citep{2019ApJ...883..191R}, and we have shown that the rotational energy extraction from the BH leads to the discrete and quantized MeV and GeV radiations. These results as well as the ones here presented open unprecedented opportunities for reaching new results in the determination of the physical laws in nature.

The structure of this article is as follows:

In section~\ref{sec:data} it is presented the detailed time-resolved data analysis procedure. We have fully considered the spectral contribution from thermal components. Our approach of the spectral analysis is based on fitting Bayesian models by using MCMC technique, which is superior compared to previous techniques (e.g., Frequency method). Based on the Bayesian analysis and MCMC technique, the preferred model is the one with the lowest DIC score.

In section~\ref{sec:GRB160625B}, we derive our complete spectral analysis for all the episodes of GRB 160625B. 

In section~\ref{sec:GRB160509A}, we derive complete spectral analysis for GRB 160509A. 

In section~\ref{sec:GRB130427A}, the corresponding analysis for GRB 130427A, which is the only case that the UPE analysis is hampered by the pile-up problem.

In section~\ref{sec:GRB180728A}, we recall the result of the BdHNe II GRB 180720A. 

In section~\ref{sec:discussion}, we summarize the  results on the analysis of the SN-rise of BdHNe I and II, and present the implications of these results in the physical and astrophysical scenario of BdHNe.

In section~\ref{sec:conclusion}, we draw the general conclusions of this work. 

\section{Data Analysis} \label{sec:data}

\subsection{BATSE Data}\label{sub:BATSE}

In the BATSE era, the observation of GRB spectra is limited to $25$-- $1800$~keV \citep{2000ApJS..126...19P}. \cite{1993ApJ...413..281B} studied $54$ BATSE GRBs. It was there analyzed the time-integrated spectra without addressing the specific physical process operating at any given moment. The software used to perform the spectral analysis was the Burst Spectral Analysis Software (BSAS; \citealp{schaefer1991batse}), and the basic fitting algorithm is based on CURFIT of \citet{1969drea.book.....B} (see page 237 therein), which finds the optimum spectral parameters by minimizing $\chi^2$. For the fitting of the spectra, they proposed and adopted the Band function \citep{1993ApJ...413..281B}, which they found well described all the BATSE spectra. The parameters of the Band function are the low-energy spectral index $\alpha$, the high-energy spectral index $\beta$, and the peak energy $E_{\rm p}$. They pointed out that these parameters vary for each GRB without reaching universal values, so they were not able to open up the GRB taxonomy\footnote{Quote from \citet{1993ApJ...413..281B}: ``BATSE spectra are well described by this spectral form, but that $\alpha$, $\beta$ and $E_0$ all vary; there are no universal values. Such diversity must be addressed by physical models of the burst process ...  we do not find any striking characteristics upon which to base a classification taxonomy.''}. 

\subsection{{\it Fermi} Data}\label{sub:Fermi}

The {\it Fermi} satellite, launched in 2008, provides a wider observational window in energy ({\it Fermi}-GBM: $8$~keV to $40$~MeV, {\it Fermi}-LAT: $100$~MeV to $100$~GeV), as well as a higher time resolution (as low as $5~\mu$s for time-tagged event data) \citep{2009ApJ...702..791M}. \citet{2014ApJS..211...12G} presented the catalog of spectral analysis of GRBs by {\it Fermi}-GBM during its first four years of operation. 
They studied two types of spectra, the time-integrated spectrum and the spectrum of the brightest time bin. The software of RMfit (version 4.0rcl) was employed, which applies a modified, forward-folding Levenberg-Marquardt algorithm for spectral fitting. Four different spectral models were adopted: Band, Comptonized cut-off power-law (CPL), power-law (PL), and smoothly broken power-law models (SBPL). 
The PL and CPL models are preferred for most GRBs, the popularity of the simple PL model was interpreted as an observational effect.

In our approach since 2018, we implement the data from the \textit{Neil Gehrels Swift} and \textit{Fermi} satellite, our priority of having bright GRBs has already been stated in the introduction.

\subsection{Spectral Models}\label{sub:spectral_models}

There are several basic spectral components have been proposal in previous literature \citep[e.g.,][]{2011ApJ...730..141Z}. The observed GRB spectrum in keV-Mev band usually can be fitted by a  non-thermal component, namely, the Band (or CPL) function \citep{1993ApJ...413..281B}. The Band function defined as
\begin{equation}\label{n}
    f_{\rm BAND}(E)=A 
    \begin{cases}
    (\frac{E}{E_{\rm piv}})^{\alpha} \exp(-\frac{{\it E}}{{\it E_{0}}}), & E \le (\alpha-\beta)E_{0}\\
    \lbrack\frac{(\alpha-\beta)E_{0}}{E_{\rm piv}}\rbrack^{(\alpha-\beta)} \exp(\beta-\alpha)(\frac{{\it E}}{{\it E_{\rm piv}}})^{\beta}, & E\ge (\alpha-\beta)E_{0}
    \end{cases}
\end{equation}
where
\begin{equation}
E_{\rm p}=(2+\alpha)E_{0},
\end{equation}
has two power-law photon indices: the low-energy power-law photon spectral index $\alpha $ (typically $\sim $ -1.0), and the high-energy power-law photon spectral index $\beta $ (typically $\sim $ -2.2), they are connected at the peak energy $E_{\mathrm{p}}$ (typically $\sim 300$~keV) in the $\nu F_{\nu}$ space (e. g., \citealt{2000ApJS..126...19P, 2006ApJS..166..298K}), $A$ is the normalization factor at $100$~keV in units of ph~cm$^{-2}$keV$^{-1}$s$^{-1}$, $E_{\rm piv}$ is the pivot energy fixed at $100$~keV, the break energy $E_{0}$ in units of keV.

For the UPE phase, we mainly adopt the CPL model, or the so-called Comptonized model (COMP), which is given by
\begin{equation}
f_{\rm COMP}(E) =A \left(\frac{E}{E_{\rm piv}}\right)^{\alpha} e^{-E/E_{0}}
\label{CPL}
\end{equation}
where $A$ is the normalization factor at $100$~keV in units of ph cm$^{-2}$keV$^{-1}$s$^{-1}$, $E_{\rm piv}$ is the pivot energy fixed at $100$~keV, $\alpha$ is the low-energy power-law photon spectral index, and $E_{0}$ is the break energy in units of keV.

Some bursts have an additional thermal component, and generally fitted with Planck blackbody (BB) function. The Planck function, which is given by
\begin{equation}
f_{\rm BB}(E,t)=A(t)\frac{E^{2}}{\exp\lbrack\frac{E}{kT(t)}\rbrack-1},
\end{equation}
where $A(t)$ is the normalization, $k$ is the Boltzmann constant and $kT(t)$ the blackbody temperature.

For the high-energy \textit{Fermi}-LAT emission, the best-fit spectral model is usually a power-law model \citep[e.g.,][]{2010ApJ...712..558A, 2011ApJ...730..141Z,2019ApJ...878...52A} in the $0.1$--$100$~GeV energy band, i.e.,
\begin{equation}
f_{\rm PL} (E) = A \left(\frac{E}{E_{\rm piv}}\right)^{\Gamma},
\label{fplgev}
\end{equation}
where $A$ is the normalization, and $\Gamma$ is the power-law index.

In the spectral fitting for the MeV-UPE phase, we adopt a Bayesian analysis and model comparison using $\Delta {\rm DIC}$ value \cite[e.g.,][]{2017arXiv171008362B, 2019ApJS..242...16L,2019arXiv190502340L, 2019arXiv190809240L}. For the GeV emission, a Maximum Likelihood Estimate (MLE) analysis is used to obtain the best fitting \citep[e.g.,][]{2012ApJS..199...19G, 2013ApJS..209...11A, 2014ApJS..211...12G, 2016ApJS..223...28N, 2019ApJ...878...52A, 2019ApJ..884...109L}.

\subsection{Bayesian Analysis for \textit{Fermi}-GBM Data}\label{sub: GBM}

The temporal and spectral analysis of \textit{Fermi}-GBM data is applied by the Bayesian approach package, namely, the Multi-Mission Maximum Likelihood Framework (3ML, \citealt{2015arXiv150708343V}). The GBM \citep{2009ApJ...702..791M} carries $14$ detectors that includes $12$ sodium iodide (NaI, $8$~keV--$1$~MeV) and $2$ bismuth germinate (BGO, $200$~keV--$40$~Mev) scintillation detectors. 
The pre-source and the post-source data are used to fit the background by a $0$-$4$ order polynomial function. The time interval of source is selected longer than the duration of bursts ($T_{90}$), in order to cover the entire background subtracted emission. During the fitting procedure, the likelihood-based statistics, the so-called Pgstat is used, given by a Poisson (observation, \citealt{1979ApJ...228..939C})-Gaussian (background) profile likelihood.
The spectral analysis is performed by employing a Markov Chain Monte Carlo (MCMC) technique to fit Bayesian models, and the model parameters in the Monte Carlo iteration vary in the following range: PL model, index: $[-5,1]$; BB model, $kT$ (keV): $[1,10^{3}]$; CPL model, $\alpha$: $[-5, 1]$, $E_{\rm c}$ (keV): $[1, 10^{4}]$. We use the typical spectral parameters from $Fermi$-GBM catalogue as the informative priors: $\alpha \sim \mathcal{N} (\mu=-1.,\sigma=0.5)$; $E_{\rm c} \sim \mathcal{N} (\mu=200,\sigma=300)$; $\beta \sim \mathcal{N} (\mu=-2.2,\sigma=0.5)$. Each time we perform $20$ chains, each chain includes $10,000$ time iterations. The final value and its uncertainty ($68\%$ (1$\sigma$) Bayesian credible level) are calculated from the last $80\%$ of the iterations. In this paper, we adopt the deviance information criterion (DIC) to select the best one from two different models, defined as ${\rm DIC}=-2\log[p({\rm data}\mid\hat{\theta})]+2p_{\rm DIC}$, where $\hat{\theta}$ is the posterior mean of the parameters, and $p_{\rm DIC}$ is the effective number of parameters. The preferred model is the model with the lowest DIC score. We define $\Delta {\rm DIC}={\rm DIC (CPL+BB)-DIC(CPL)}$, for instance, if $\Delta {\rm DIC}$ is negative, indicating the CPL+BB is better than CPL. These methods have be applied in each Episode. The best-fit parameters for each spectrum ($\alpha$, $E_{\rm c}$), along with its time interval, $\Delta$DIC, blackbody temperature $kT$, blackbody flux ($F_{\rm BB}$), total flux ($F_{\rm total}$), thermal to total flux ratio, and the total energy are summarized in Table~\ref{tab:160625B} for GRB 160625B and in Table~\ref{tab:160509A} for GRB 160509A.

\subsection{Calculation of Luminosity and Energetics}
\label{subsec:energy}

The observed flux, $\Phi(E_1, E_2, z)$, integrated between the minimum energy $E_1$  and the maximum energy $E_2$ where $\Phi(E_1, E_2, z)$ is defined as
\begin{equation}
\Phi(E_1, E_2, z) = \displaystyle\int_{{E_{\,1}/{(1 + z)}}}^{{E_{\,2}}/{(1 + z)}} E \, f_{\rm obs} (E)\, dE.
\label{fbolo}
\end{equation}

In principle, for different models and different energy bands  the values of $E_{\, 1}$, $E_{\, 2}$ and $f_{\rm obs}$ would be different. For instance for GeV radiation $E_1=0.1~$~GeV and $E_2=100$~GeV and $f_{\rm obs}=f_{\rm PL}$ is obtained from Eq.~(\ref{fplgev}) with typical value of $\Gamma \approx -2.5$ \citep{2019ApJ...878...52A}. 

We adopt a flat FLRW universe model with $\Omega_{\Lambda} = 0.714$, $\Omega_{M} = 0.286$ and $H_0 = 69.6$~km~s$^{-1}$~Mpc$^{-1}$ \citep{Bennett_2014, 2016A&A...594A..13P}, and then the luminosity distance is given by \citep{1972gcpa.book.....W}
\begin{equation}
d_L(z,\Omega_{\Lambda},\Omega_M) = (1+z)\,\frac{c}{H_0}\int_0^z \frac{dz'}{\sqrt{\Omega_{M}\,(1+z')^3+\Omega_{\Lambda}}}\,.
\label{eq_dL}
\end{equation} 

The isotropic radiated luminosity is
\begin{equation}
L_{\,\rm iso} = \frac{4\pi \,d_{\,L}^{\,2}}{1 + z} \;\, \Phi(E_1, E_2, z)\,,  
\label{Liso_sbolo}
\end{equation}
where $d_{\,L}$ is the luminosity distance, and $z$ is the redshift. The observed fluence $S$ is given by
\begin{equation}
S(E_1, E_2, z) = \Delta T_{\rm i}  \Phi(E_1, E_2, z),
\label{sbolo}
\end{equation}
where $\Delta T_{\rm i}$ is the duration of the time interval in which the analysis is made, see \citet{2019ApJ...878...52A} for details.

The isotropic radiated energy which is assumed to be isotropically radiated is defined as
\begin{equation}
E_{\,\rm iso} = \frac{4\pi \,d_{\,L}^{\,2}}{1 + z} \;\, S(E_1, E_2, z).
\label{Eiso_sbolo}
\end{equation}

\section{GRB 160625B}\label{sec:GRB160625B}

\begin{figure*}
\centering
\includegraphics[width=0.95\hsize,clip]{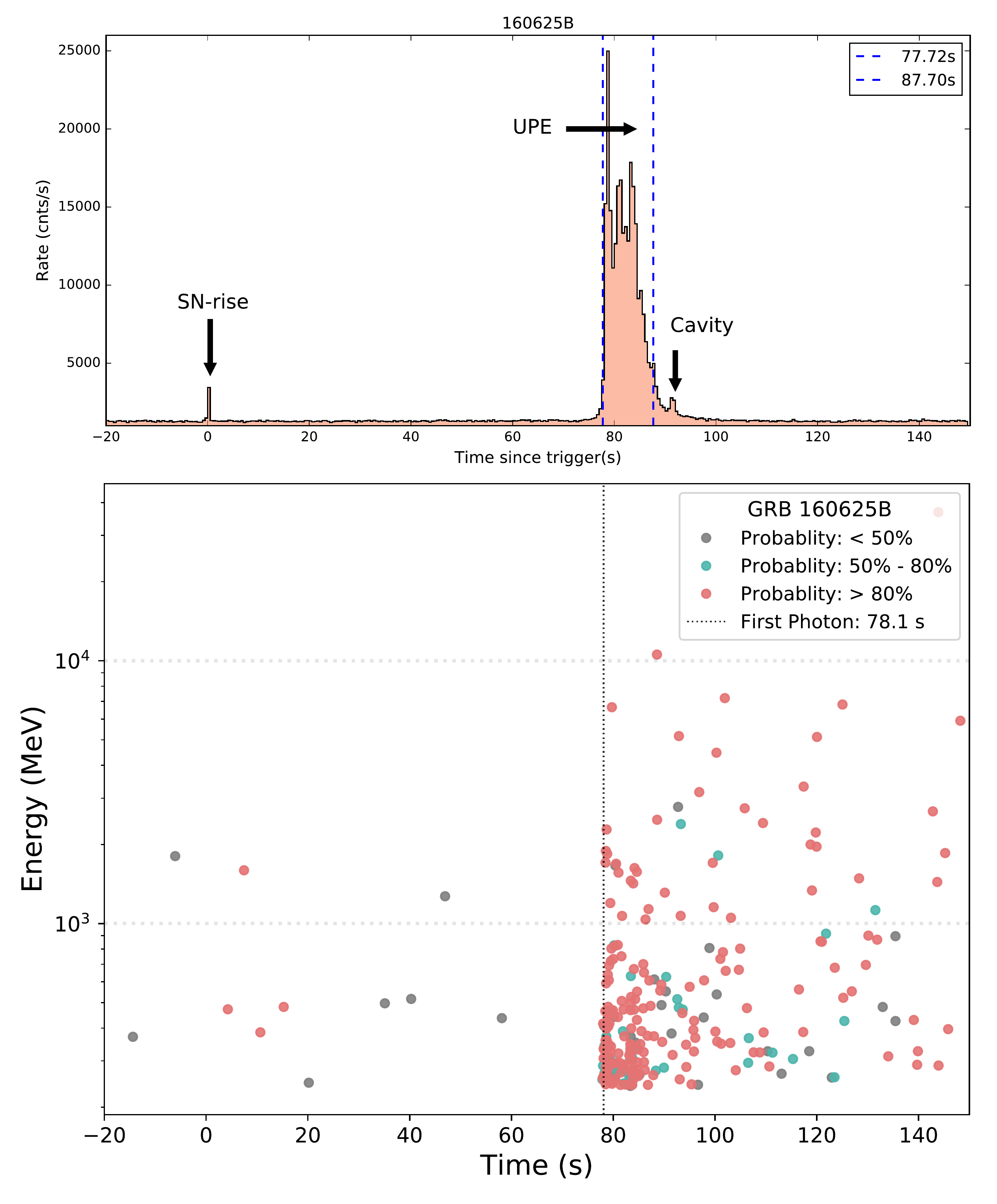}
\caption{\textbf{Upper panel}: the proposed three new Episodes of GRB 160625B as a function of the rest-frame time. Episode 1 occurs $t_{\rm rf} = 0$~s and $t_{\rm rf} = 0.83$~s. The blue dashed vertical lines represents the SN-rise. Episode 2 occurs from $t_{\rm rf} = 77.72$~s to $t_{\rm rf} = 87.70$~s, and includes the UPE emission. Episode 3 occurs at times after $t_{\rm rf} = 87.70$~s, staring at $t_{\rm rf} =87.70$~s and ending at $t_{\rm rf} =92.27$~s. The redshift for GRB 160625B is $1.406$ \citep{2016GCN.19600....1X}. The light-curve consists of two clear spikes, the isotropic energy in the first one is ($1.09\pm 0.20)\times 10^{52}$~erg. The total energy is $\approx 3\times 10^{54}$~erg. \textbf{Lower panel}: the rest-frame time and energy of {\it Fermi}-LAT photons in the energy band $0.1$--$100$~GeV. The first photon of the GeV emission occurs at $t_{\rm rf} =$ $78.1$~s. {The onset of the GeV radiation coincides with the onset of the UPE. The detailed information for each episode {(SN-rise, UPE phase, Cavity, GeV, and afterglow emission)}, see Section \ref{sec:GRB160625B} and Table \ref{table:episodes_160625B}, which include the typical starting time, the duration, the isotropic energy, and the preferred model.}}
\label{fig:lc_160625B}
\end{figure*}

On 25 June 2016 at 22:40:16.28 UT, GRB 160625B triggered Gamma-ray Burst Monitor (GBM) onboard the NASA \textit{Fermi} Gamma-ray Space Telescope \citep{2016GCN.19581....1B}. \textit{Fermi}-LAT starts the observation $188.54$~s after the trigger \citep{2016GCN.19586....1D}, and detected more than $300$ photons with energy  $>100$~ MeV, the highest energy photon is about $15$~GeV \citep{2017ApJ...849...71L}. {\it Swfit}-XRT starts the observation at late time ($> 10^4$~s), a power-law behaviour with decaying index $\sim -1.25$ \citep{2016GCN.19585....1M}. GRB 160625B is one of the most energetic GRBs with an isotropic energy $\approx 3\times 10^{54}$~erg \citep{2017Natur.547..425T, 2018NatAs...2...69Z}. The redshift $z=1.406$ is reported in \citet{2016GCN.19600....1X}.  GRB 160625B is a luminous GRB with the clear detected polarisation \citep{2017HEAD...1630505T}. 
There is no supernova confirmation due to its high redshift; $z>1$ \citep{2006ARA&A..44..507W}.

GRB 160625B has been extensively analyzed in  \citet{2017Natur.547..425T} and \citet{2018NatAs...2...69Z}. Both papers define the early emission as three episodes: a short precursor (G1), a main burst (G2), and a long lasting tail (G3).

\begin{figure*}[h!]
\includegraphics[width=0.49\hsize,clip]{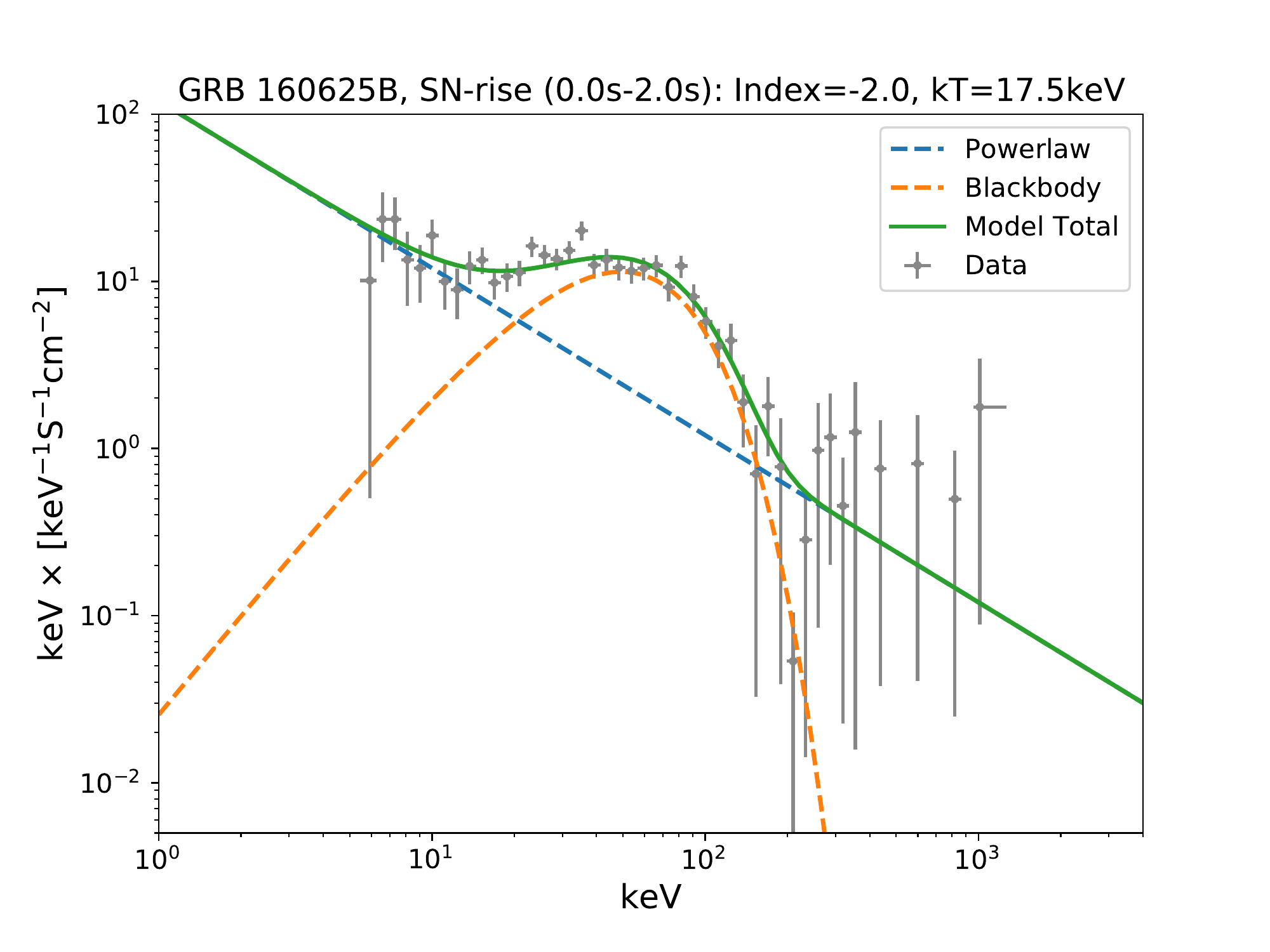}\includegraphics[width=0.49\hsize,clip]{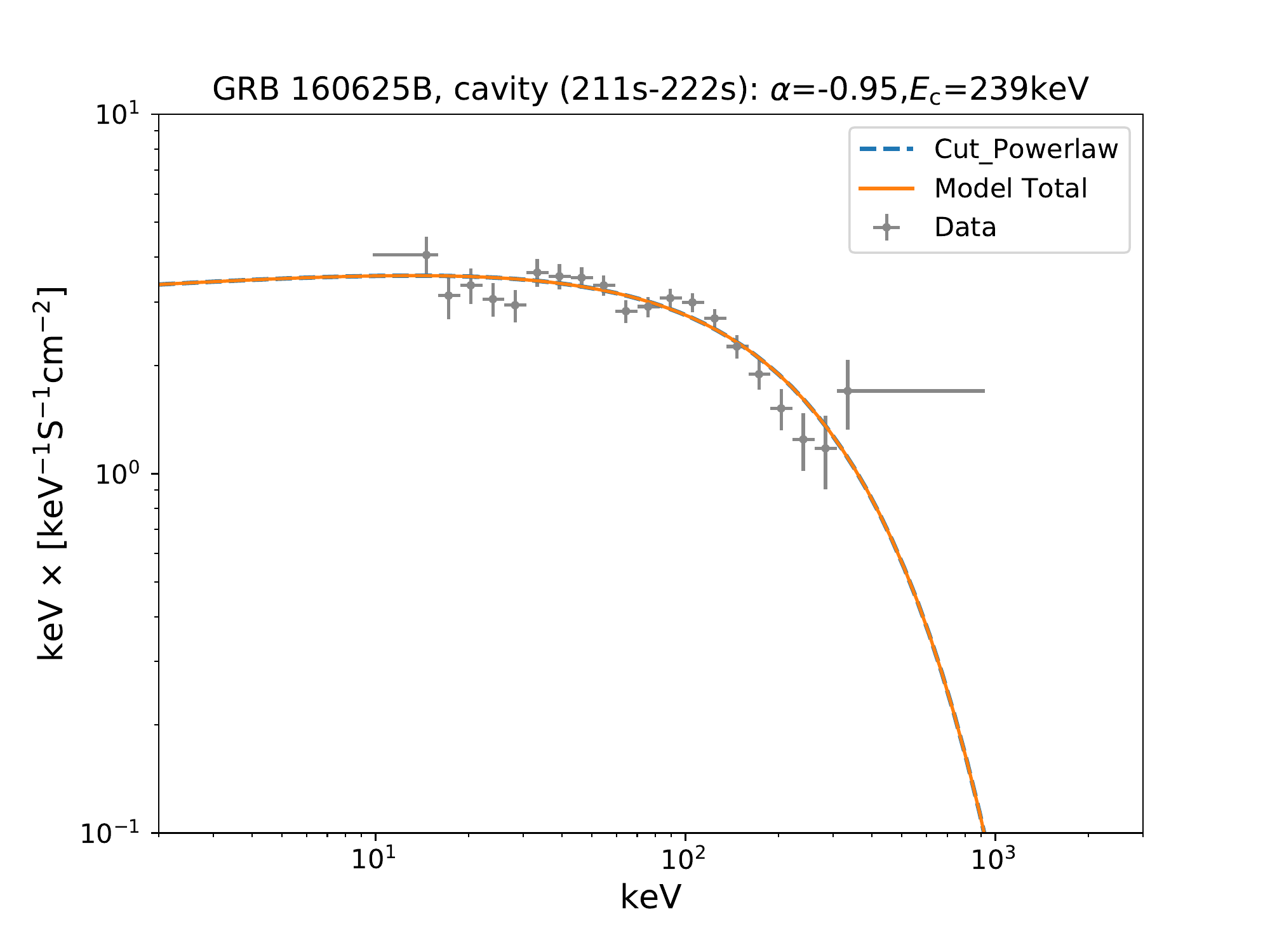}
\includegraphics[width=0.49\hsize,clip]{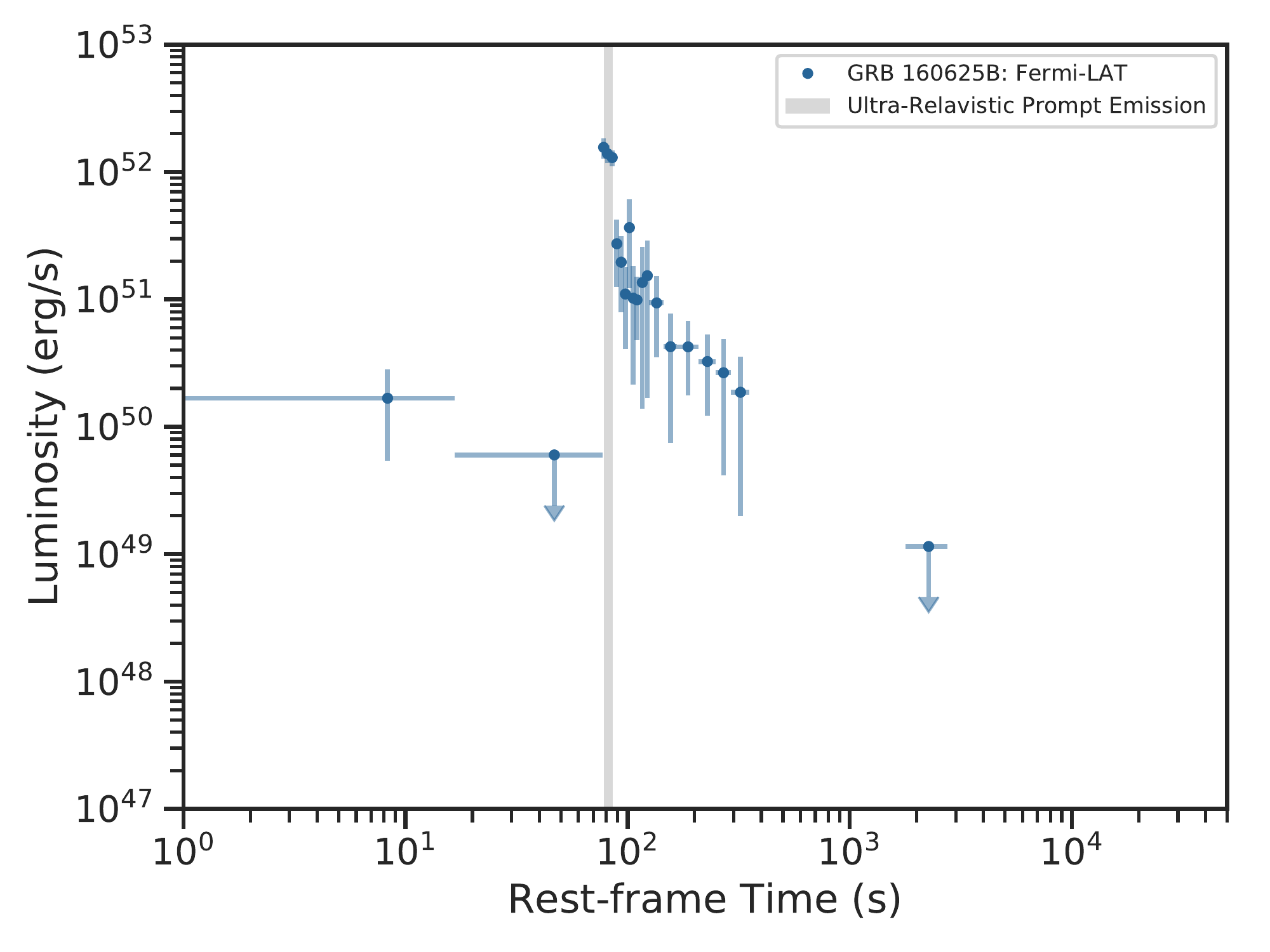}
\includegraphics[width=0.49\hsize,clip]{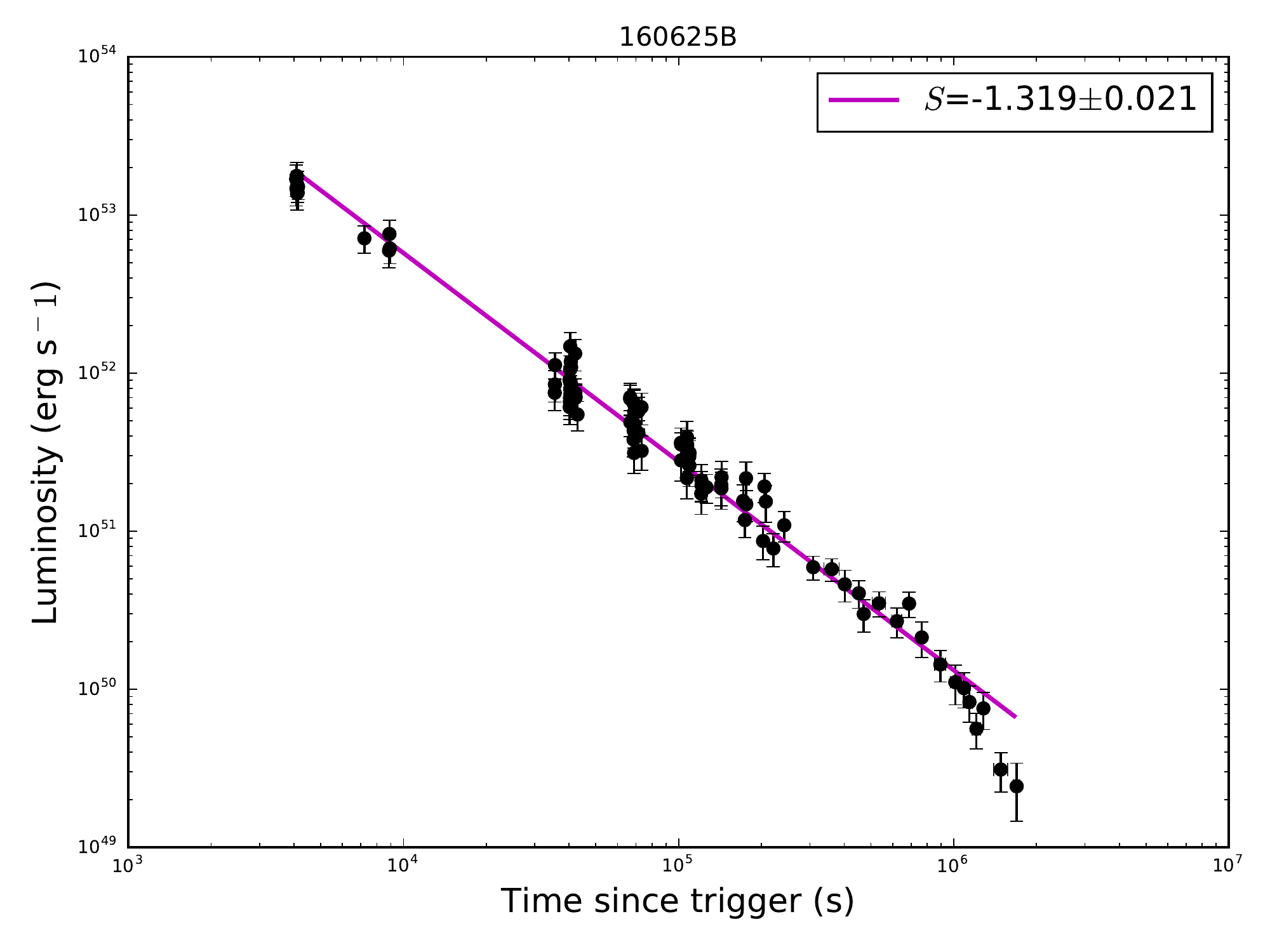}
\caption{SN-rise, Cavity, GeV, and Afterglow of GRB 160625B; see also Table \ref{table:episodes_160625B} that includes, for each Episode, the starting time, the duration, the isotropic energy, and the model that best fits the spectrum. \textbf{Upper left}: The spectrum of the SN-rise from $0$~s to $\approx 2.0$~s ($t_{\rm rf} \approx 0.83$~s). The spectrum is fitted by a blackbody of temperature $17.5$~keV (in the observer's frame) plus a power-law of index $-2.0$. \textbf{Upper right}: The cavity spectra, from $\approx 211$~s ($t_{\rm rf} = 87.70$~s) to $\approx 222$~s ($t_{\rm rf} = 92.27$~s), are well fitted by a CPL, with the photon index $\alpha$ is $-1.67$ and the cutoff energy $251$~keV in the observer's frame. \textbf{Lower left}: {\it Fermi}-LAT rest-frame luminosity in the $100$~MeV to $100$~GeV energy band (the UPE region is marked with the grey shadow). \textbf{Lower right}: k-corrected X-ray afterglow luminosity observed by {\it Swift}-XRT in the $0.3$--$10$~keV band, as a function of the rest-frame time. It is best fitted by a power-law with index $1.319\pm 0.021$.}\label{fig:joinedGRB160625B}
\end{figure*}

Based on the temporal and spectral analysis, we confirm that the gamma-ray light curve of GRB 160625B has three different episodes, shown in Fig.~\ref{fig:lc_160625B} (see also Table~\ref{table:episodes_160625B}).
{Three different physical episodes have been identified in the keV-MeV energy range 
(see Fig.~\ref{fig:lc_160625B}, Fig.~\ref{fig:joinedGRB160625B}, and Table~\ref{table:episodes_160625B}):
(1) SN-rise, the time-interval ranges from $t_{\rm rf}=0.00$~s to $t_{\rm rf}=0.83$~s.
(2) UPE phase, the time-interval ranges from $t_{\rm rf}=77.72$~s to $t_{\rm rf}=87.70$~s.
(3) Cavity, the time-interval ranges from $t_{\rm rf}=87.70$~s to $t_{\rm rf}=92.27$~s.}

\begin{table*}[ht!]
\caption{The Episodes of GRB 160625B, including the starting time, the duration, the energy (isotropic), the preferred spectral model, and the references. For the starting time of GeV emission, we take the time of the first GeV photon form the BH. The GeV emission may last for a very long duration, but the observational time is limited due to \textit{Fermi}-LAT is not capable to resolve the late-time low flux emission, therefore the ending time of GeV observation in the table gives the lower limit of the ending time of GeV emission. The starting time of X-ray afterglow in the table is taken from the starting time of \textit{Swift}-XRT. The energy in the afterglow is integrated from  $10^2$~s to $10^6$~s. All times are given in the rest frame.}             
\label{table:episodes_160625B}
\centering 
\begin{tabular}{ccccccc}       
\hline\hline
&&&&&\\
Episode &Starting Time & Ending Time&Energy&Spectrum&Reference\\ &Rest-frame&Rest-frame\\  
&(s)& (s)& (erg)&&\\
&&&&&\\
\hline                        
&&&&&\\
SN-rise&$  0 $&$ 0.83 $&$ 1.09 \times10^{52}$&CPL+BB&New in this paper\\
UPE& $77.72$& $87.70$&$ 4.53 \times10^{54}$&CPL+BB&New in this paper\\
Cavity&$87.70$&$92.27$&$ 2.79\times 10^{52}$&CPL&New in this paper\\
GeV&$ 78.1$&  $> 300$&$ 2.99 \times 10^{53}$&PL&New in this paper\\
Afterglow&$ 4082$&$ >10$ days&$1.08 \times 10^{53}$&PL&New in this paper\\
\hline 
\end{tabular}
\end{table*}

In a BdHN I, the ``inner engine'' starts at the moment of formation of the BH, accelerating charged particles that radiate photons in a wide energy band, generating the UPE phase and the GeV photons. The onset of the UPE phase is indicated by the appearance of the thermal component since the plasma is originally optically thick. Since the count rate of GeV photons observed in the onset phase is a few per second, it requires to have the discrepancy of at most fractions of a second between the observed starting time of the UPE and the GeV. Indeed, as for GRB 160625B, the starting time of its thermal emission is just $0.38$~s ahead of the observational time of the first GeV photon, which for the above reasons can be considered temporally coincident. This time coincidence is also observed in the other BdHNe I studied in this article. 

--- \textit{SN-rise}. Figure~\ref{fig:joinedGRB160625B} (upper left panel) shows the fit of the SN-rise spectrum during its rest-frame time interval of occurrence, i.e. from $0$ to $t_{\rm rf} \simeq 0.83$~s. It is best fitted by PL+BB model with temperature $17.5$~keV and power-law index $-2.0$.

\begin{figure*}[ht!]
\centering
\includegraphics[angle=0, scale=0.6]{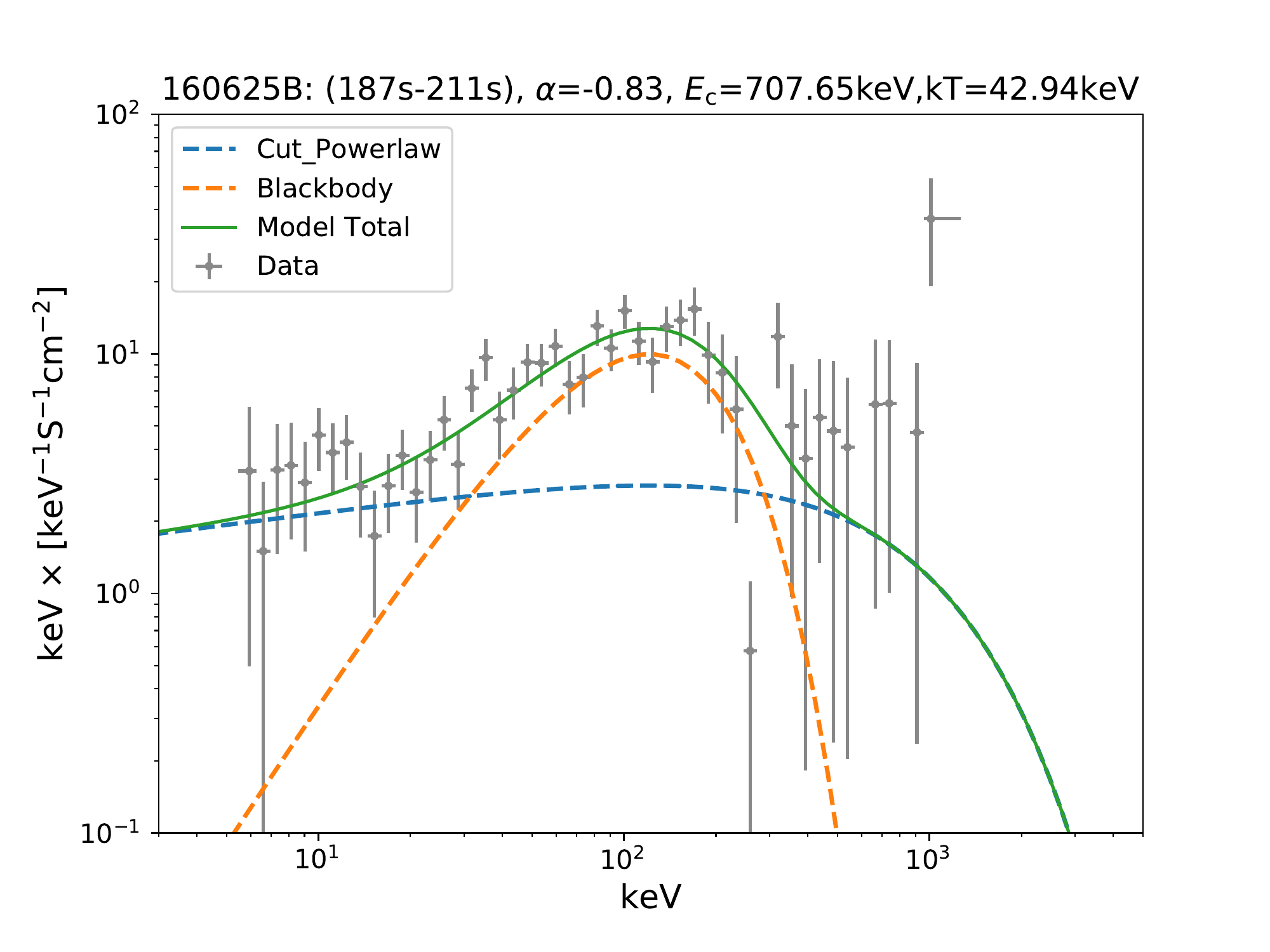}\\
\includegraphics[angle=0, scale=0.45]{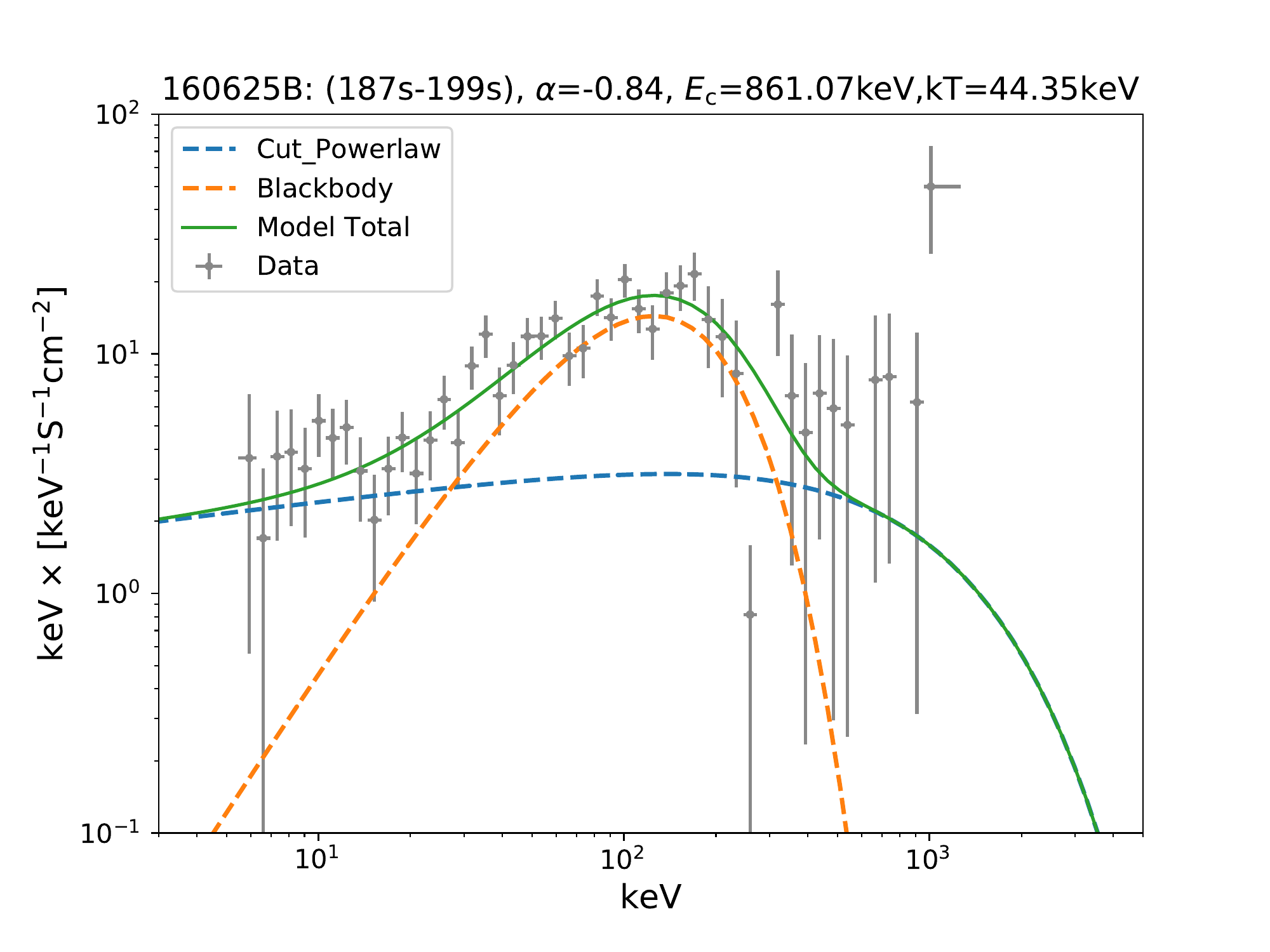}
\includegraphics[angle=0, scale=0.45]{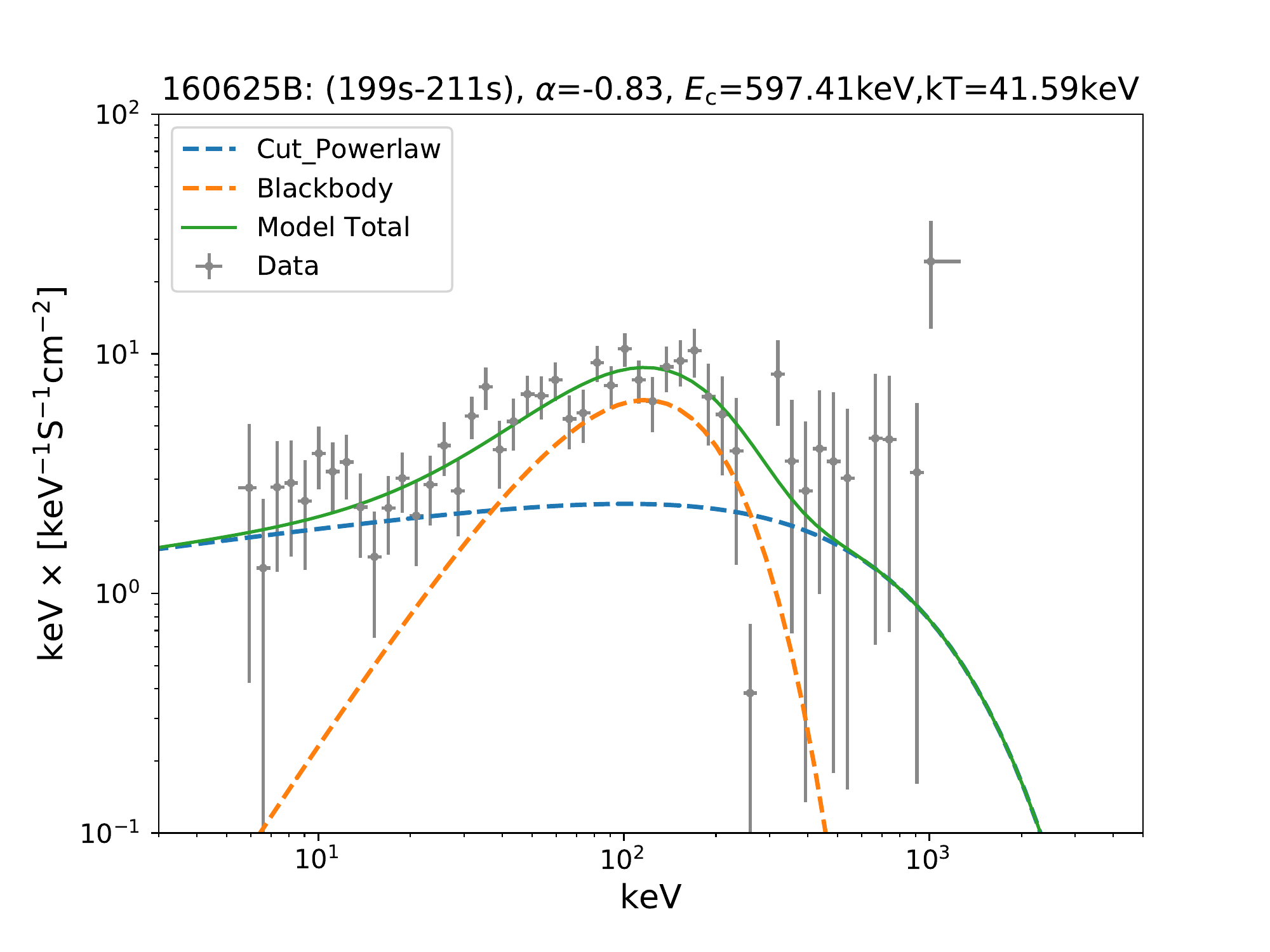}\\
\includegraphics[angle=0, scale=0.215]{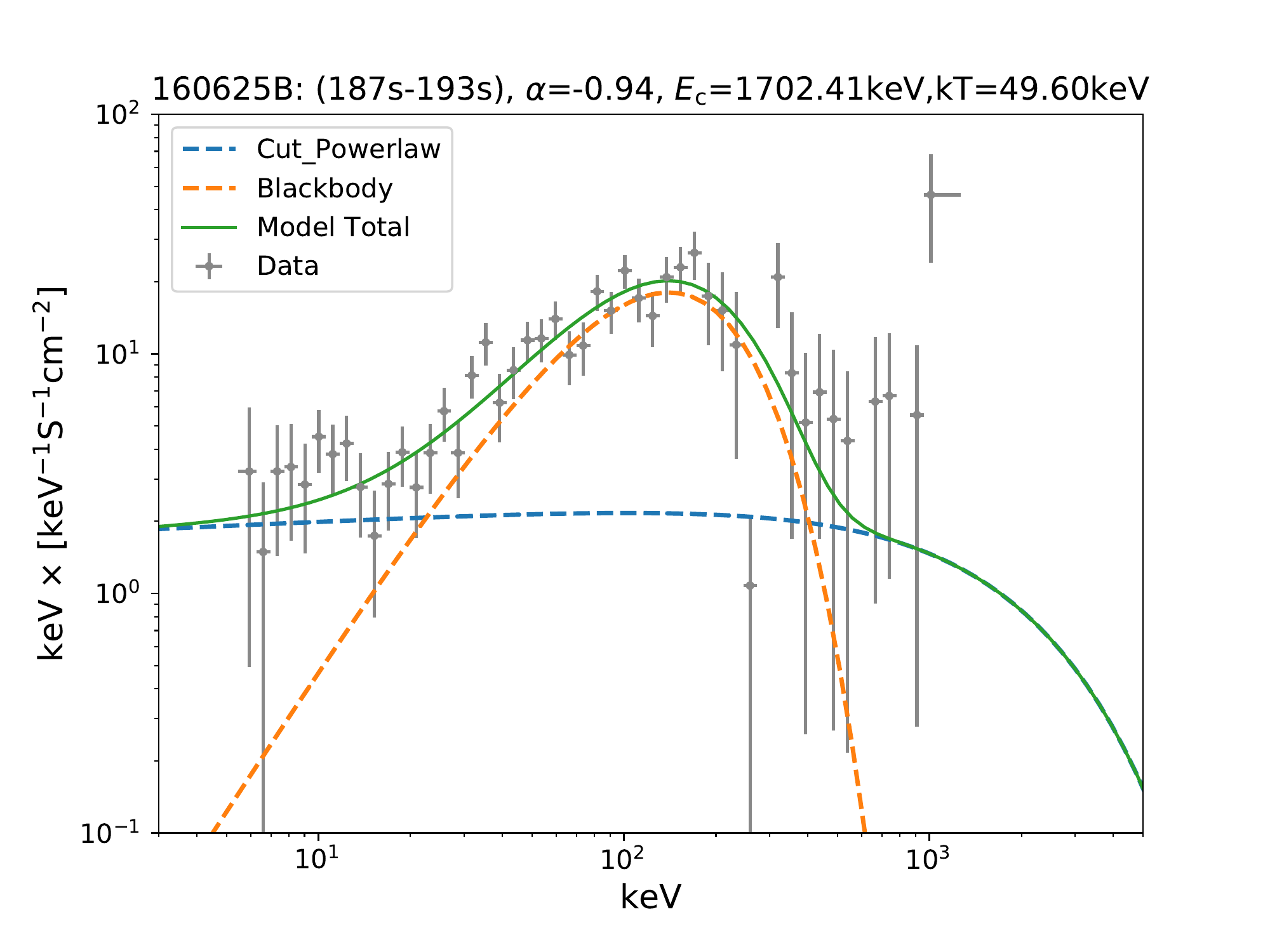}
\includegraphics[angle=0, scale=0.215]{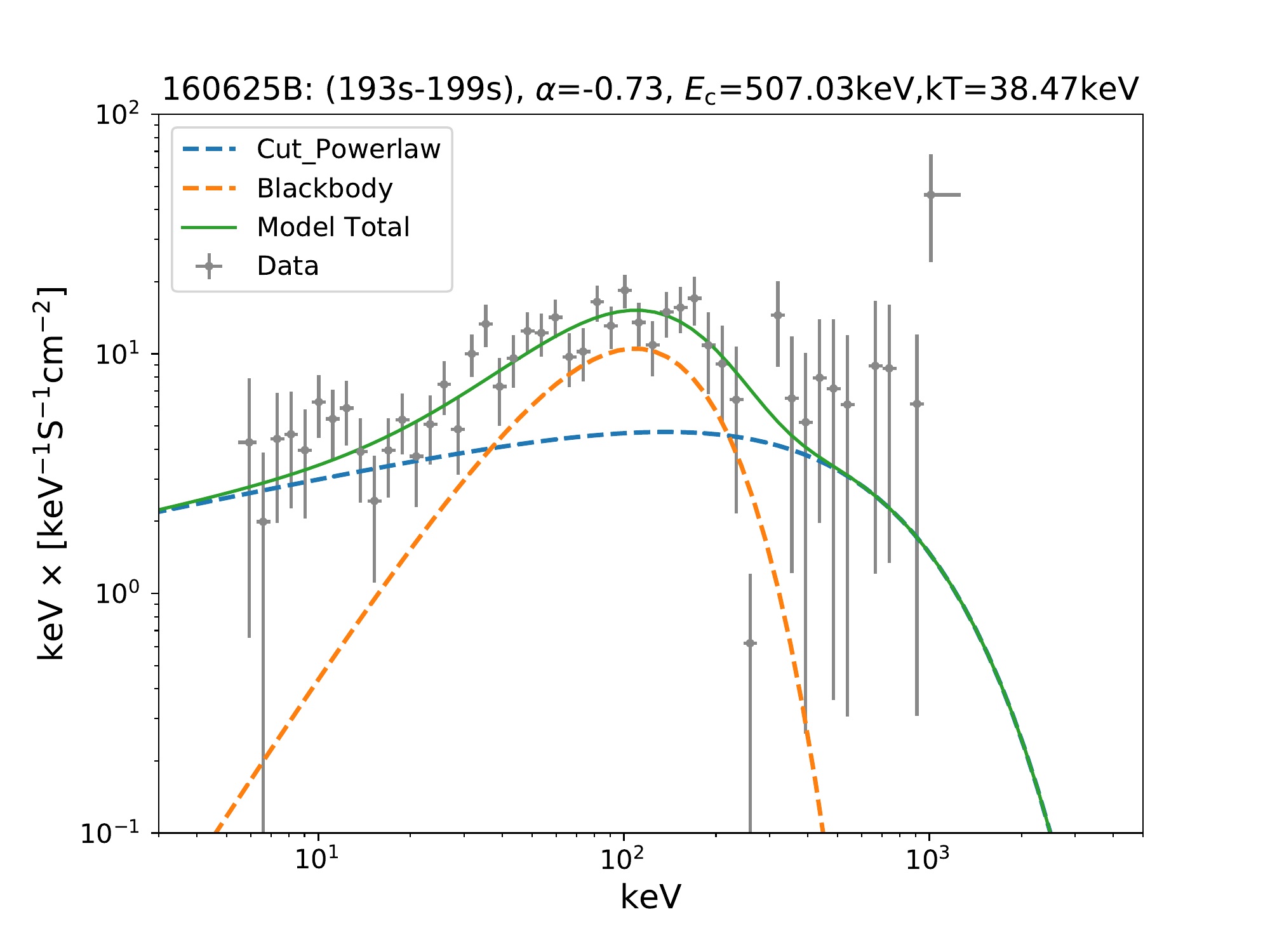}
\includegraphics[angle=0, scale=0.215]{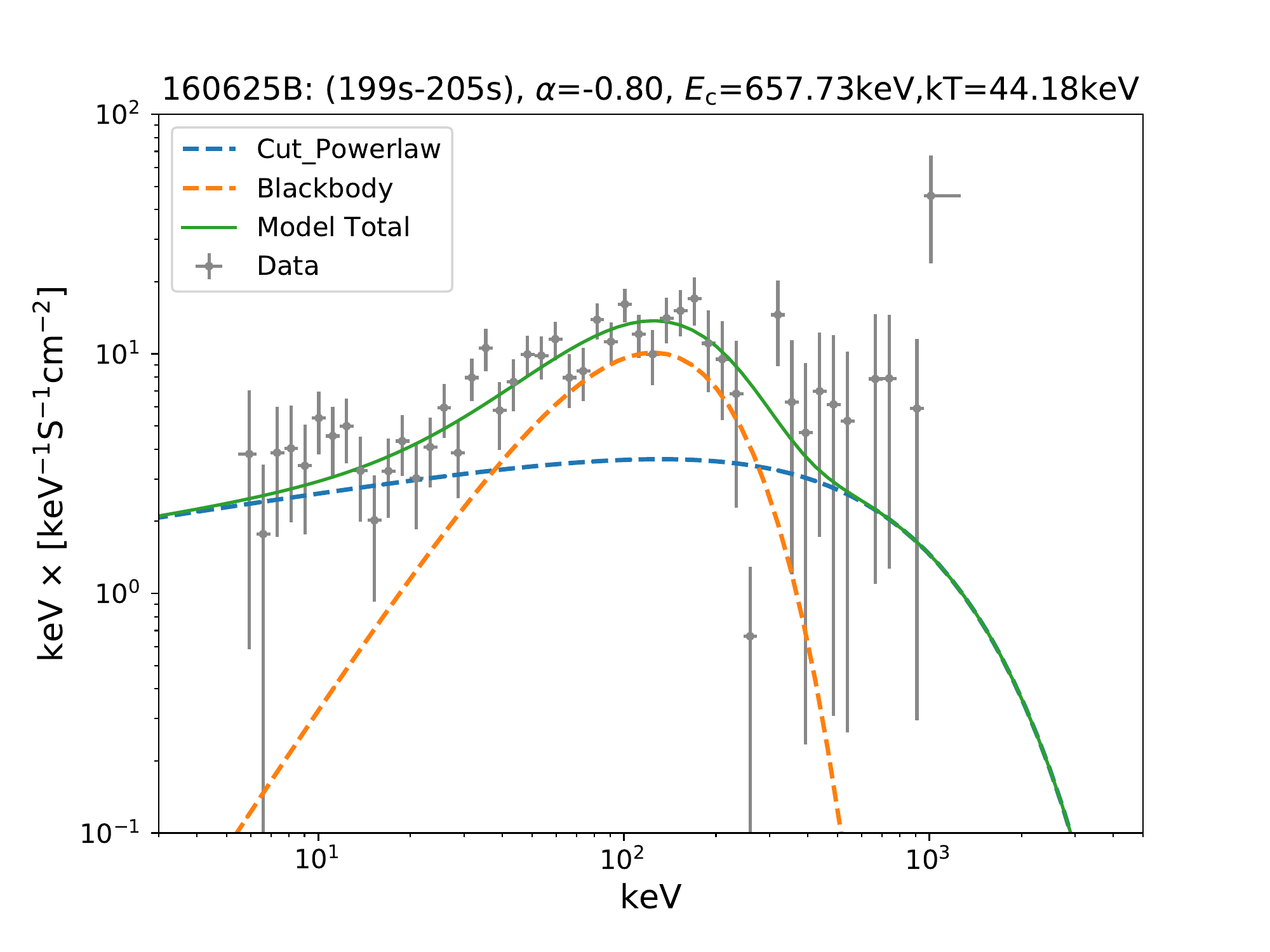}
\includegraphics[angle=0, scale=0.215]{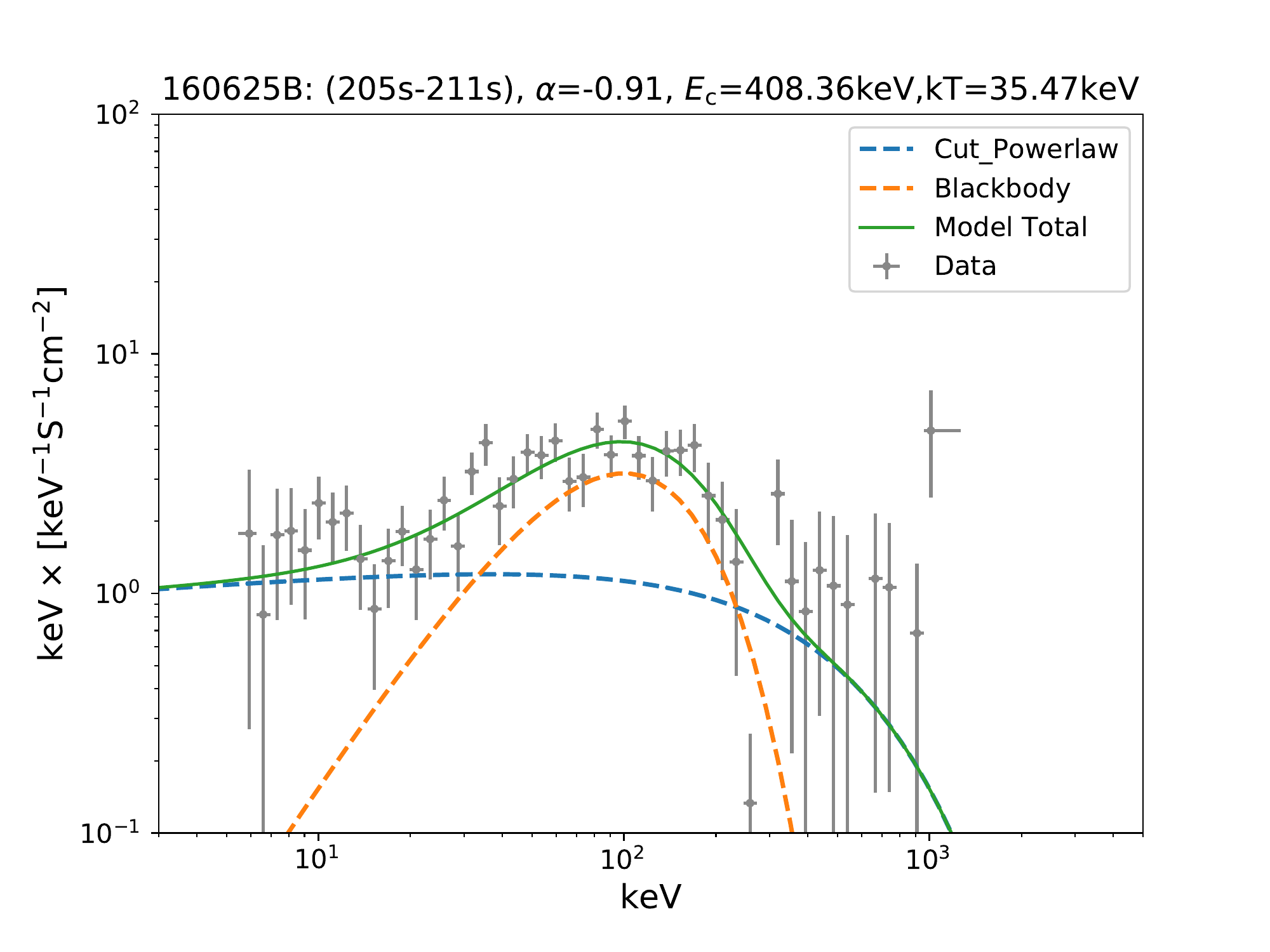}\\
\includegraphics[angle=0, scale=0.105]{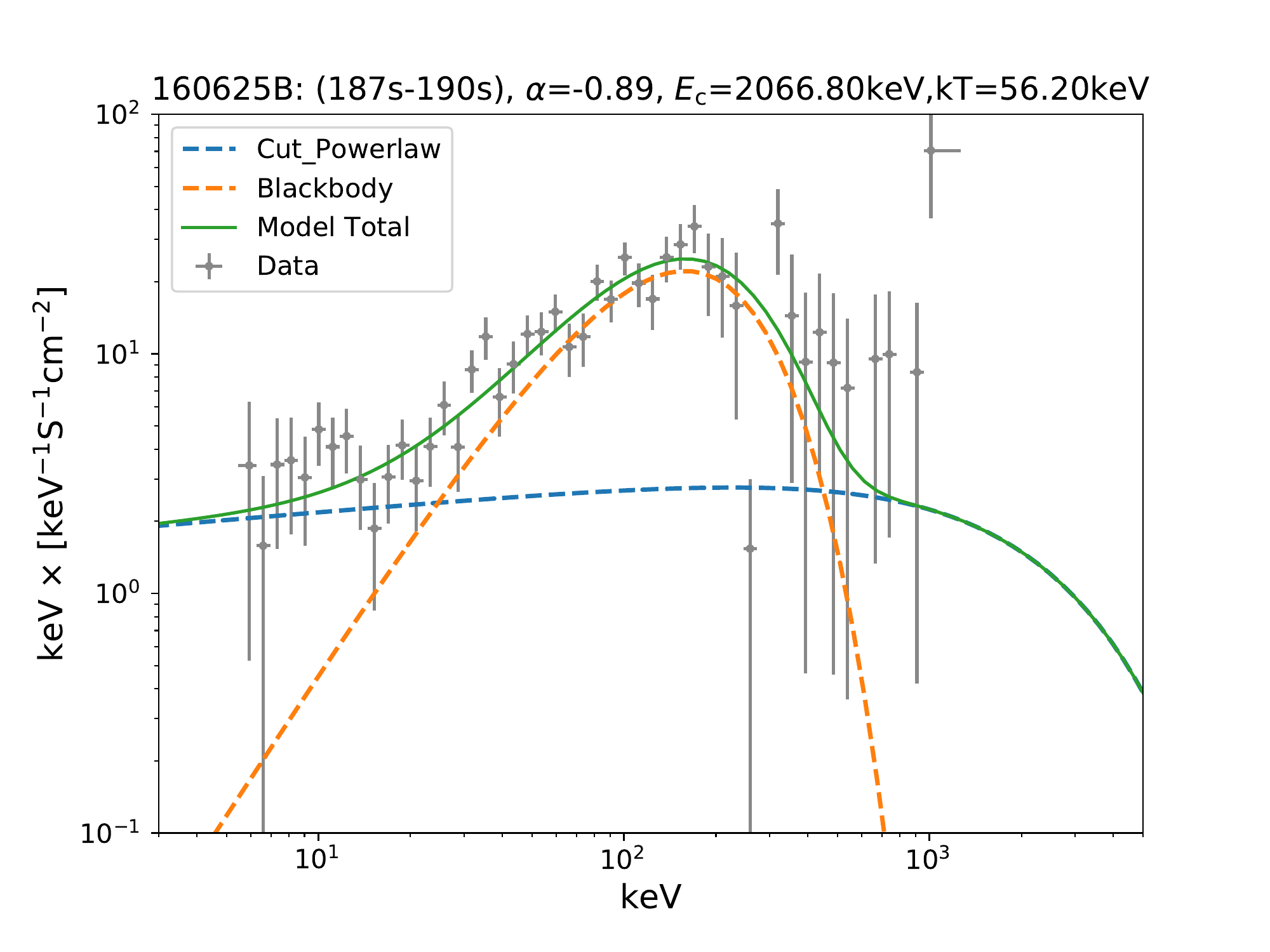}
\includegraphics[angle=0, scale=0.105]{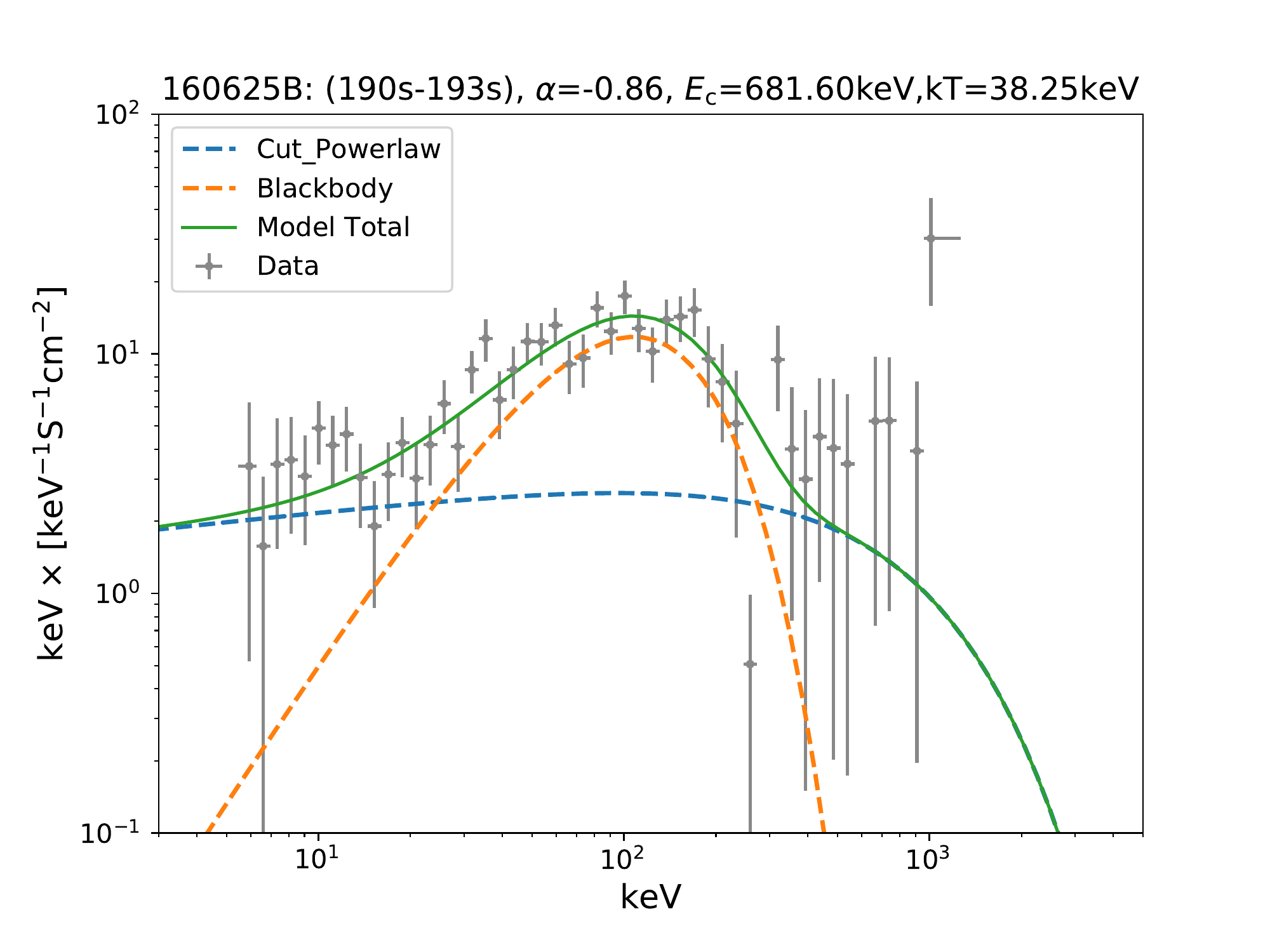}
\includegraphics[angle=0, scale=0.105]{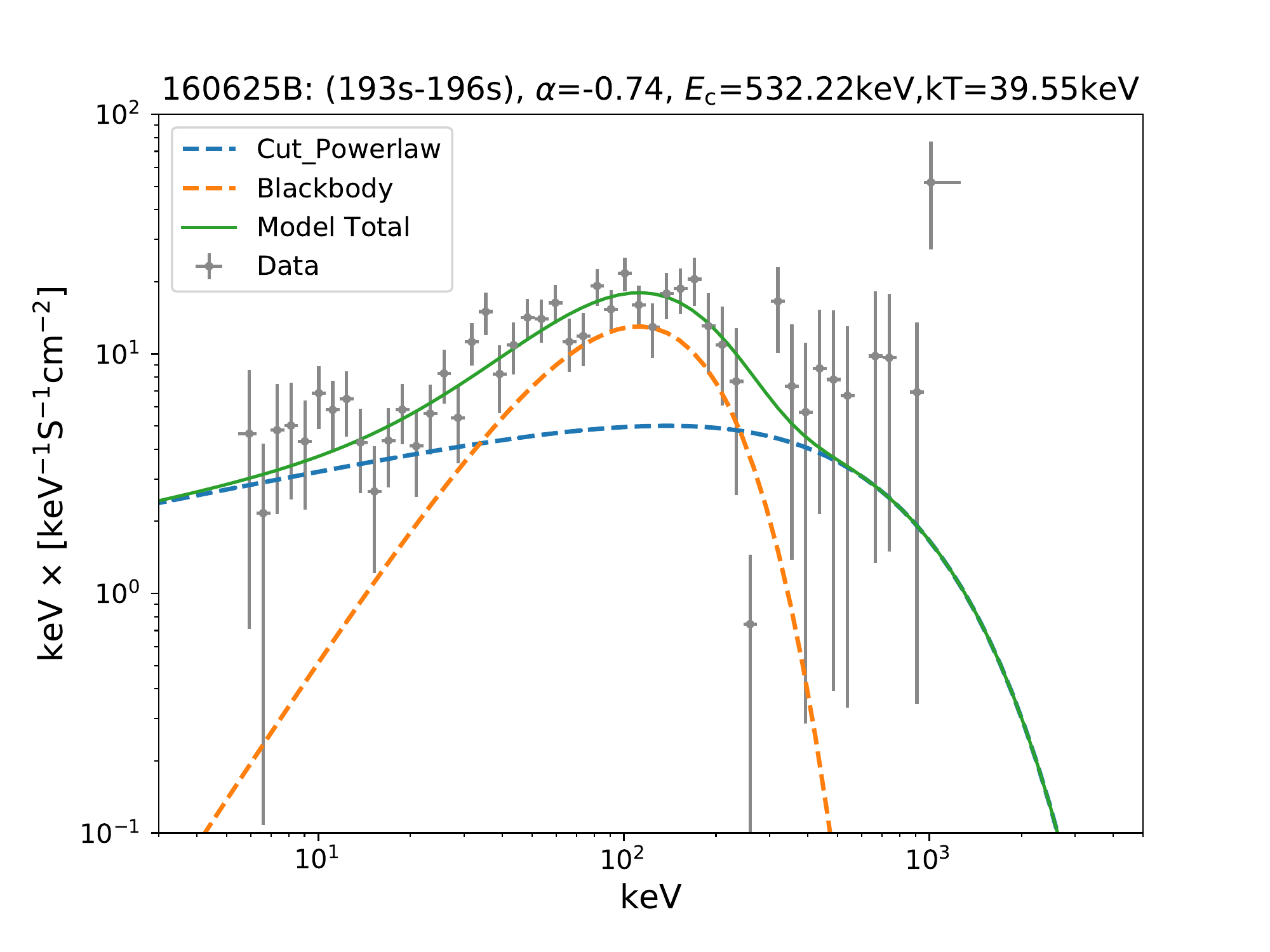}
\includegraphics[angle=0, scale=0.105]{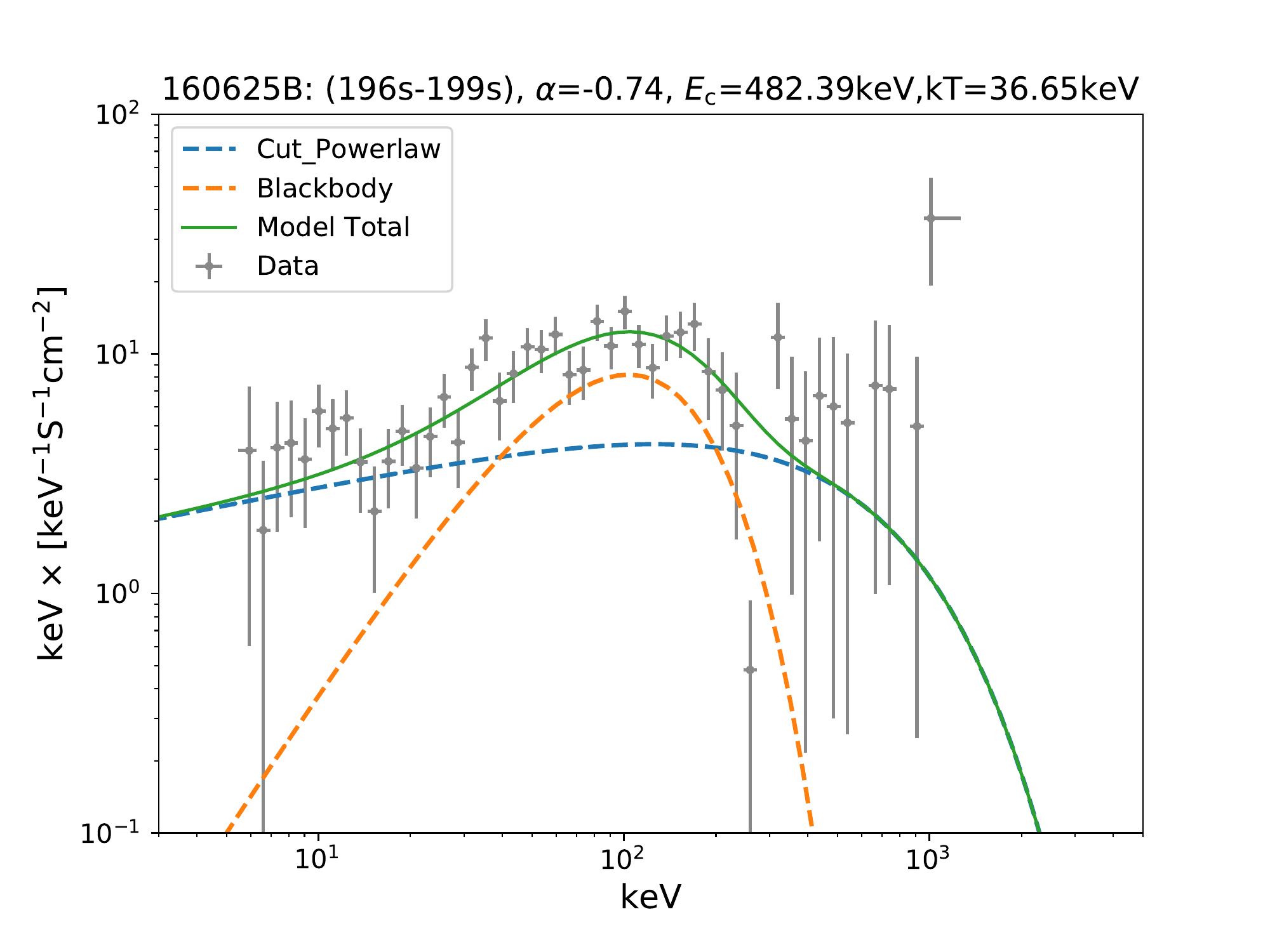}
\includegraphics[angle=0, scale=0.105]{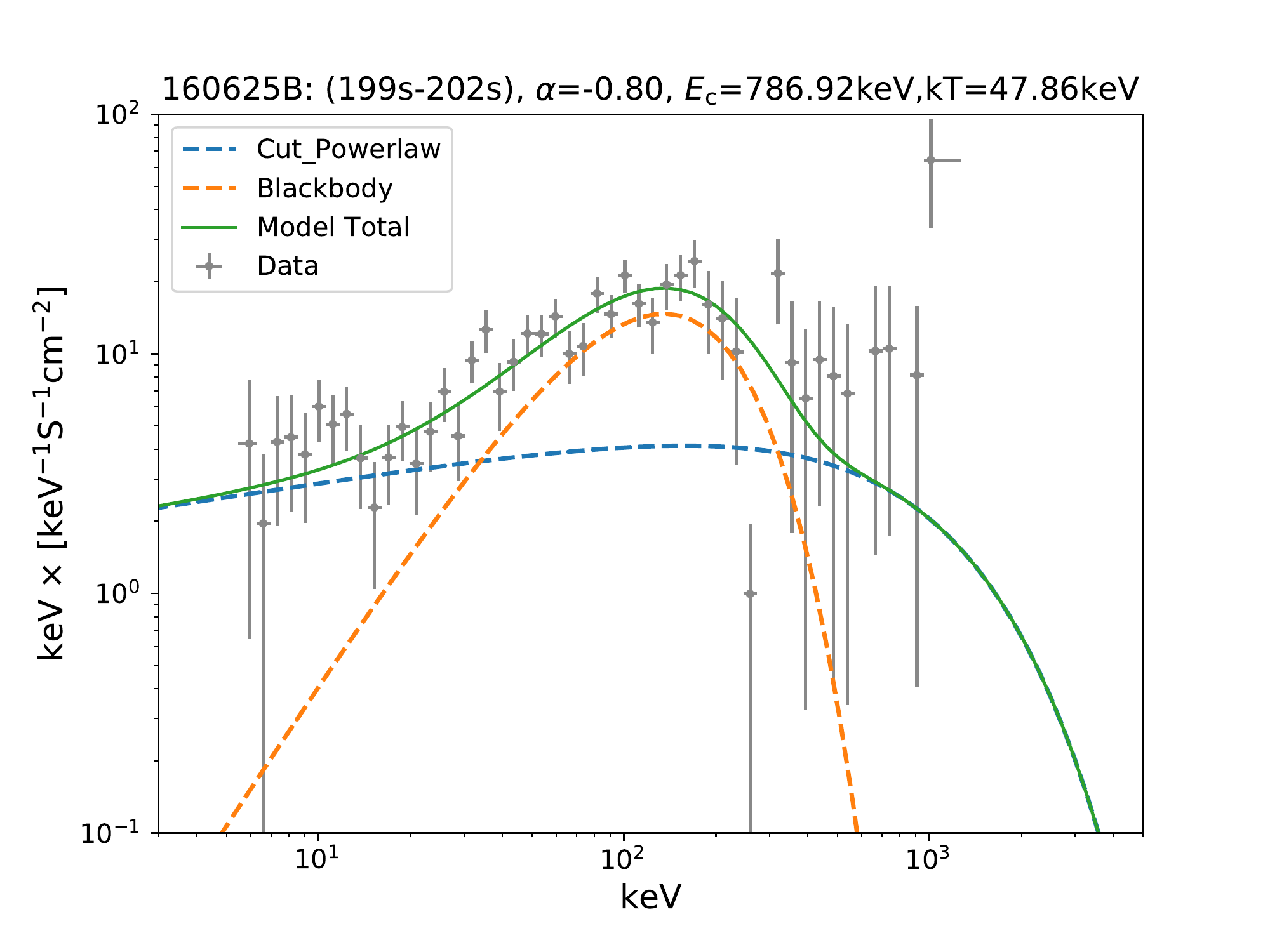}
\includegraphics[angle=0, scale=0.105]{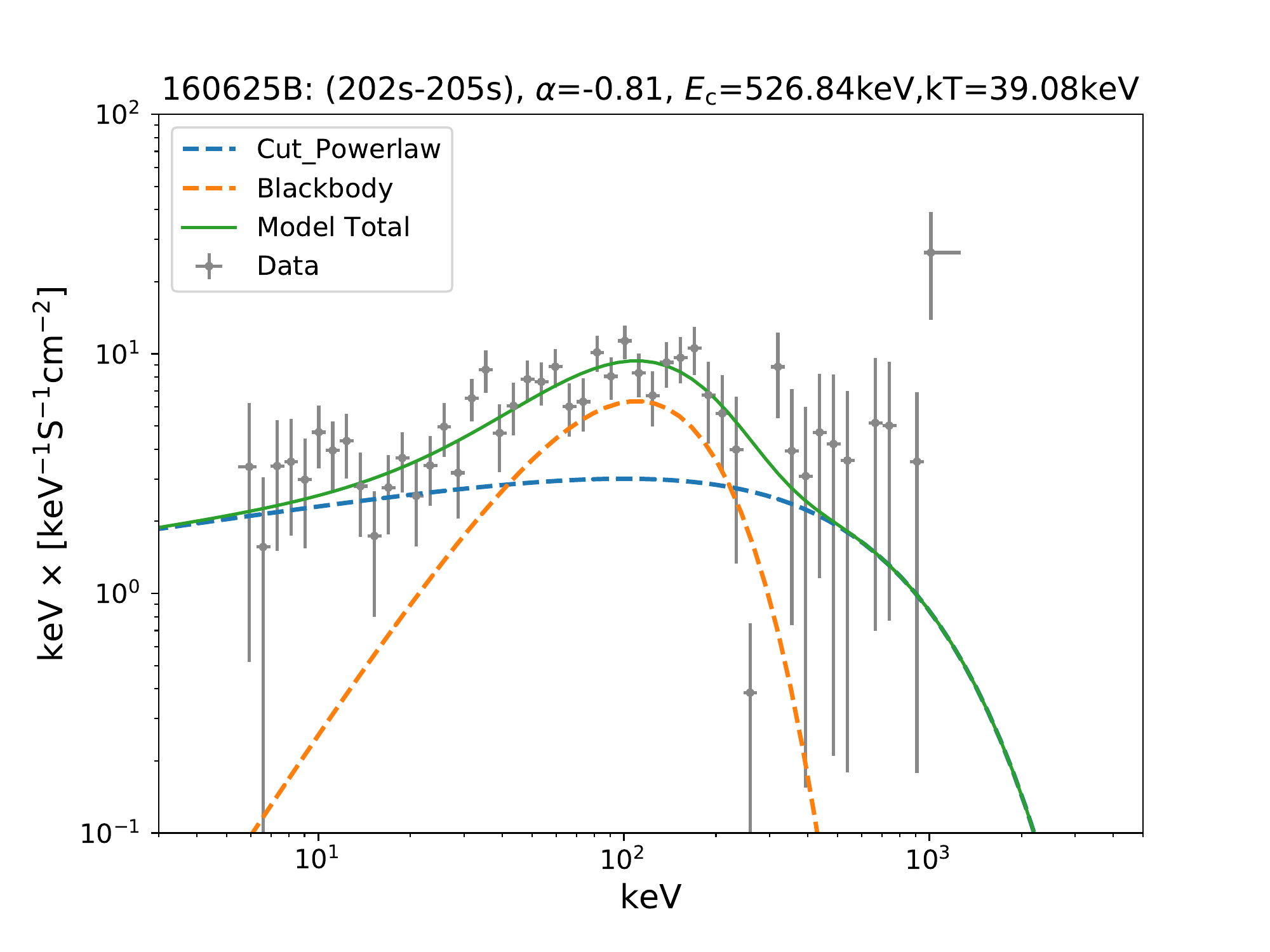}
\includegraphics[angle=0, scale=0.105]{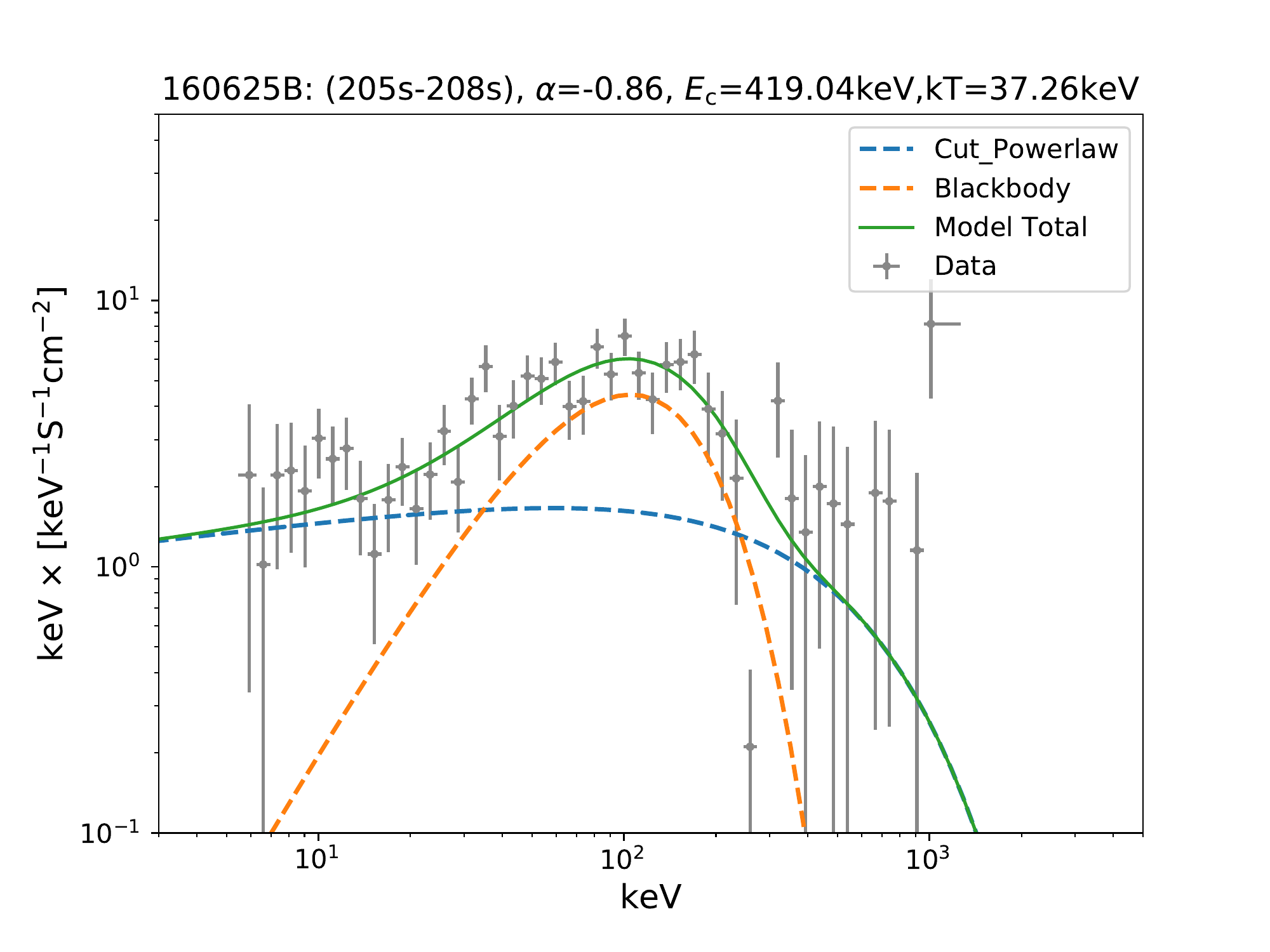}
\includegraphics[angle=0, scale=0.105]{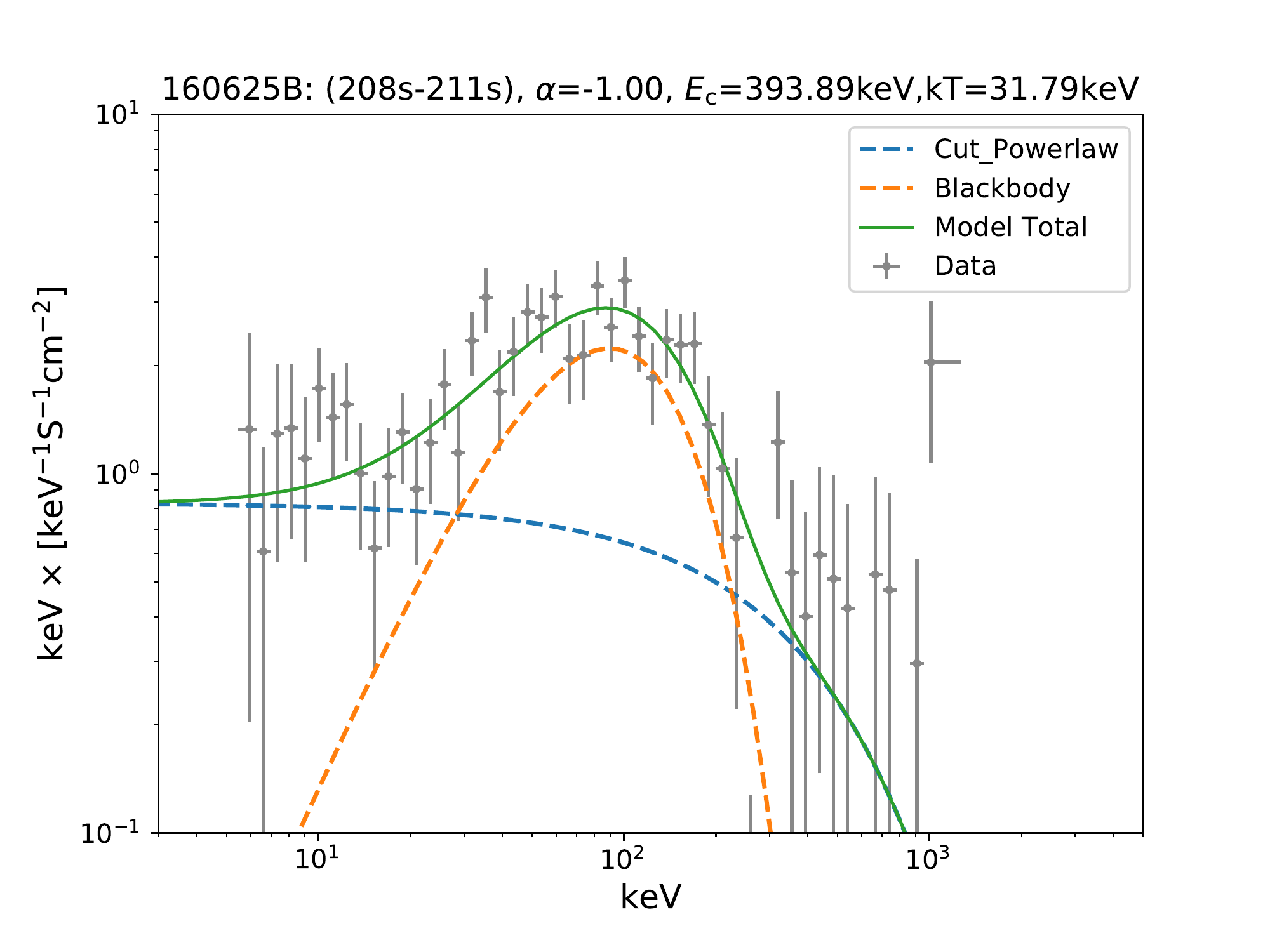}\\
\includegraphics[angle=0, scale=0.05]{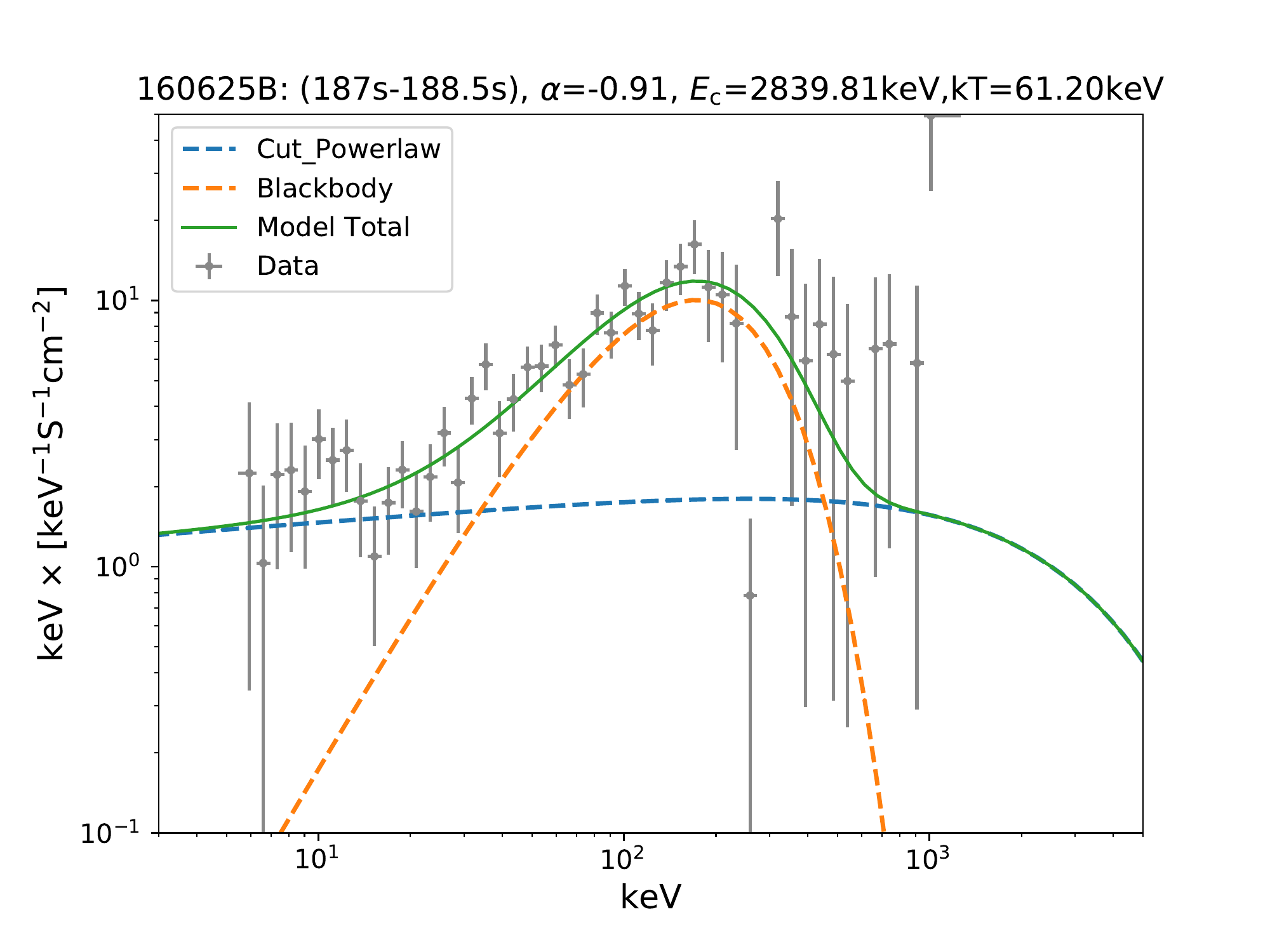}
\includegraphics[angle=0, scale=0.05]{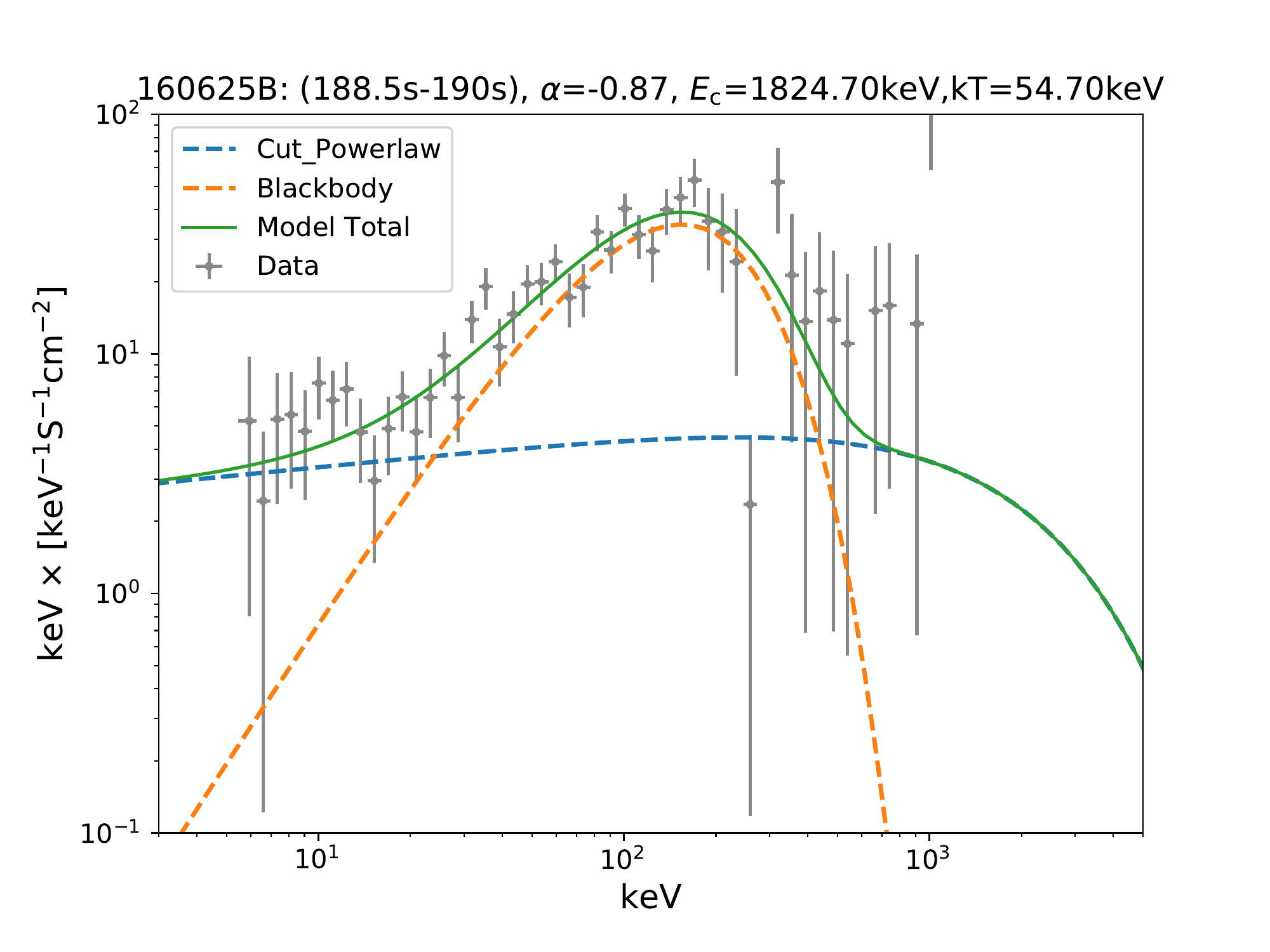}
\includegraphics[angle=0, scale=0.05]{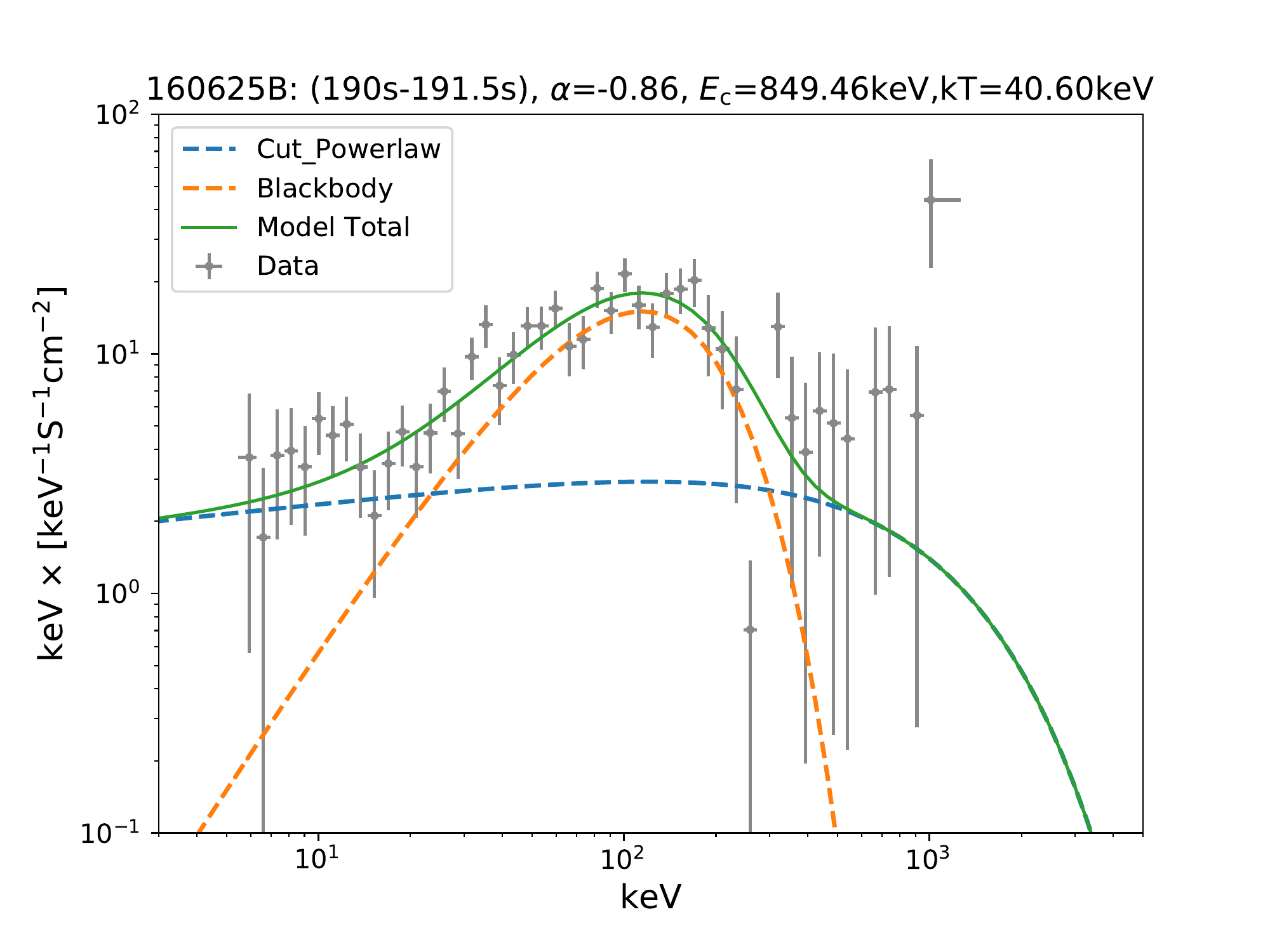}
\includegraphics[angle=0, scale=0.05]{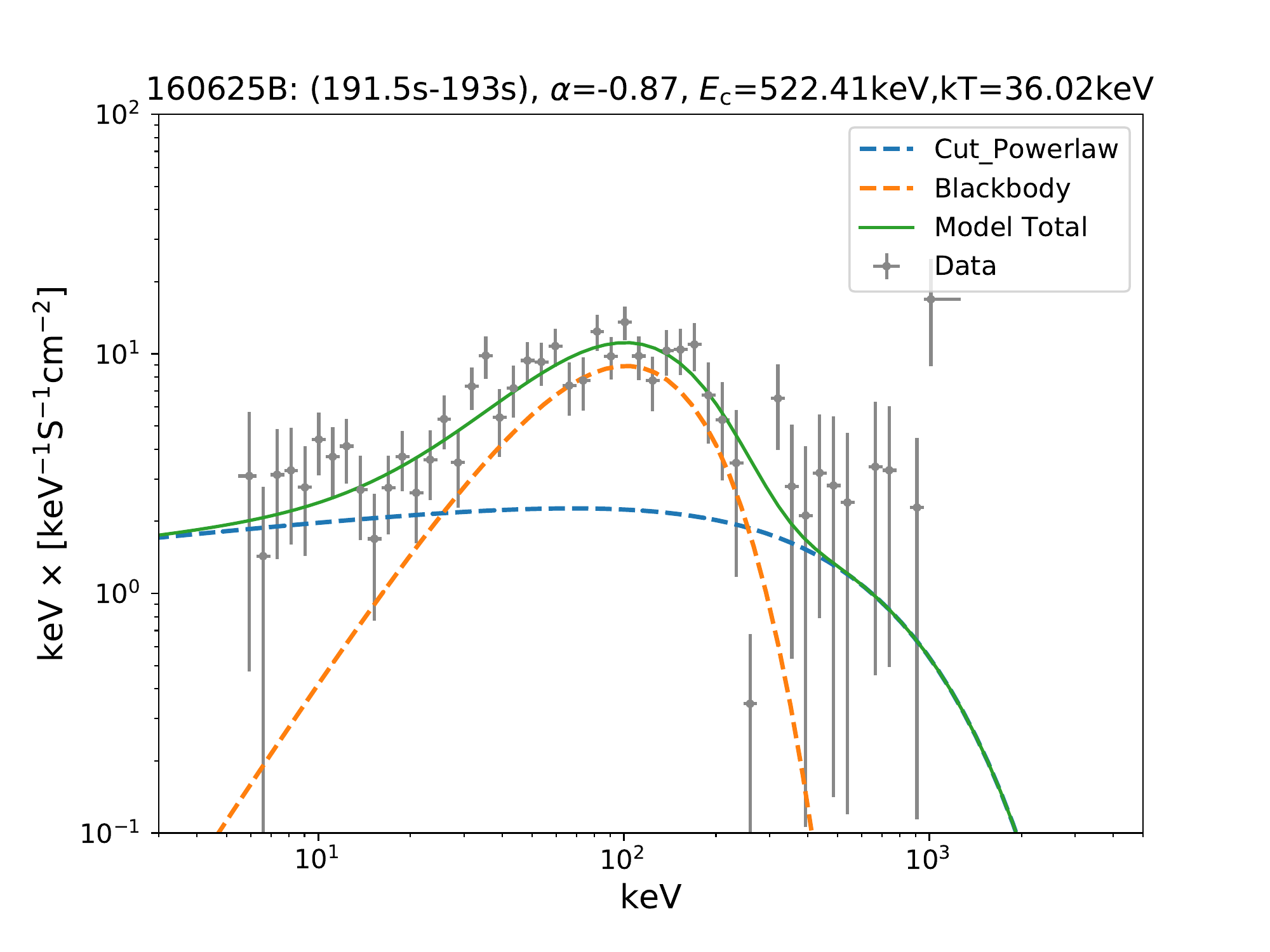}
\includegraphics[angle=0, scale=0.05]{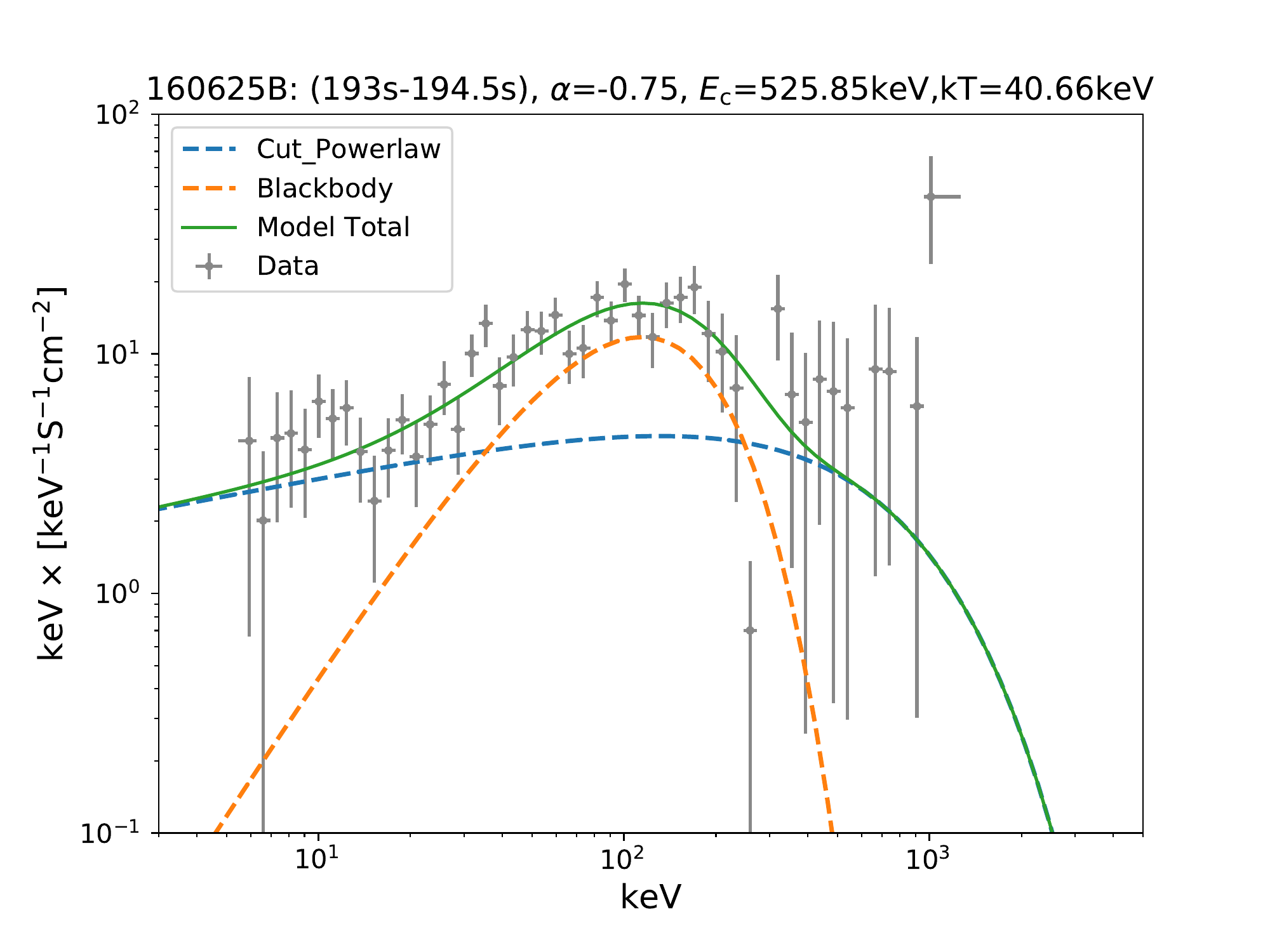}
\includegraphics[angle=0, scale=0.05]{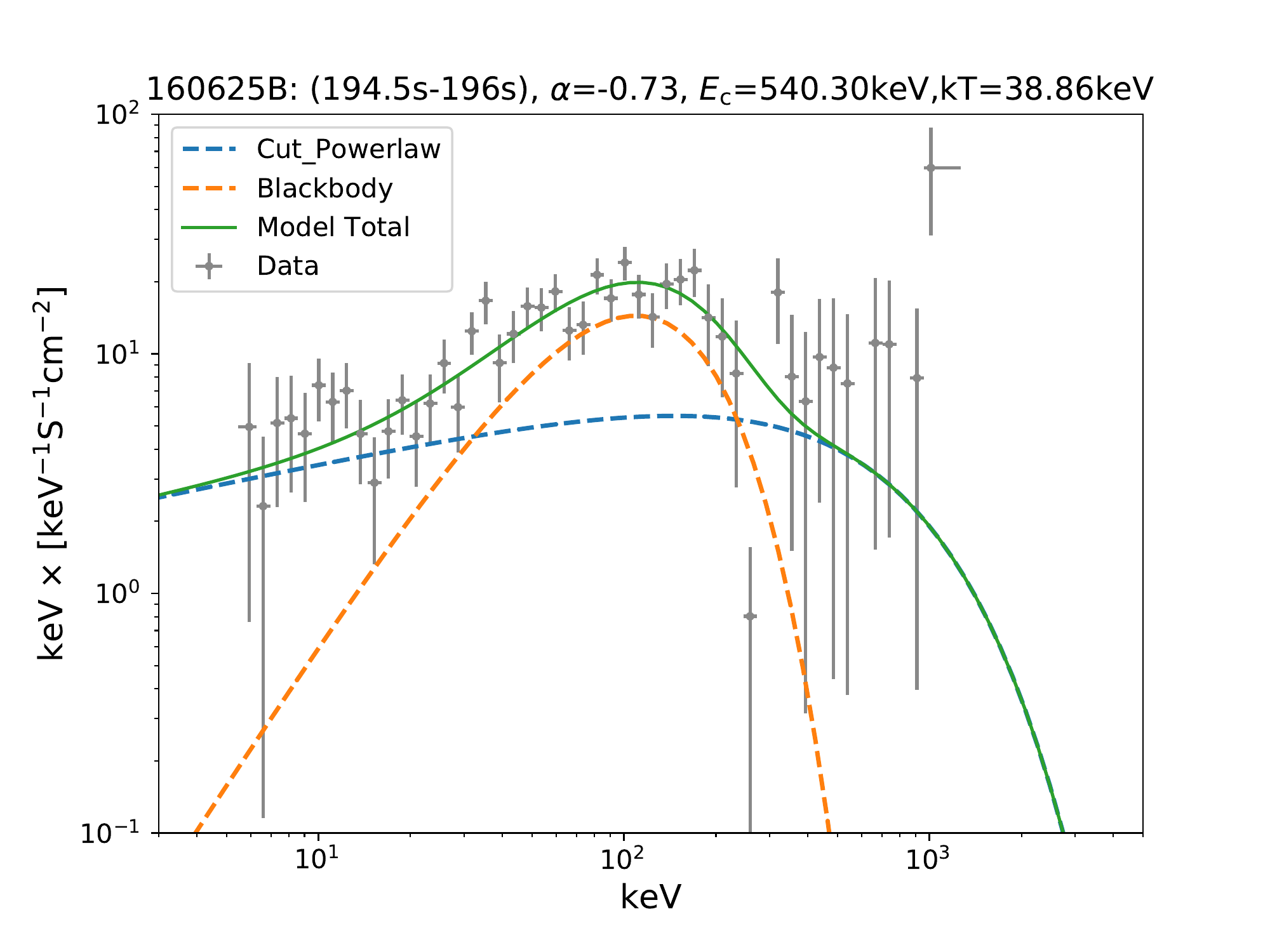}
\includegraphics[angle=0, scale=0.05]{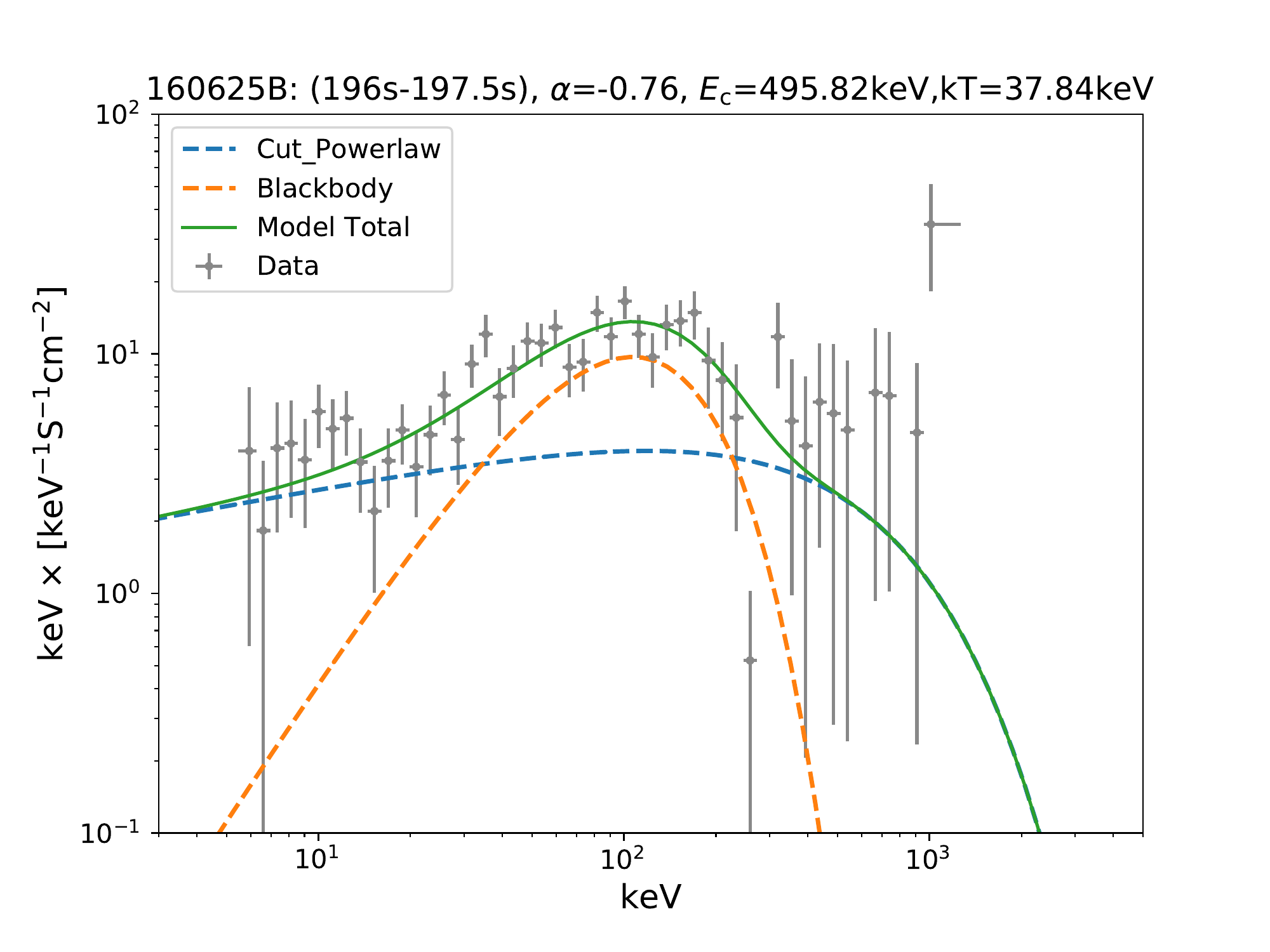}
\includegraphics[angle=0, scale=0.05]{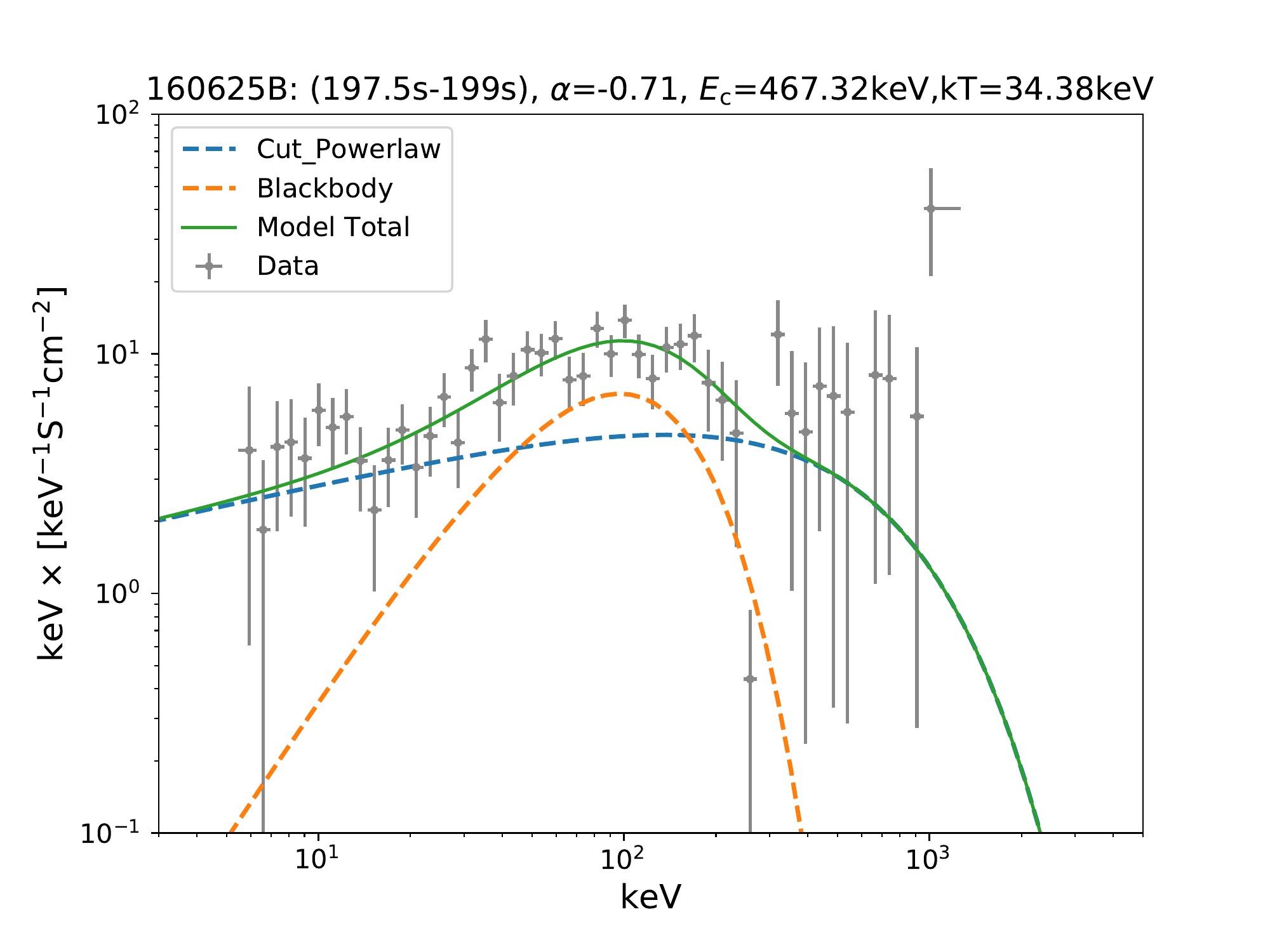}
\includegraphics[angle=0, scale=0.05]{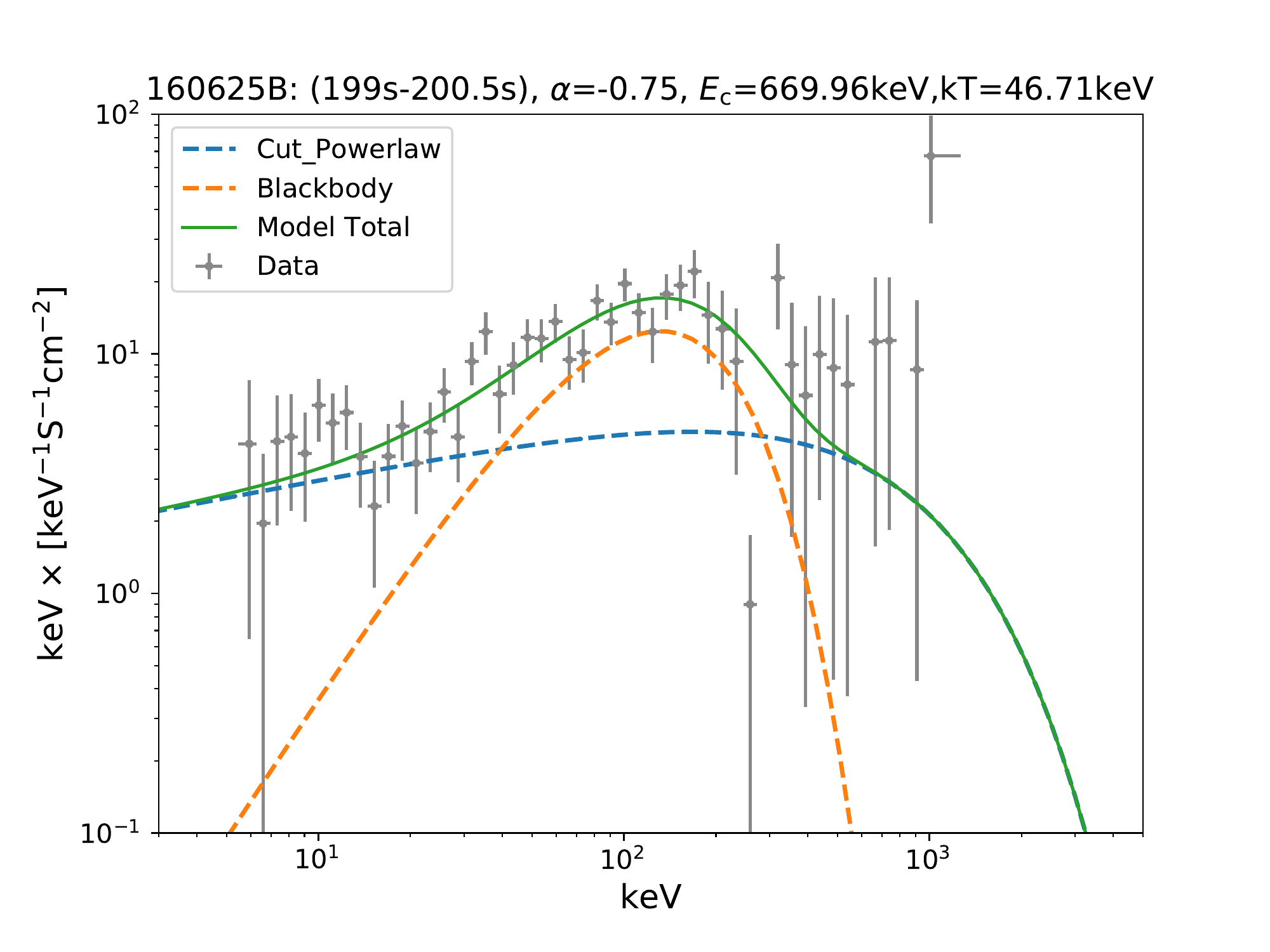}
\includegraphics[angle=0, scale=0.05]{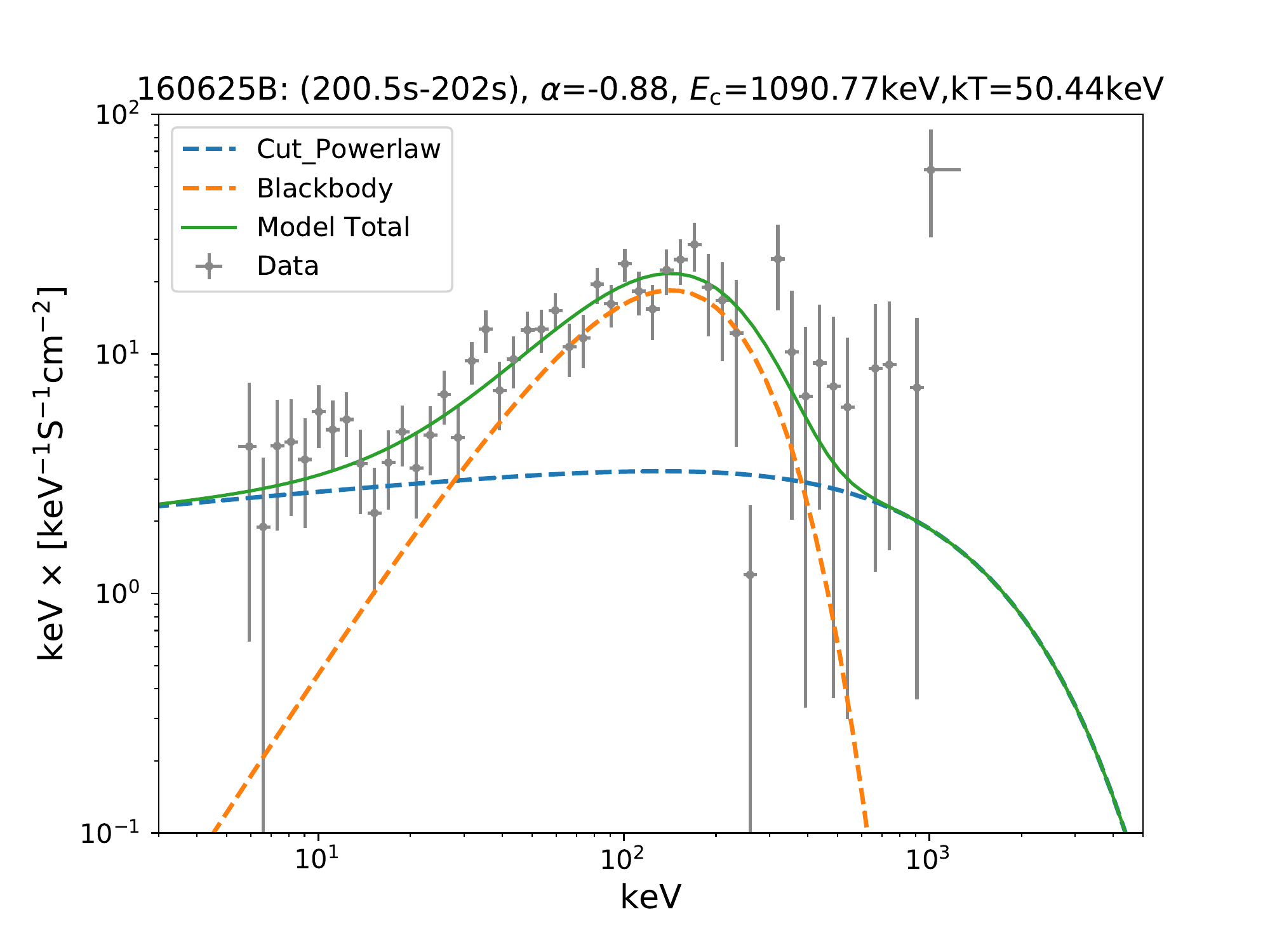}
\includegraphics[angle=0, scale=0.05]{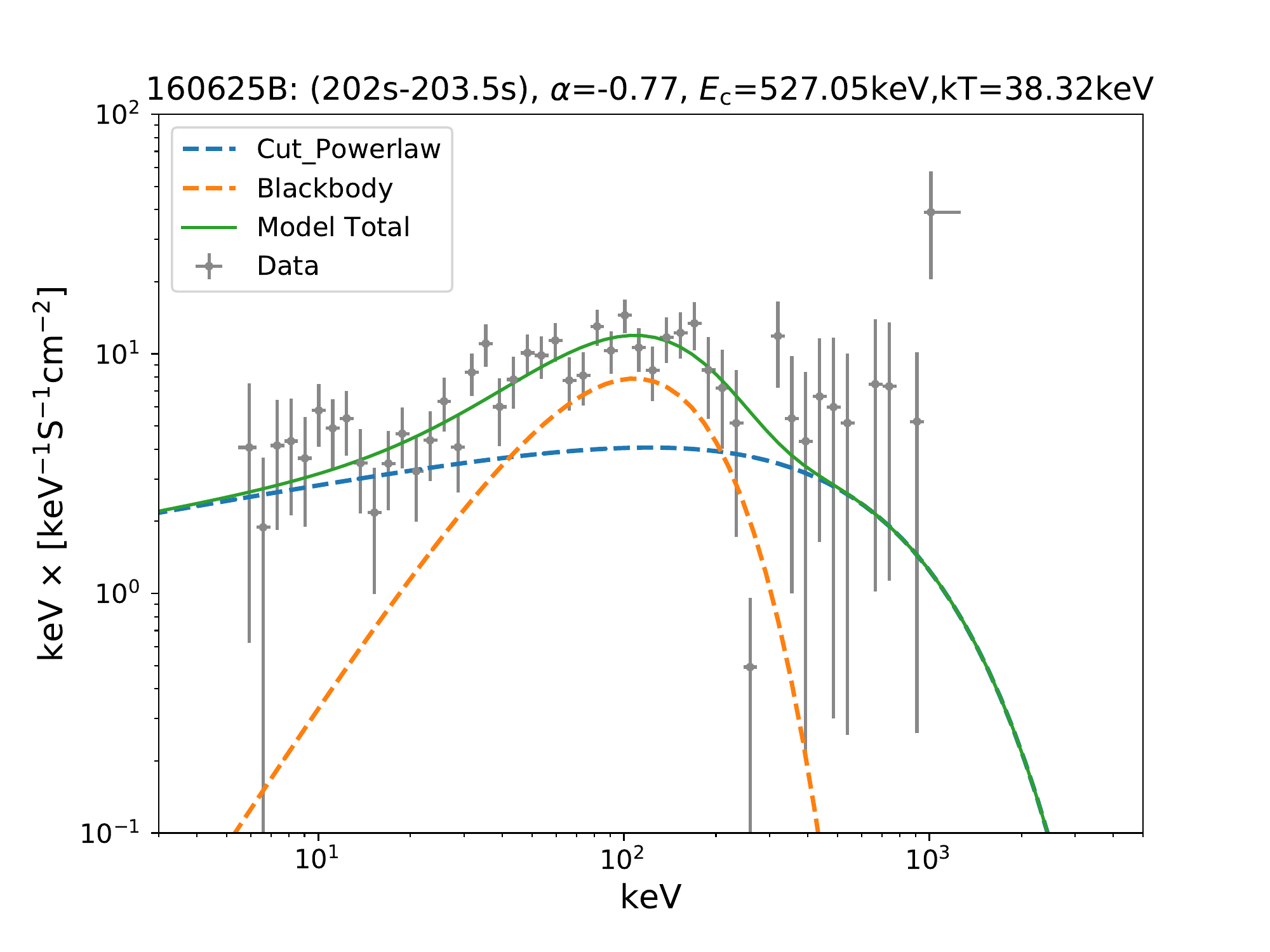}
\includegraphics[angle=0, scale=0.05]{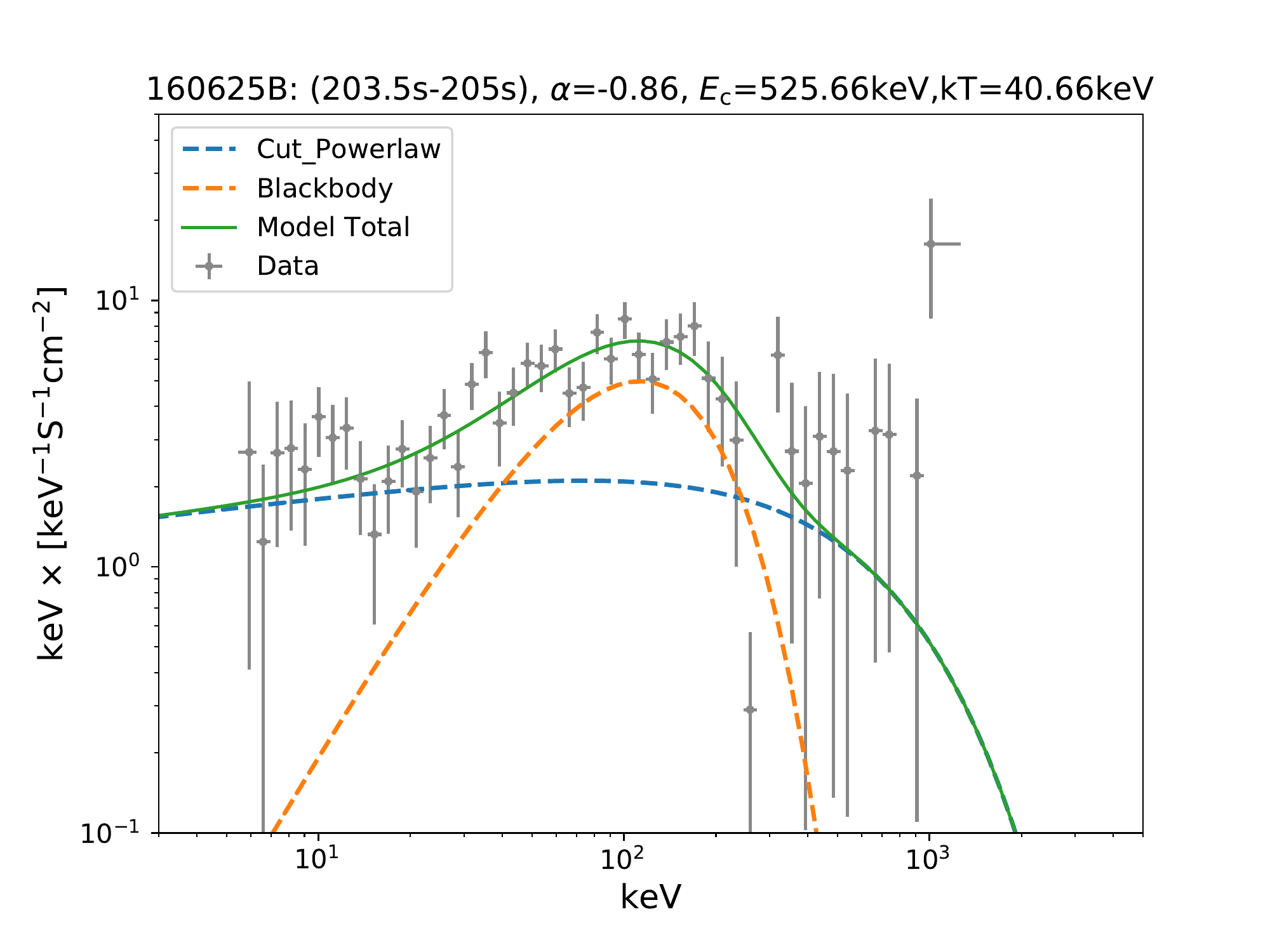}
\includegraphics[angle=0, scale=0.05]{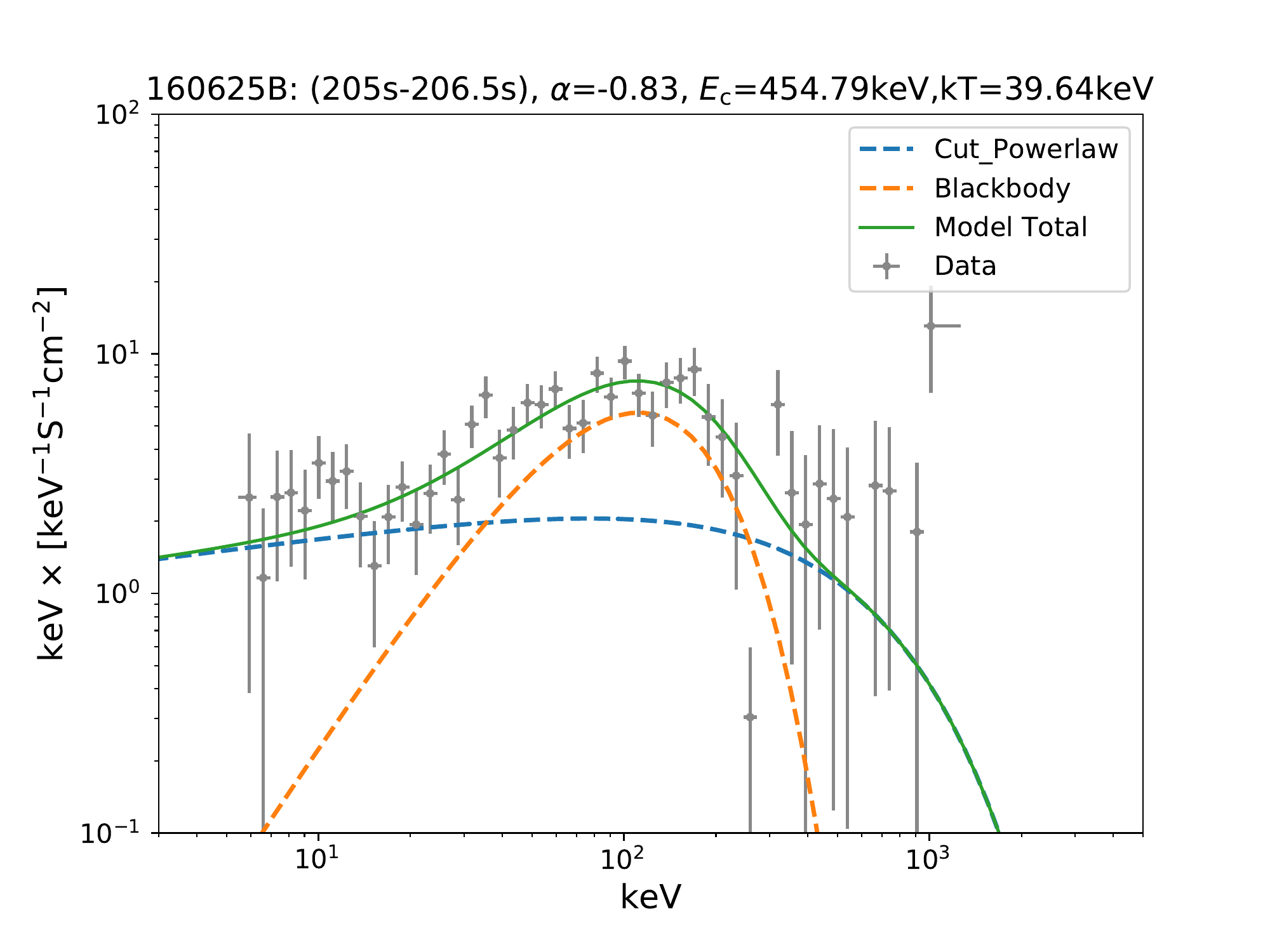}
\includegraphics[angle=0, scale=0.05]{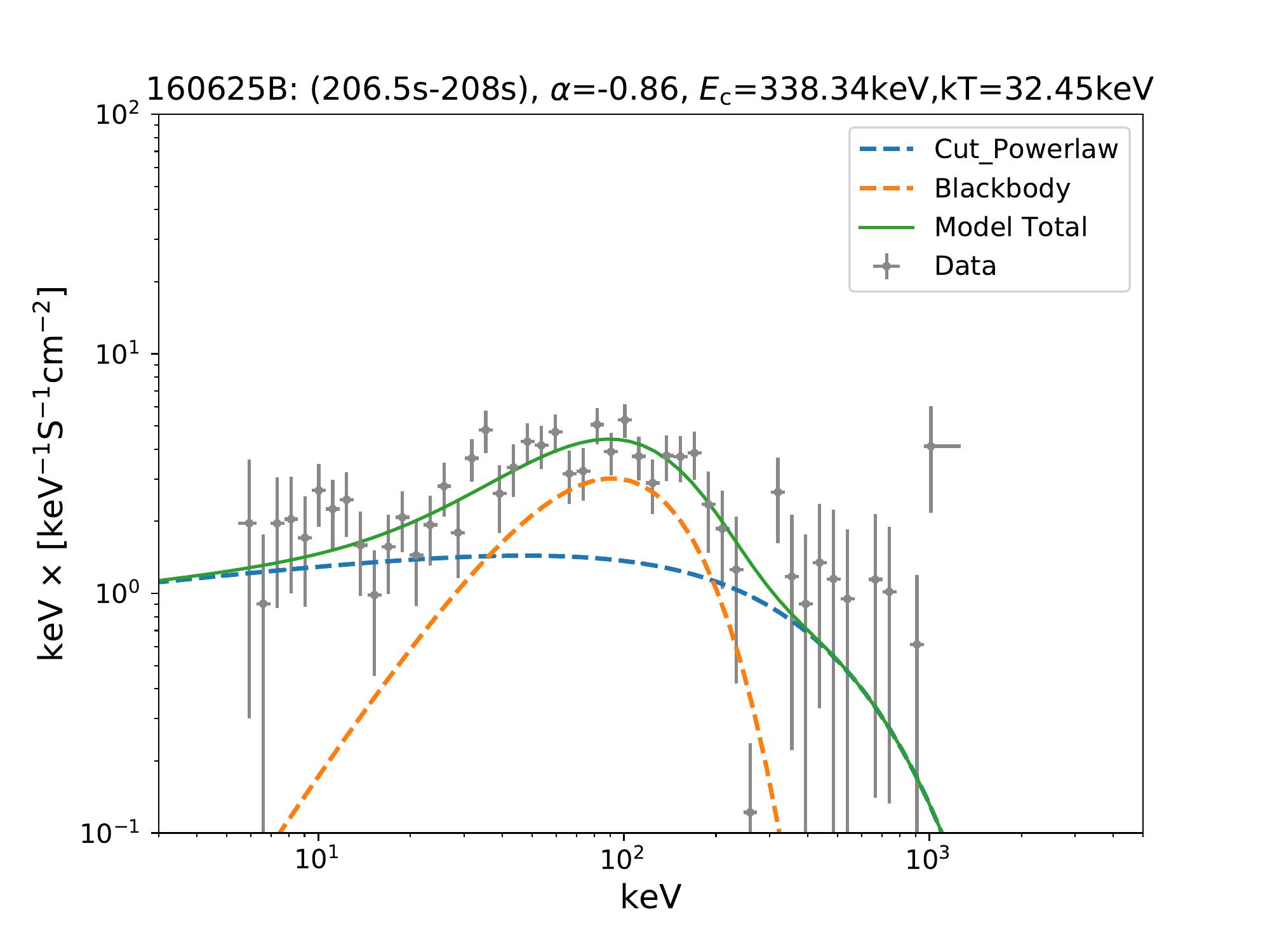}
\includegraphics[angle=0, scale=0.05]{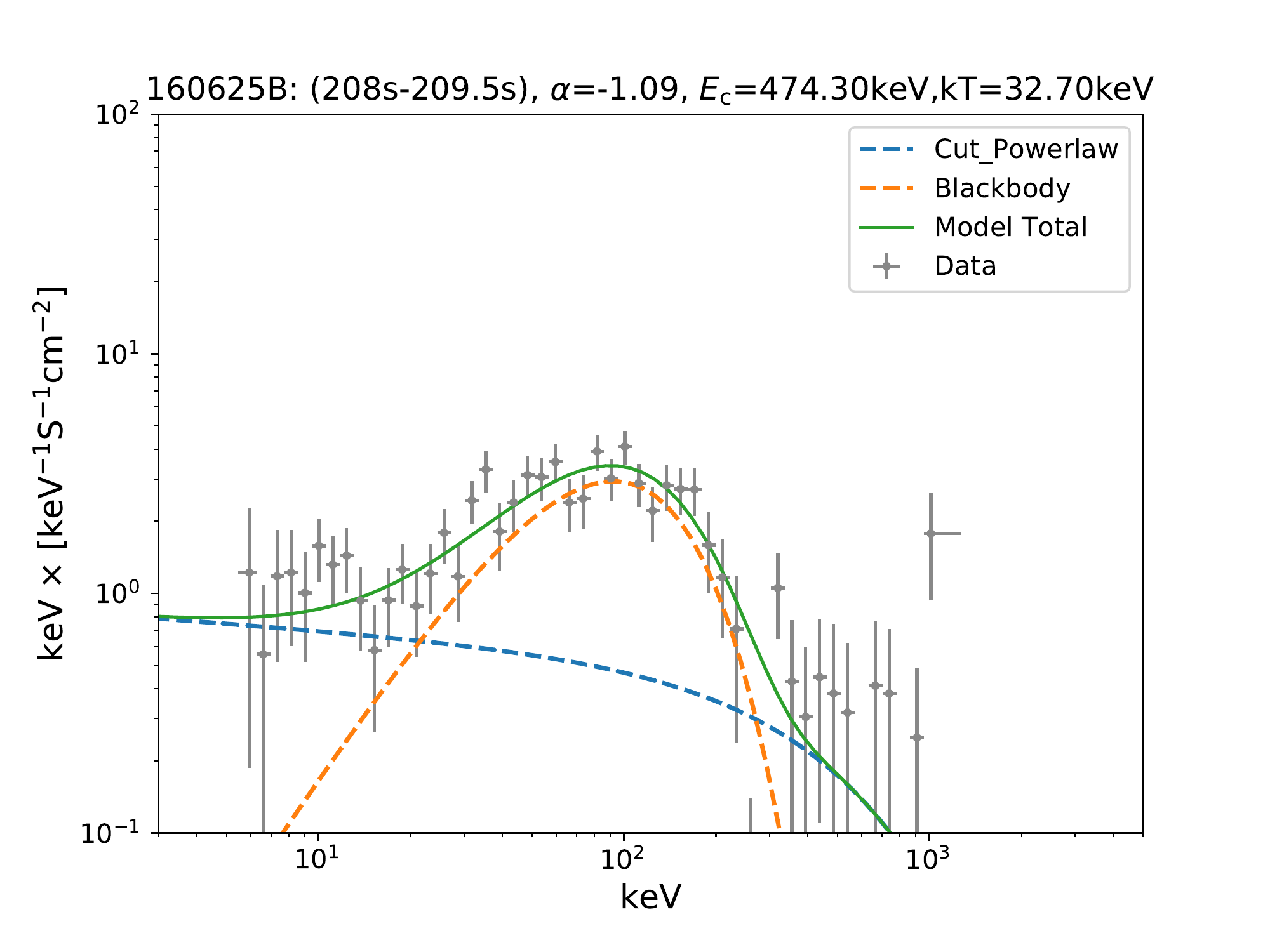}
\includegraphics[angle=0, scale=0.05]{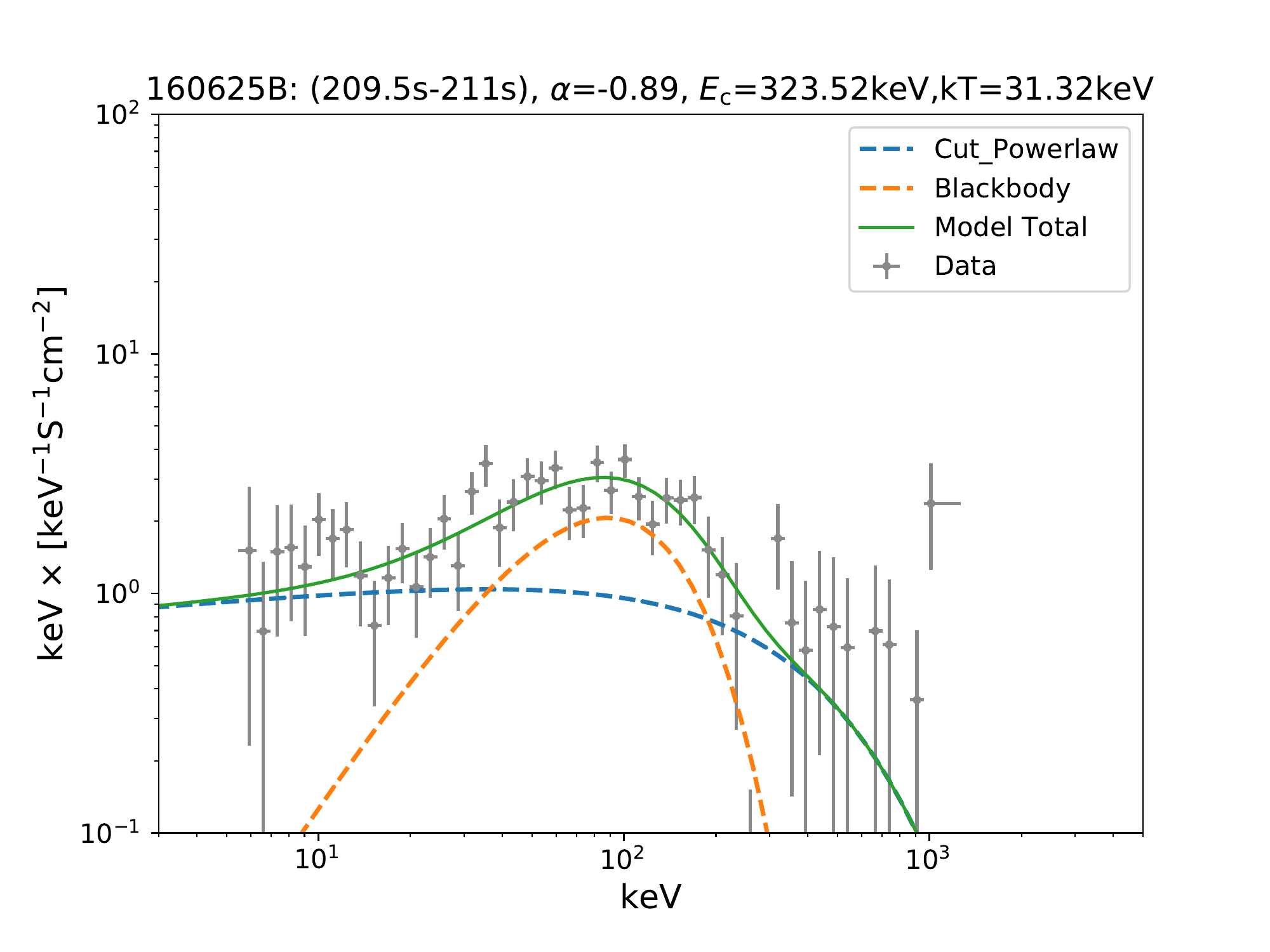}
\caption{Time-resolved spectral analysis of GRB 160625B. All the layers have the same time coverage, from $\approx 187$~s ($t_{\rm rf} = 77.72$~s) to $\approx 211$~s ($t_{\rm rf} =87.70$~s), but with different time divisions: one interval (top layer), two equal parts (second layer), four equal parts (third layer), eight equal parts (fourth layer), and sixteen equal parts (bottom layer), respectively. The results of spectral analysis including time duration, temperature and cutoff energy are obtained in the observed frame, as shown in this figure. We have converted to have their corresponding value in the rest-frame, see Table \ref{tab:160625B}, where rest-frame time in column 2, and rest-frame temperature in column 6.}
\label{fig:spectra_resolved_160625B}
\end{figure*}

--- \textit{UPE phase}. 
Similarly to GRB 190114C, we also find a self-similarity in the UPE phase for GRB 160625B after carrying out the detailed time-resolved spectral analysis, with a cutoff powerlaw + blackbody (CPL+BB) model, for five successive iteration process on shorter and shorter time scales (expressed in the laboratory and in the rest frame).
For the first iteration, Fig.~\ref{fig:spectra_resolved_160625B} (first layer) shows the best-fit of the spectrum of the UPE entire duration from $t_{\rm rf}=77.72$~s to $t_{\rm rf}=87.70$~s.

We then divide the rest-frame time interval in half and perform again the same spectral analysis for the two intervals, each of $4.99$s, namely [77.72s-82.71s] and [82.71s-87.70s], obtaining the results shown in Fig.~\ref{fig:spectra_resolved_160625B} (second layer). In the third iteration, we divide each of these half intervals again in half. We continue this procedure up to five iterations, i.e up to dividing the UPE in $16$ time sub-intervals. 
For each iterative step, we give the duration and the spectral parameters of CPL+BB model, including: the low-energy photon index $\alpha$, the peak energy $E_{\rm c}$, the BB temperature $k T$ ($k$ is the Stefan-Boltzmann constant), the model comparison parameter (DIC), the BB flux, the total flux, the BB to total flux ratio, and the total energy. The results are summarized in Fig.~\ref{fig:spectra_resolved_160625B} and Table~\ref{tab:160625B}, which confirm the validity, also in GRB 160625B, of the self-similar structure first discovered in GRB 190114C. 

\begin{figure*}[h!]
\centering
\includegraphics[width=0.49\hsize,clip]{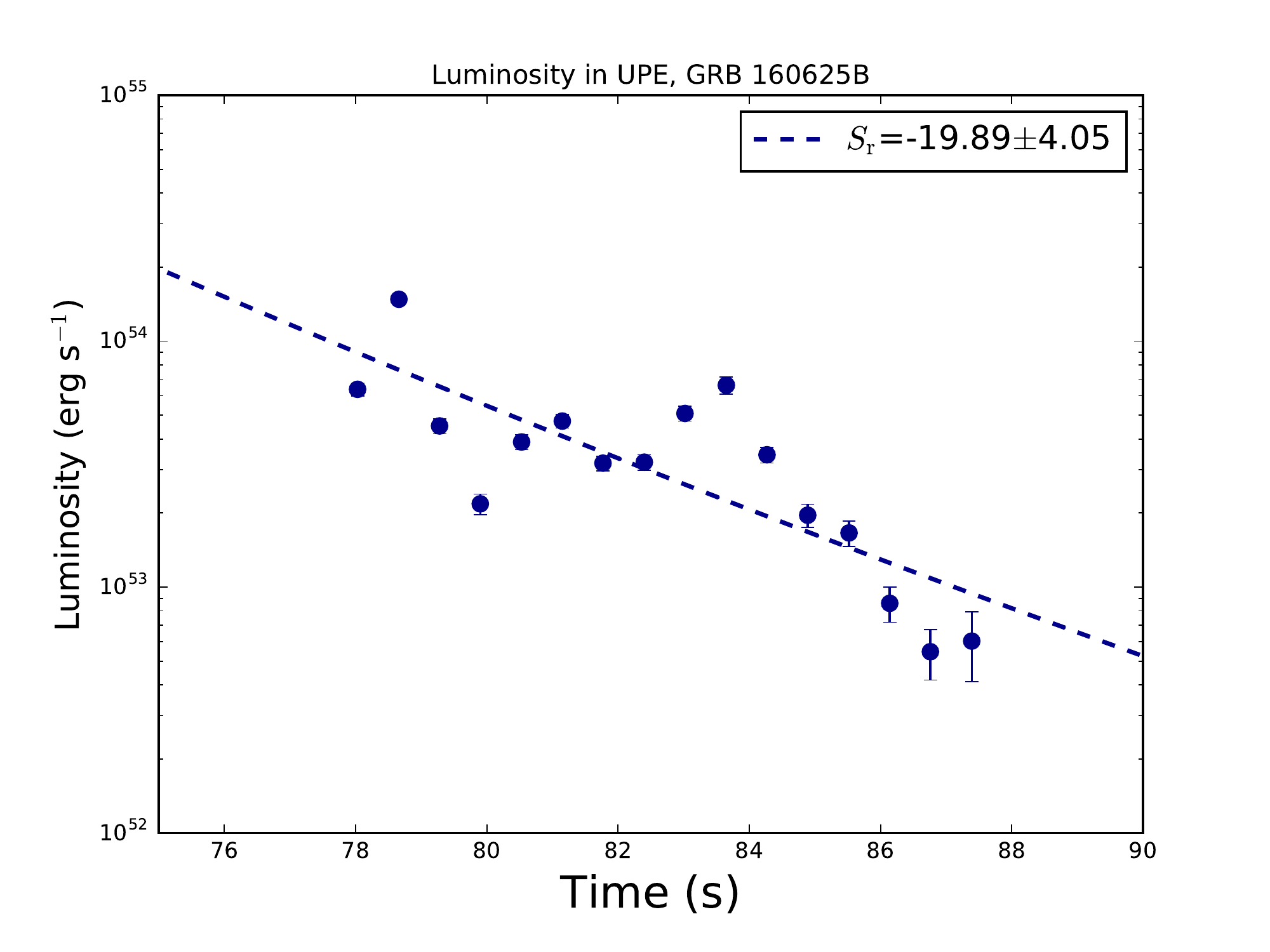}
\includegraphics[width=0.49\hsize,clip]{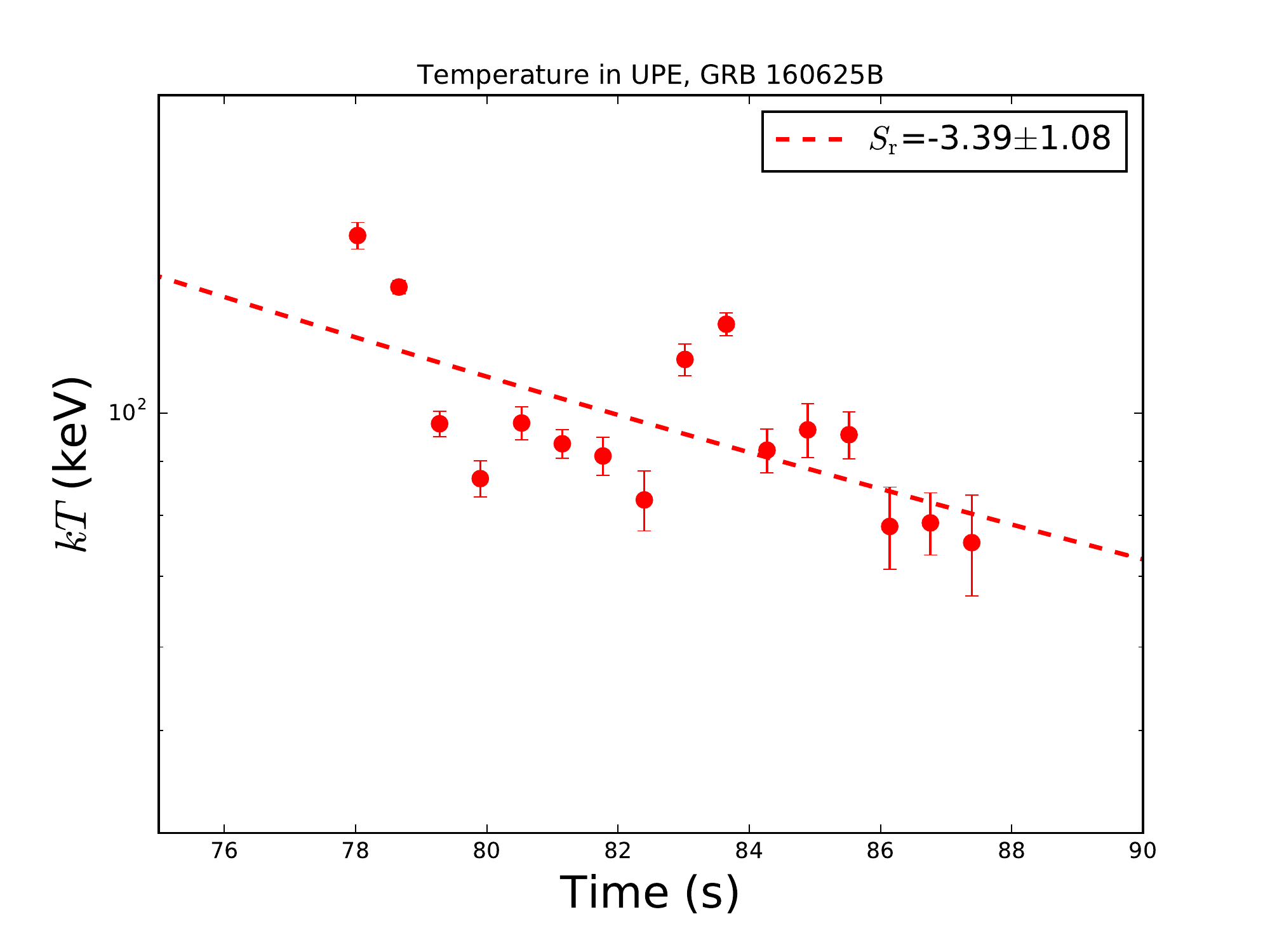}
\caption{\textbf{Left}: GRB 160625B light-curve of the UPE derived from the fifth iteration with 16 sub-intervals. The values of the best-fit parameters from Table~\ref{tab:160625B} are used to apply the k-correction and plot the rest-frame luminosity as a function of rest-frame time. The power-law index of the luminosity is at $-19.89 \pm 4.05$. For more information about GeV luminosity behavior see \citep{2019arXiv190107505W} and the companion paper (Ruffini, Moradi et al, 2019, submitted). \textbf{Right}: Corresponding rest-frame temperature of the UPE as a function of the rest-frame time.}
\label{iterativ}
\end{figure*}

Figure~\ref{iterativ} shows the luminosity of the {\it Fermi}-GBM as a function of the rest-frame time, derived from the fifth iteration (see Table~\ref{tab:160625B}). We also show the corresponding time-evolution of the rest-frame temperature.

--- \textit{Cavity}. Figure~\ref{fig:joinedGRB160625B} (upper right panel) shows the spectrum of cavity for GRB 160625B, from $t_{\rm rf}=87.70$~s to $t_{\rm rf}=92.27$~s, can be well fitted by a featureless CPL model, with the photon index $\alpha= 0.95$ and the cutoff energy is at $239$~keV.

--- \textit{GeV emission}. Figure~\ref{fig:joinedGRB160625B} (lower left panel) shows the luminosity of the GeV emission in the rest-frame as a function of the rest-frame time. 

--- \textit{Afterglow}. Figure~\ref{fig:joinedGRB160625B} (lower right panel) shows the (k-corrected) afterglow luminosity ({\it Swift}/XRT data) in the rest-frame as a function of rest-frame time, and obtained a best fit parameters with the power-law index of $1.319\pm 0.021$.

\begin{table*}[ht!]
\small\addtolength{\tabcolsep}{-2pt}
\caption{Results of the time-resolved spectral fits of GRB 160625B (CPL+BB model) from the $t_{\rm rf}=77.72$~s to $t_{\rm rf}=87.70$~s. This table reports: the time intervals both in rest-frame and observer frame, the significance ($S$) for each time interval, the power-law index, cut-off energy, temperature, $\Delta$DIC, BB flux, total flux, the BB to total flux ratio, $F_{\rm BB}/F_{\rm tot}$ and finally the isotropic energy. To select the best model from two different given models, we adopt the deviance information criterion (DIC), defined as DIC=-2log[$p$(data$\mid\hat{\theta}$)]+2$p_{\rm DIC}$, where $\hat{\theta}$ is the posterior mean of the parameters, and $p_{\rm DIC}$ is the effective number of parameters. The preferred model is the model with the lowest DIC score. Here we define $\Delta$DIC=(CPL+BB)-CPL, if $\Delta$DIC is negative, indicating the CPL+BB is better. After comparing the DIC, we find the CPL+BB model is the preferred model than the CPL and other model. The $\Delta$DIC scores are reported in column 6.}  
\label{tab:160625B}
\centering                         
\begin{tabular}{ccccccccccc}       
\hline\hline                  
$t_{1}$$\sim$$t_{2}$&$t_{rf,1}$$\sim$$t_{rf,2}$&$S$&$\alpha$&$E_{\rm c}$&$kT$&$\Delta$DIC&$F_{\rm BB}$&$F_{\rm tot}$&$F_{\rm ratio}$&$E_{\rm tot}$\\
\hline
(s)&(s)&&&(keV)&(keV)&&(10$^{-6}$)&(10$^{-6}$)&&(erg)\\
Obs&Rest-frame&&&&{Rest-frame}&&(erg~cm$^{-2}$~s$^{-1}$)&(erg~cm$^{-2}$~s$^{-1}$)\\ 
\hline                        
187.00$\sim$211.00&77.72$\sim$87.70&649.12&-0.83$^{+0.01}_{-0.01}$&707.6$^{+13.0}_{-12.9}$&42.9$^{+0.4}_{-0.4}$&-2840.2&3.13$^{+0.16}_{-0.15}$&35.50$^{+0.81}_{-0.87}$&0.09$^{+0.0}_{-0.0}$&4.53e+54\\
\hline
187.00$\sim$199.00&77.72$\sim$82.71&566.19&-0.84$^{+0.01}_{-0.01}$&861.1$^{+20.8}_{-20.9}$&44.4$^{+0.5}_{-0.5}$&-2789.1&4.67$^{+0.23}_{-0.24}$&48.44$^{+1.41}_{-1.40}$&0.10$^{+0.01}_{-0.01}$&3.09e+54\\
199.00$\sim$211.00&82.71$\sim$87.70&421.10&-0.83$^{+0.01}_{-0.01}$&597.4$^{+15.8}_{-15.9}$&41.6$^{+0.8}_{-0.8}$&-716.6&1.95$^{+0.18}_{-0.18}$&24.53$^{+0.83}_{-0.87}$&0.08$^{+0.01}_{-0.01}$&1.57e+54\\
\hline
187.00$\sim$193.00&77.72$\sim$80.22&426.56&-0.94$^{+0.01}_{-0.01}$&1702.4$^{+42.5}_{-42.7}$&49.6$^{+0.5}_{-0.5}$&-2935.0&6.51$^{+0.35}_{-0.32}$&69.51$^{+1.80}_{-1.92}$&0.09$^{+0.01}_{-0.01}$&2.22e+54\\
193.00$\sim$199.00&80.22$\sim$82.71&421.75&-0.73$^{+0.01}_{-0.01}$&507.0$^{+12.2}_{-12.4}$&38.5$^{+0.8}_{-0.8}$&-784.9&2.95$^{+0.30}_{-0.28}$&37.46$^{+1.33}_{-1.31}$&0.08$^{+0.01}_{-0.01}$&1.19e+54\\
199.00$\sim$205.00&82.71$\sim$85.20&409.24&-0.80$^{+0.01}_{-0.01}$&657.7$^{+18.1}_{-18.6}$&44.2$^{+0.9}_{-0.9}$&-729.7&3.25$^{+0.35}_{-0.28}$&40.93$^{+1.57}_{-1.59}$&0.08$^{+0.01}_{-0.01}$&1.31e+54\\
205.00$\sim$211.00&85.20$\sim$87.70&205.28&-0.91$^{+0.02}_{-0.02}$&408.4$^{+25.3}_{-25.8}$&35.5$^{+1.4}_{-1.4}$&-105.6&0.82$^{+0.17}_{-0.15}$&9.08$^{+0.75}_{-0.68}$&0.09$^{+0.02}_{-0.02}$&2.90e+53\\
\hline
187.00$\sim$190.00&77.72$\sim$78.97&344.58&-0.89$^{+0.01}_{-0.01}$&2066.8$^{+50.1}_{-50.0}$&56.2$^{+0.7}_{-0.7}$&-2860.2&9.08$^{+0.63}_{-0.55}$&105.00$^{+3.03}_{-3.29}$&0.09$^{+0.01}_{-0.01}$&1.67e+54\\
190.00$\sim$193.00&78.97$\sim$80.22&282.28&-0.86$^{+0.01}_{-0.01}$&681.6$^{+31.2}_{-31.7}$&38.2$^{+0.8}_{-0.8}$&-603.9&3.30$^{+0.37}_{-0.35}$&32.41$^{+1.93}_{-1.63}$&0.10$^{+0.01}_{-0.01}$&5.17e+53\\
193.00$\sim$196.00&80.22$\sim$81.46&333.07&-0.74$^{+0.01}_{-0.01}$&532.2$^{+17.1}_{-17.0}$&39.5$^{+0.9}_{-1.0}$&-546.1&3.76$^{+0.51}_{-0.42}$&43.09$^{+2.07}_{-1.84}$&0.09$^{+0.01}_{-0.01}$&6.87e+53\\
196.00$\sim$199.00&81.46$\sim$82.71&287.45&-0.74$^{+0.01}_{-0.01}$&482.4$^{+16.9}_{-16.5}$&36.6$^{+1.3}_{-1.3}$&-287.5&2.17$^{+0.45}_{-0.34}$&32.03$^{+1.67}_{-1.50}$&0.07$^{+0.01}_{-0.01}$&5.11e+53\\
199.00$\sim$202.00&82.71$\sim$83.96&341.22&-0.80$^{+0.01}_{-0.01}$&786.9$^{+29.1}_{-29.2}$&47.9$^{+1.0}_{-1.0}$&-661.0&5.16$^{+0.56}_{-0.50}$&56.34$^{+3.11}_{-2.55}$&0.09$^{+0.01}_{-0.01}$&8.99e+53\\
202.00$\sim$205.00&83.96$\sim$85.20&258.65&-0.81$^{+0.02}_{-0.02}$&526.8$^{+21.7}_{-21.7}$&39.1$^{+1.5}_{-1.5}$&-181.9&1.79$^{+0.34}_{-0.31}$&26.95$^{+1.52}_{-1.45}$&0.07$^{+0.01}_{-0.01}$&4.30e+53\\
205.00$\sim$208.00&85.20$\sim$86.45&182.22&-0.86$^{+0.03}_{-0.03}$&419.0$^{+28.9}_{-28.9}$&37.3$^{+1.7}_{-1.6}$&-90.1&1.20$^{+0.27}_{-0.27}$&12.55$^{+1.16}_{-1.06}$&0.10$^{+0.02}_{-0.02}$&2.00e+53\\
208.00$\sim$211.00&86.45$\sim$87.70&116.10&-1.00$^{+0.04}_{-0.04}$&393.9$^{+46.2}_{-47.4}$&31.8$^{+2.1}_{-2.1}$&-37.9&0.51$^{+0.19}_{-0.15}$&5.63$^{+0.84}_{-0.67}$&0.09$^{+0.04}_{-0.03}$&8.97e+52\\
\hline
187.00$\sim$188.50&77.72$\sim$78.35&147.15&-0.91$^{+0.01}_{-0.01}$&2839.8$^{+140.7}_{-141.4}$&61.2$^{+1.8}_{-1.8}$&-706.0&4.47$^{+0.74}_{-0.59}$&63.65$^{+3.89}_{-3.71}$&0.07$^{+0.01}_{-0.01}$&5.08e+53\\
188.50$\sim$190.00&78.35$\sim$78.97&354.91&-0.87$^{+0.01}_{-0.01}$&1824.7$^{+49.3}_{-49.4}$&54.7$^{+0.8}_{-0.8}$&-2291.1&13.77$^{+1.02}_{-0.93}$&147.60$^{+4.86}_{-5.17}$&0.09$^{+0.01}_{-0.01}$&1.18e+54\\
190.00$\sim$191.50&78.97$\sim$79.59&227.35&-0.86$^{+0.02}_{-0.02}$&849.5$^{+52.9}_{-53.5}$&40.6$^{+1.1}_{-1.1}$&-465.8&4.46$^{+0.63}_{-0.58}$&45.19$^{+3.19}_{-3.25}$&0.10$^{+0.02}_{-0.01}$&3.60e+53\\
191.50$\sim$193.00&79.59$\sim$80.22&181.28&-0.87$^{+0.03}_{-0.03}$&522.4$^{+37.1}_{-37.6}$&36.0$^{+1.4}_{-1.4}$&-178.9&2.34$^{+0.48}_{-0.42}$&21.81$^{+2.08}_{-1.89}$&0.11$^{+0.02}_{-0.02}$&1.74e+53\\
193.00$\sim$194.50&80.22$\sim$80.84&229.41&-0.75$^{+0.02}_{-0.02}$&525.9$^{+25.5}_{-25.2}$&40.7$^{+1.5}_{-1.5}$&-223.5&3.48$^{+0.69}_{-0.62}$&38.84$^{+2.72}_{-2.41}$&0.09$^{+0.02}_{-0.02}$&3.10e+53\\
194.50$\sim$196.00&80.84$\sim$81.46&254.52&-0.73$^{+0.02}_{-0.02}$&540.3$^{+23.0}_{-23.0}$&38.9$^{+1.2}_{-1.2}$&-338.7&4.12$^{+0.67}_{-0.58}$&47.26$^{+3.01}_{-2.65}$&0.09$^{+0.02}_{-0.01}$&3.77e+53\\
196.00$\sim$197.50&81.46$\sim$82.09&212.08&-0.76$^{+0.02}_{-0.02}$&495.8$^{+24.6}_{-24.4}$&37.8$^{+1.6}_{-1.6}$&-188.9&2.65$^{+0.55}_{-0.49}$&31.87$^{+2.17}_{-2.17}$&0.08$^{+0.02}_{-0.02}$&2.54e+53\\
197.50$\sim$199.00&82.09$\sim$82.71&205.41&-0.71$^{+0.02}_{-0.02}$&467.3$^{+22.3}_{-22.4}$&34.4$^{+2.2}_{-2.2}$&-114.0&1.72$^{+0.60}_{-0.51}$&32.16$^{+2.31}_{-2.18}$&0.05$^{+0.02}_{-0.02}$&2.56e+53\\
199.00$\sim$200.50&82.71$\sim$83.33&239.62&-0.75$^{+0.02}_{-0.02}$&670.0$^{+31.7}_{-31.3}$&46.7$^{+1.6}_{-1.6}$&-256.6&4.24$^{+0.73}_{-0.65}$&50.78$^{+3.52}_{-3.33}$&0.08$^{+0.02}_{-0.01}$&4.05e+53\\
200.50$\sim$202.00&83.33$\sim$83.96&256.45&-0.88$^{+0.02}_{-0.02}$&1090.8$^{+73.8}_{-74.6}$&50.4$^{+1.3}_{-1.2}$&-458.5&6.88$^{+0.91}_{-0.81}$&66.21$^{+5.27}_{-4.91}$&0.10$^{+0.02}_{-0.01}$&5.28e+53\\
202.00$\sim$203.50&83.96$\sim$84.58&215.38&-0.77$^{+0.02}_{-0.02}$&527.0$^{+25.3}_{-25.2}$&38.3$^{+1.8}_{-1.8}$&-132.1&2.18$^{+0.54}_{-0.44}$&34.45$^{+2.53}_{-2.25}$&0.06$^{+0.02}_{-0.01}$&2.75e+53\\
203.50$\sim$205.00&84.58$\sim$85.20&157.84&-0.86$^{+0.03}_{-0.03}$&525.7$^{+39.3}_{-39.2}$&40.1$^{+2.4}_{-2.4}$&-63.6&1.43$^{+0.46}_{-0.37}$&19.61$^{+2.12}_{-1.73}$&0.07$^{+0.02}_{-0.02}$&1.56e+53\\
205.00$\sim$206.50&85.20$\sim$85.83&150.18&-0.83$^{+0.03}_{-0.03}$&454.8$^{+37.1}_{-37.9}$&39.6$^{+2.0}_{-2.0}$&-71.7&1.63$^{+0.47}_{-0.39}$&16.59$^{+1.96}_{-1.59}$&0.10$^{+0.03}_{-0.03}$&1.32e+53\\
206.50$\sim$208.00&85.83$\sim$86.45&112.49&-0.86$^{+0.05}_{-0.05}$&338.3$^{+41.3}_{-40.5}$&32.5$^{+2.9}_{-2.9}$&-29.1&0.70$^{+0.39}_{-0.27}$&8.59$^{+1.40}_{-1.15}$&0.08$^{+0.05}_{-0.03}$&6.85e+52\\
208.00$\sim$209.50&86.45$\sim$87.07&84.98&-1.09$^{+0.06}_{-0.06}$&474.3$^{+88.9}_{-88.7}$&32.7$^{+2.2}_{-2.2}$&-34.0&0.70$^{+0.28}_{-0.21}$&5.46$^{+1.27}_{-0.84}$&0.13$^{+0.06}_{-0.04}$&4.35e+52\\
209.50$\sim$211.00&87.07$\sim$87.70&82.67&-0.89$^{+0.06}_{-0.06}$&323.5$^{+52.4}_{-51.0}$&31.3$^{+3.4}_{-8.4}$&-58.7&0.26$^{+0.56}_{-0.19}$&6.03$^{+1.91}_{-1.15}$&0.04$^{+0.09}_{-0.03}$&4.81e+52\\
\hline                                   
\end{tabular}
\end{table*}

\section{GRB 160509A}\label{sec:GRB160509A}

\begin{figure*}[ht!]
\centering
\includegraphics[width=0.95\hsize,clip]{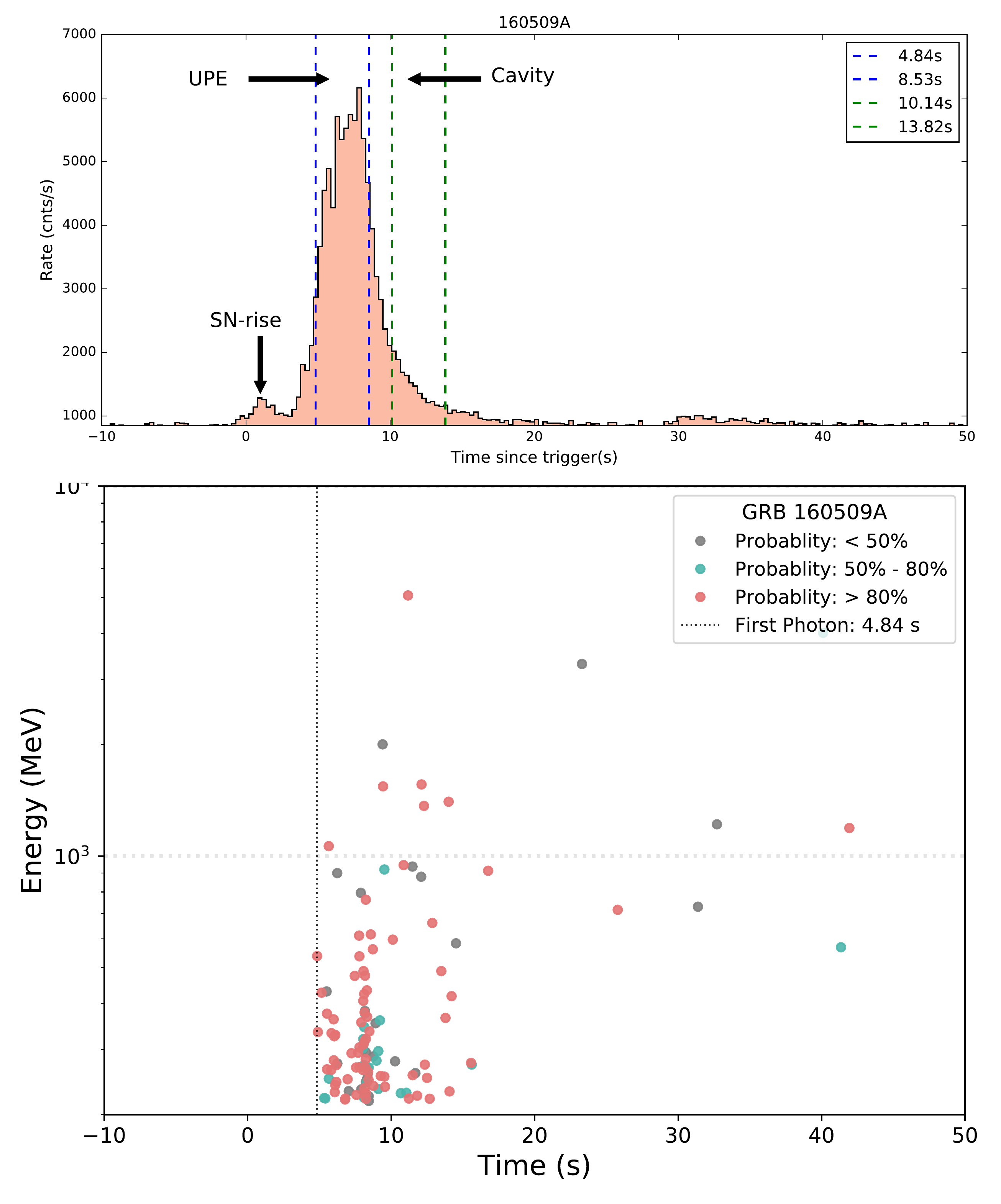}
\caption{\textbf{Upper panel}: {the proposed three new Episodes of GRB 160509A as a function of the rest-frame time. Episode 1 occurs $t_{rf}=$ 0.92~s and $t_{rf}=1.84$~s. Episode 2 including the UPE phase starts from $t_{rf}=4.84$~s and ending at $t_{rf}=8.53$~s in the rest frame. Episode 3 occurs at time after $t_{rf}=10.14$~s, staring at $t_{rf}=10.14$~s and ending at $t_{rf}=13.82$~s. The redshift for GRB 160509A is $1.17$ \citep{2016GCN.19419....1T}. The light-curve consists of two spikes, the isotropic energy in the first small one is $\sim 1.47\times 10^{52}$~erg. The total energy is $1.06 \times 10^{54}$~erg \citep{2017ApJ...844L...7T}. \textbf{Lower panel}: the energy and time of each  {\it Fermi}-LAT photon of energy $>100$~MeV. The first GeV photon occurs at $4.84$~s in the rest-frame. The onset of the GeV radiation exactly coincides with the onset of the UPE. The detailed information for each episode (SN-rise, UPE phase, Cavity, GeV, and Afterglow) see Section \ref{sec:GRB160509A} and Table \ref{table:episodes_160509A}, which include the starting time, the duration, the isotropic energy, and the preferred model.}}\label{fig:lc_160509A}
\end{figure*}

\begin{table*}[ht!]
\caption{The Episodes of GRB 160509A, the definitions of  parameters are the same as in table \ref{table:episodes_160625B}.}             
\label{table:episodes_160509A}
\centering                         
\begin{tabular}{cccccc}       
\hline\hline                  
Episode &Starting Time &Ending Time&Energy&Spectrum&Reference\\   &Rest-frame&Rest-frame\\ 
\hline                        
SN-rise&$  0.92 $~s&$ 1.84 $~s&$ 1.47 \times10^{52}$~erg&CPL+BB&New in this paper\\
UPE&$ 4.84$~s&$ 8.53$~s&$ 1.06\times10^{54}$~erg&CPL+BB&New in this paper\\
Cavity&$ 10.14$~s&$ 13.82$~s&$ 3.66\times 10^{52}$~erg&CPL&New in this paper\\
GeV&$ 4.84$~s&  $> 2\times10^4$~s&$ 3.59 \times 10^{53}$~erg&PL&New in this paper\\
Afterglow&$7287$s&$\sim 20$ days&$1.36\times 10^{52}$~erg&PL&New in this paper\\
\hline                                   
\end{tabular}
\end{table*}

GRB 160509A was observed by the \textit{Fermi} satellite on May 9, 2016, at 08:59:04.36 UT \citep{2016GCN.19403....1L}. It was a strong source of GeV photons detected by \textit{Fermi}-LAT, including a photon of $52$ GeV arrived at $77$~s, and another one of $29$ GeV, at $\sim 70$~ks \citep{2016ApJ...833...88L}. {\it Swift} has a late-time follow-up, with a total exposure time of $1700$~s starting from $7278$~s \citep{2019arXiv190603493K}. The redshift of $1.17$ is measured by Gemini North telescope  \citep{2016GCN.19419....1T}, inferring a high  isotropic energy of $1.06 \times 10^{54}$~erg \citep{2017ApJ...844L...7T}. 

Pak-Hin Thomas Tam and collaborators \citep{2017ApJ...844L...7T} analyzed in great detail the bright multi-peaked pulse from $-10$ to $30$~s, and a weaker emission period from $280$ to $420$~s. 
In Fig.~\ref{fig:lc_160509A}, we show the result of our highly time-resolved analysis applied to GRB 160509A, which further extend the result of \citet{2017ApJ...844L...7T}.

\begin{figure*}[ht!]
\centering
\includegraphics[width=0.49\hsize,clip]{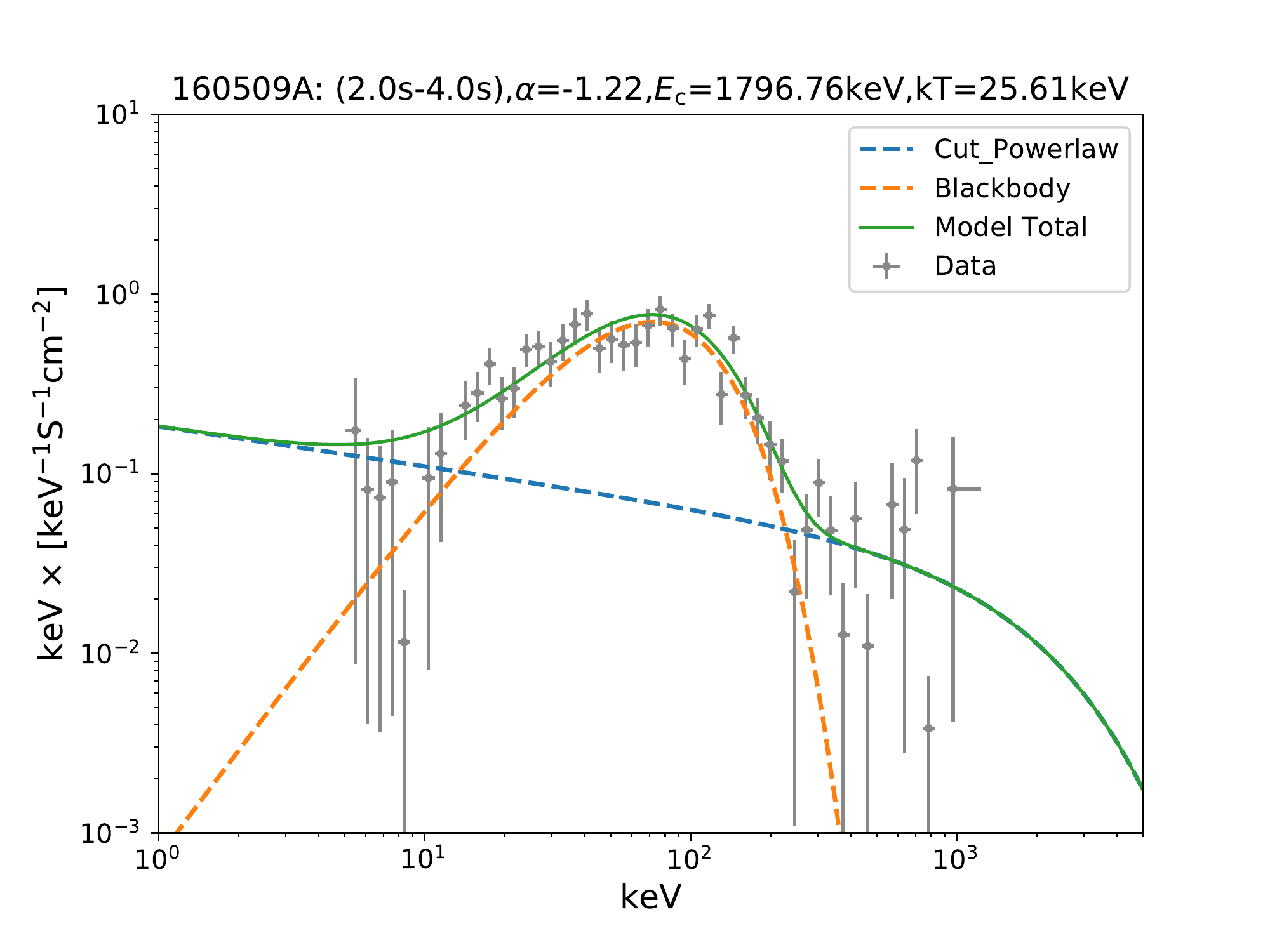}
\includegraphics[width=0.49\hsize,clip]{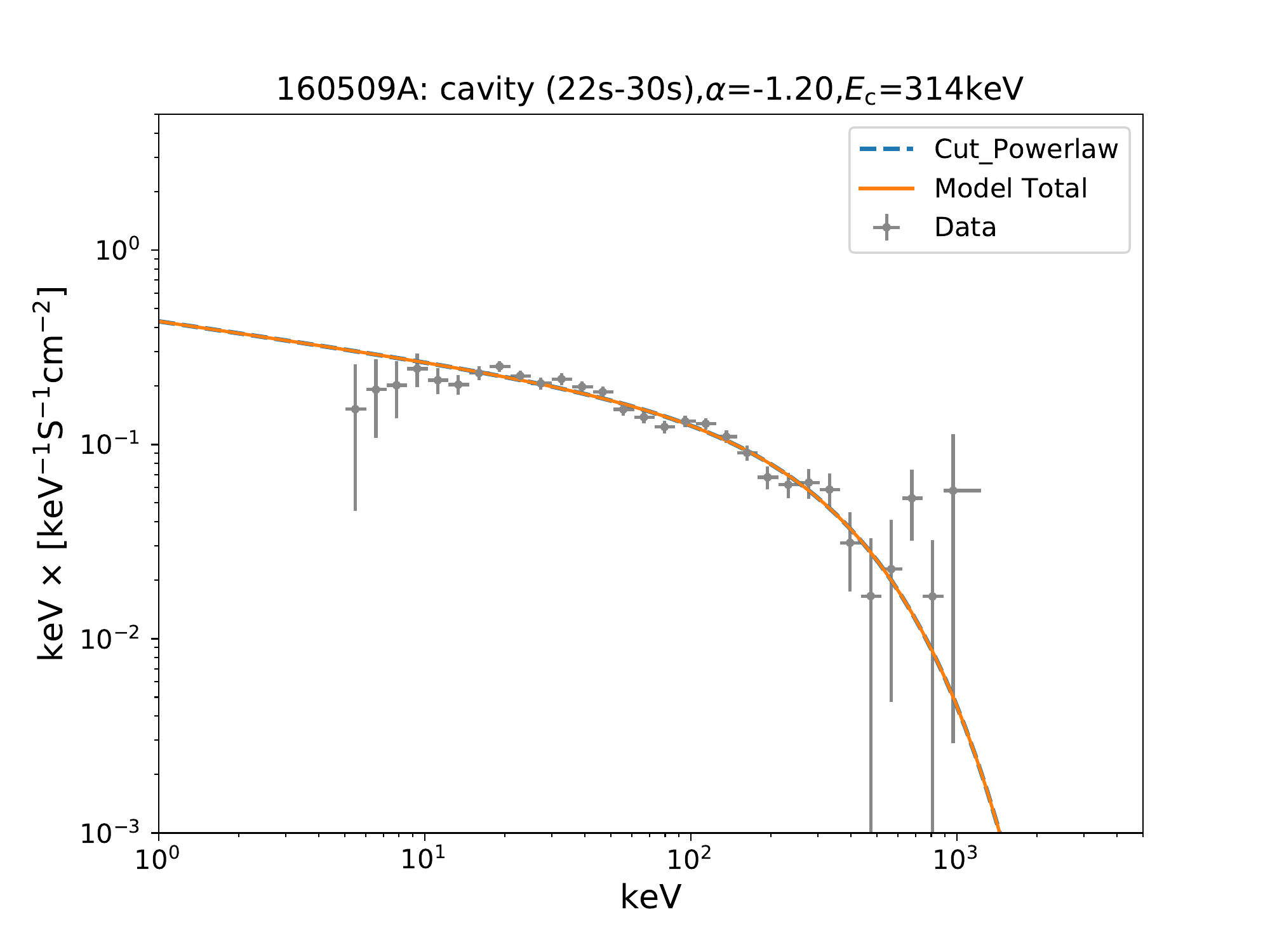}\\
\includegraphics[width=0.49\hsize,clip]{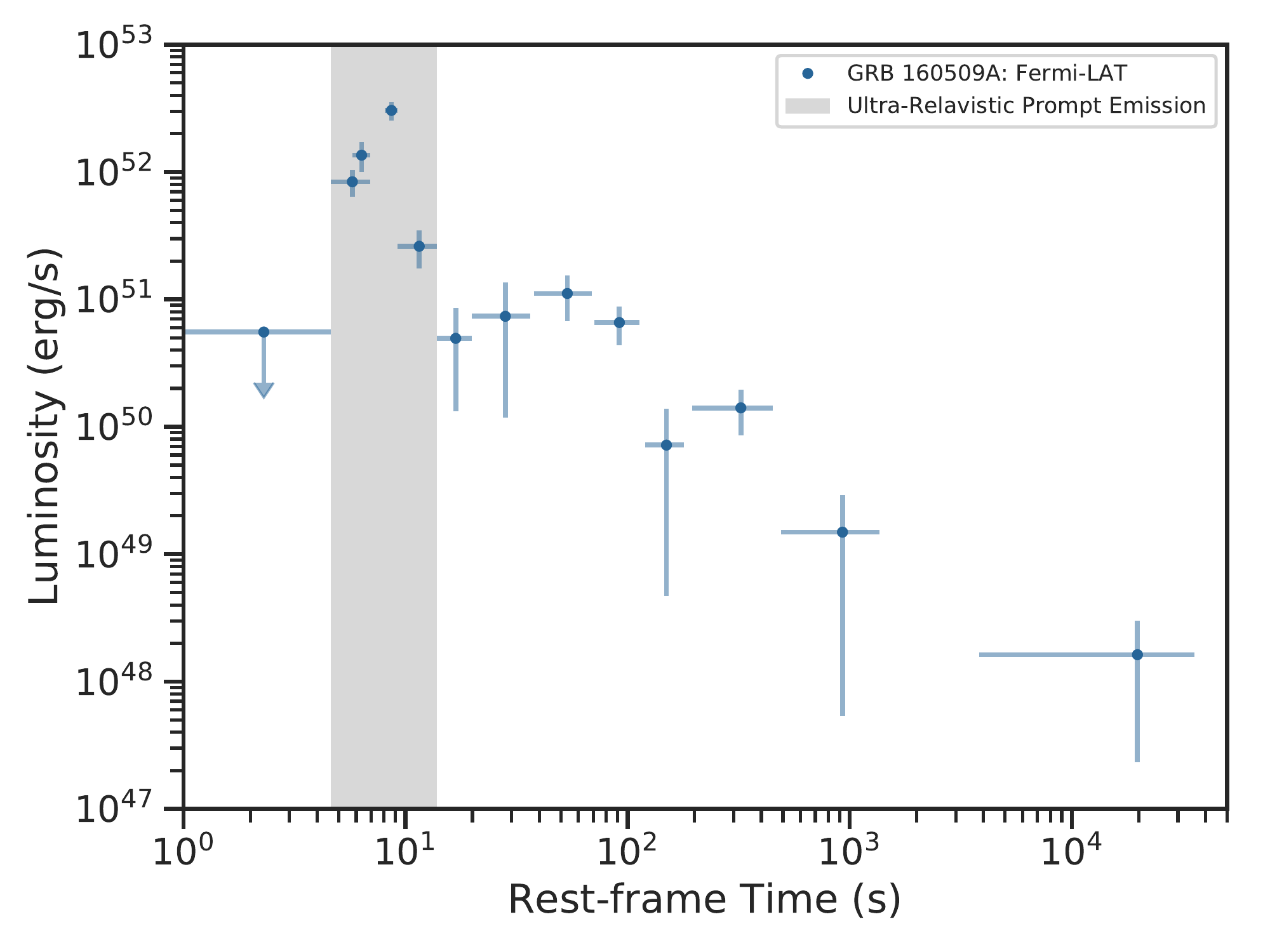}
\includegraphics[width=0.49\hsize,clip]{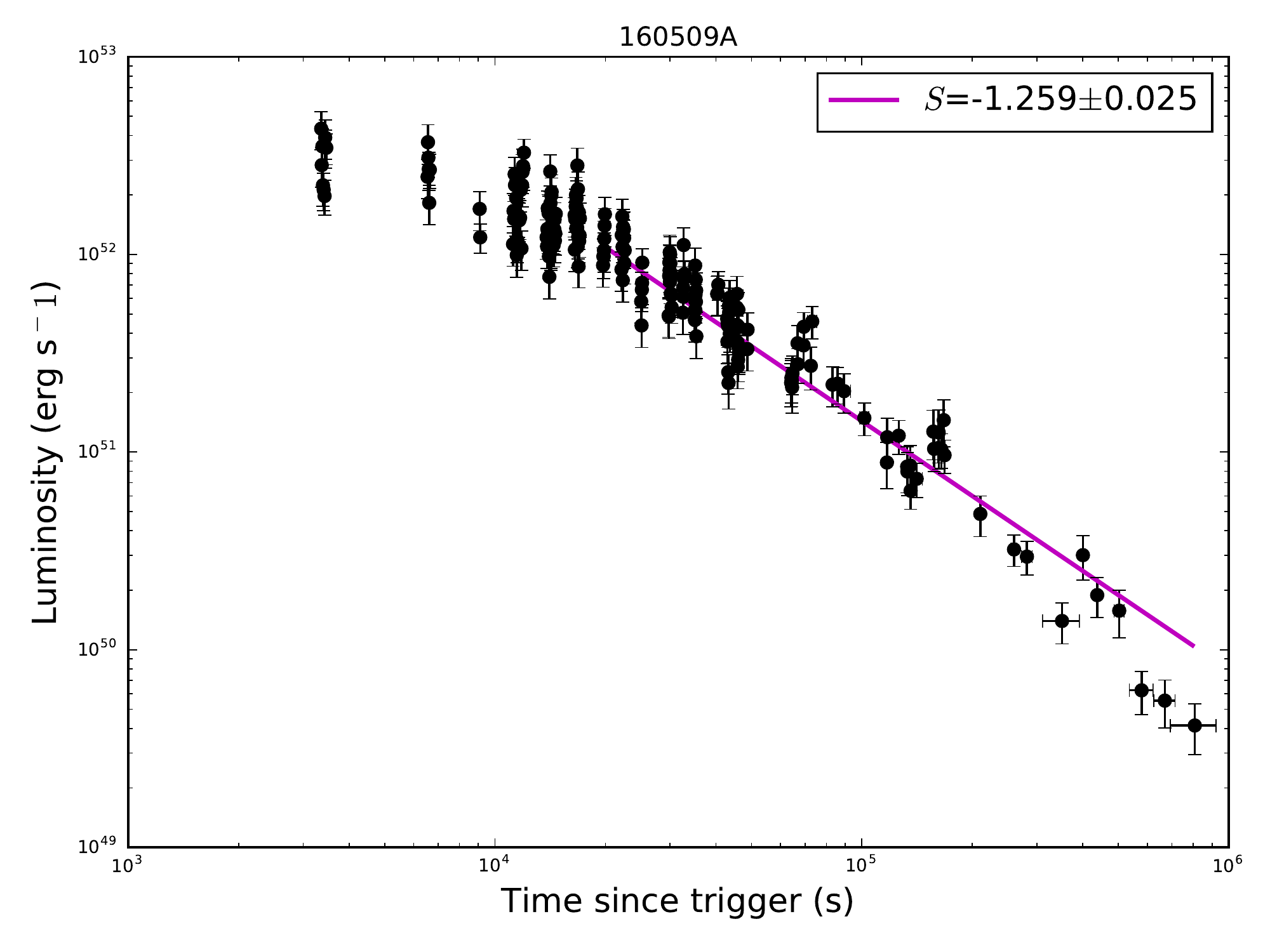}
\caption{SN-rise, Cavity, GeV, and Afterglow of GRB 160509A; see also Table \ref{table:episodes_160509A} that includes, for each Episode, the starting time, the duration, the isotropic energy, and the model that best fits the spectrum. \textbf{Upper left:} The CPL+BB spectrum of the SN-rise, for the time interval from $t=2.0$~s ($t_{\rm rf}=0.92$~s) to $t=4.0$~s ($t_{\rm rf}=1.84$~s), spectral index $\alpha=$ -1.22, cutoff energy $E_{\rm c}=$ 1769.76 ~keV, and temperature is $25.61$~keV in the observer's frame. 
\textbf{Upper right:} Featureless spectrum of the cavity emission, fitted by a CPL model, from $22$~s ($t_{\rm rf} = 10.14$~s) to $30$~s ($t_{\rm rf} = 13.82$~s), with the photon index $\alpha$ is $-1.20$ and the cutoff energy $E_c=$ $314$~keV in the observer's frame. \textbf{Lower left:} rest-frame {\it Fermi}-LAT light-curve in the $100$~MeV to $100$~GeV energy range. The UPE region is marked with the grey shadow. \textbf{Lower right:} k-corrected soft X-ray afterglow in the energy band of $0.3$--$10$~keV, observed by the {\it Swfit}-XRT satellite, as a function of rest-frame time. It is best fitted by a power-law with index $1.259\pm 0.025$.}
\label{fig:joinedGRB160509A}
\end{figure*}

\begin{figure*}[ht!]
\small\addtolength{\tabcolsep}{-3pt}
\centering
\includegraphics[angle=0, scale=0.6]{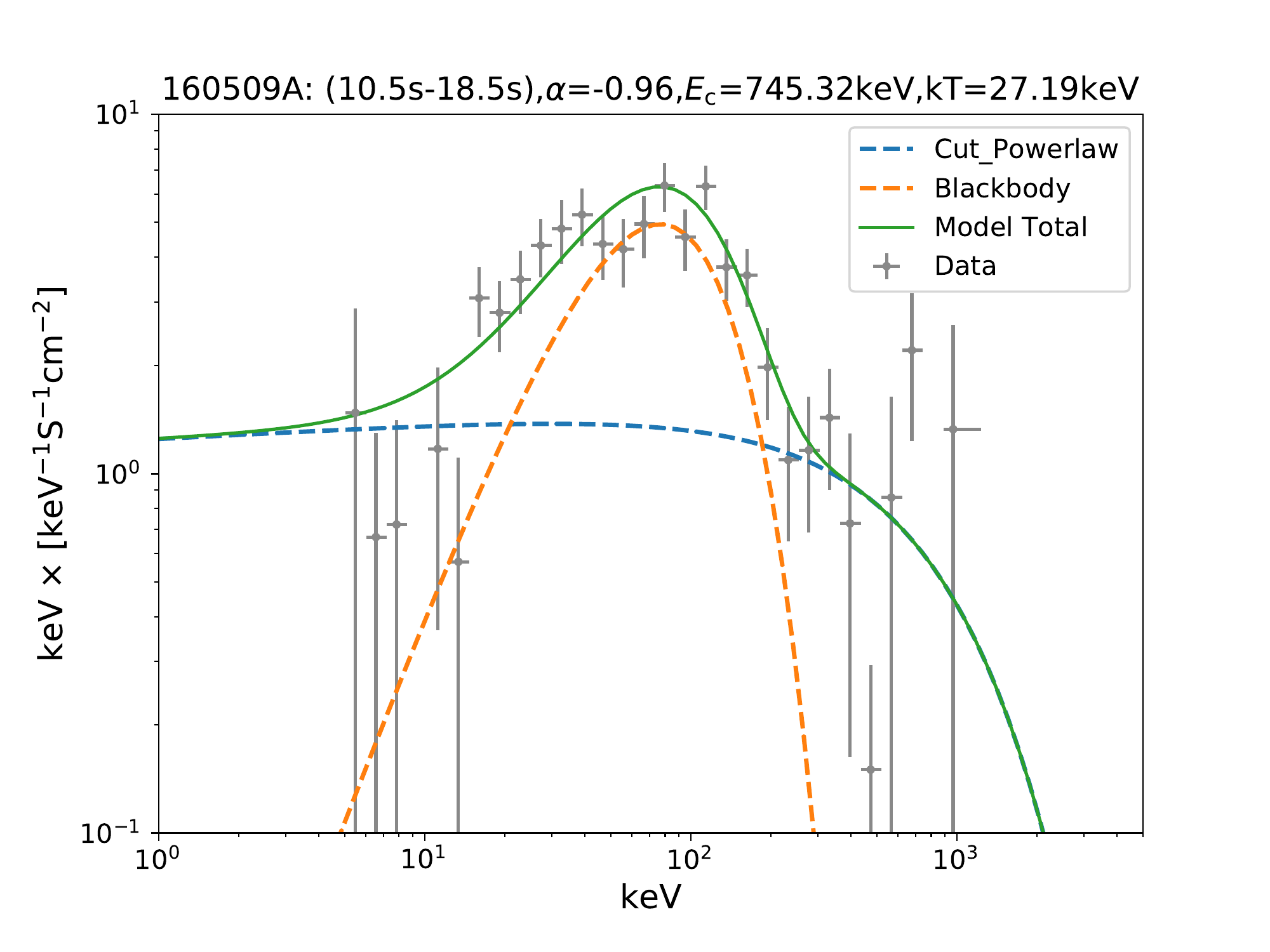}\\
\includegraphics[angle=0, scale=0.45]{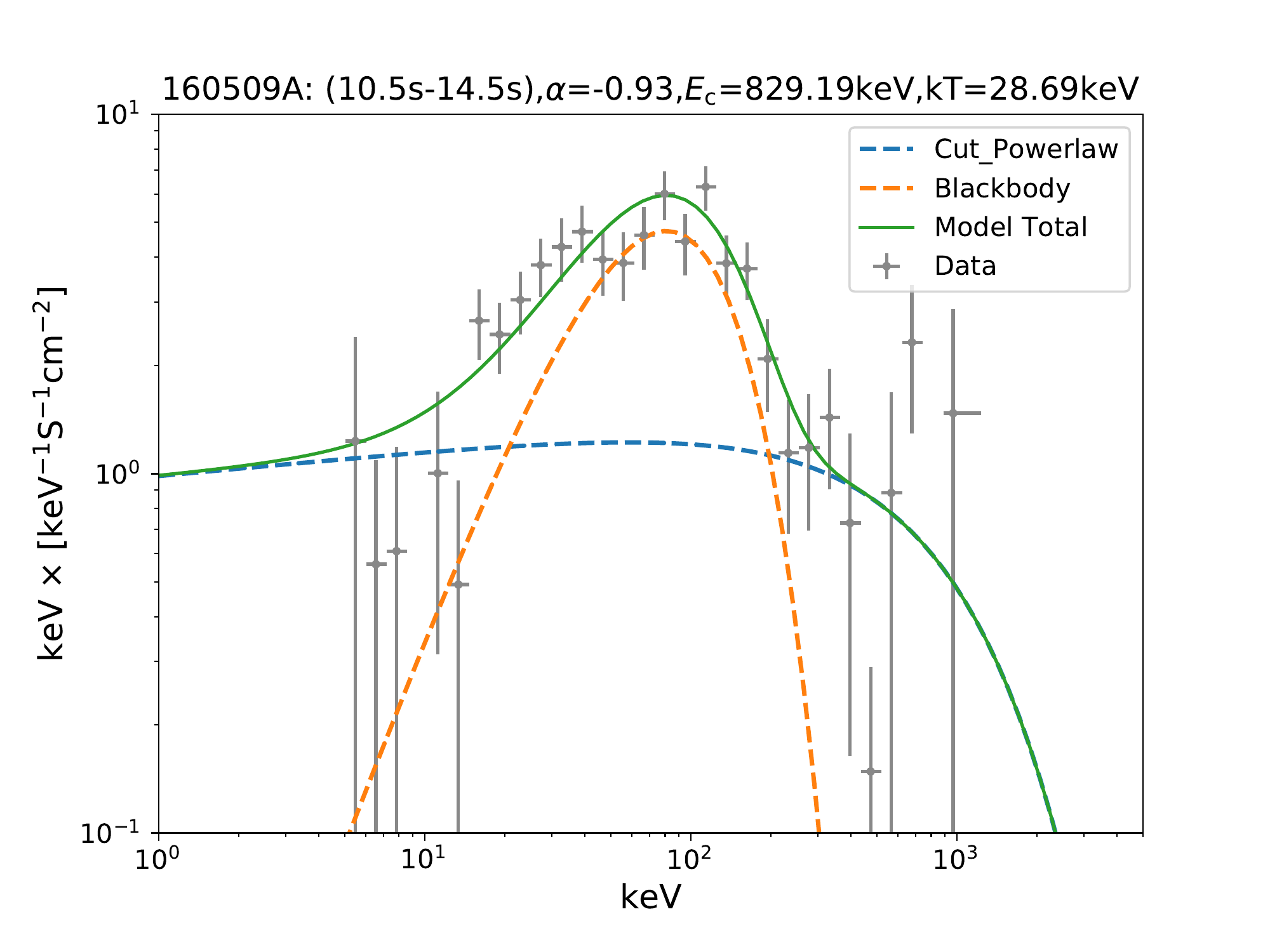}
\includegraphics[angle=0, scale=0.45]{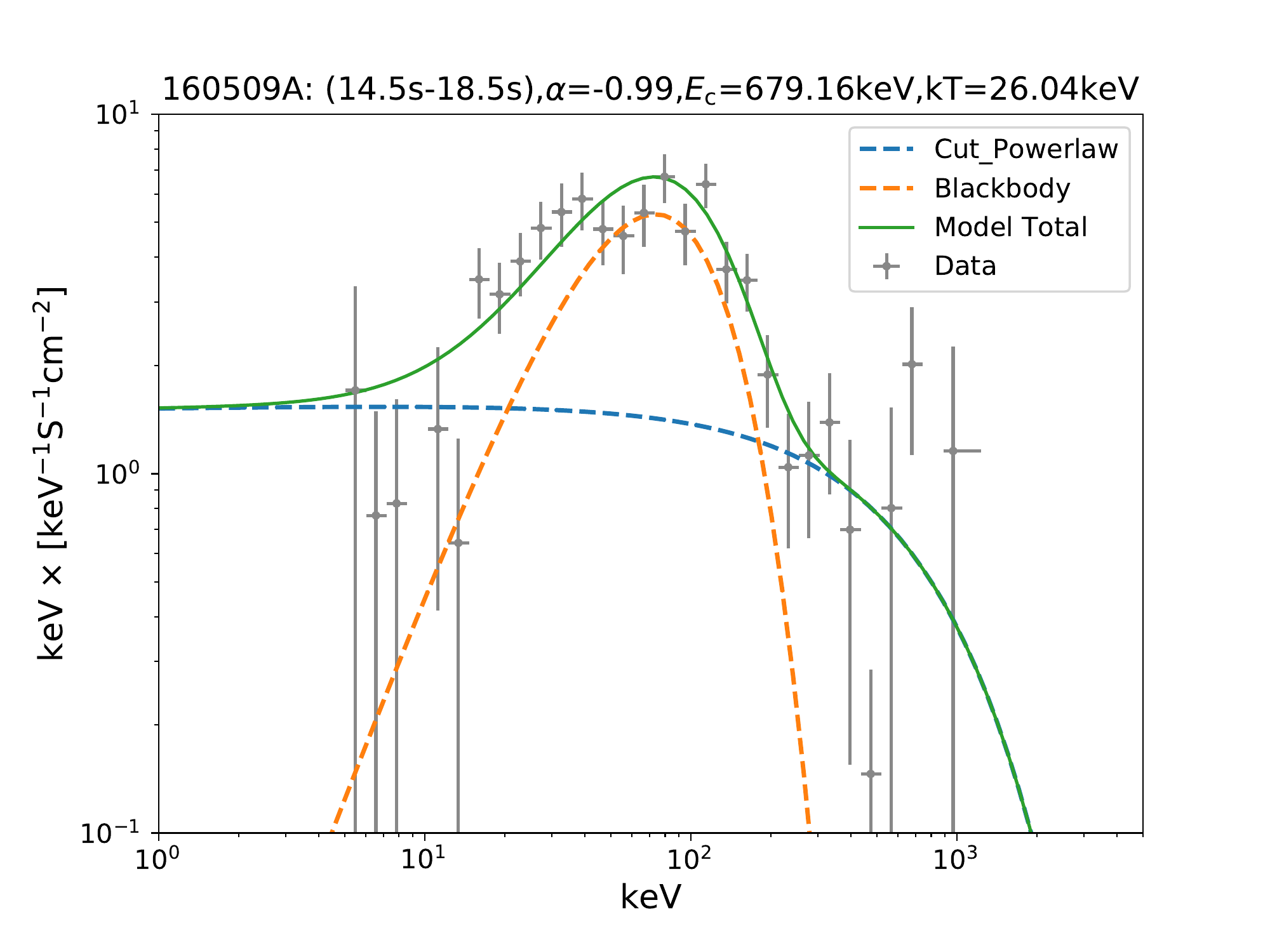}\\
\includegraphics[angle=0, scale=0.215]{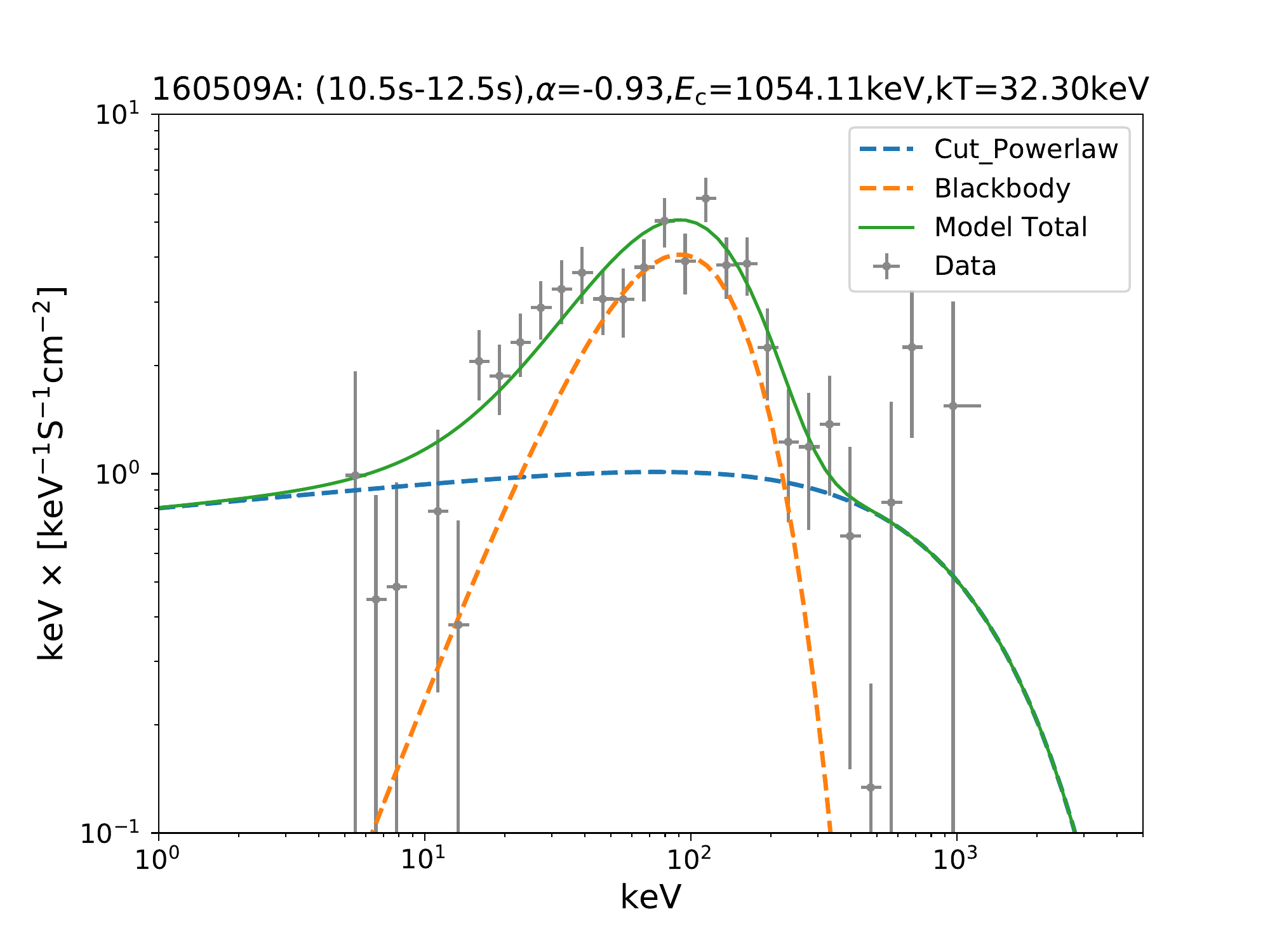}
\includegraphics[angle=0, scale=0.215]{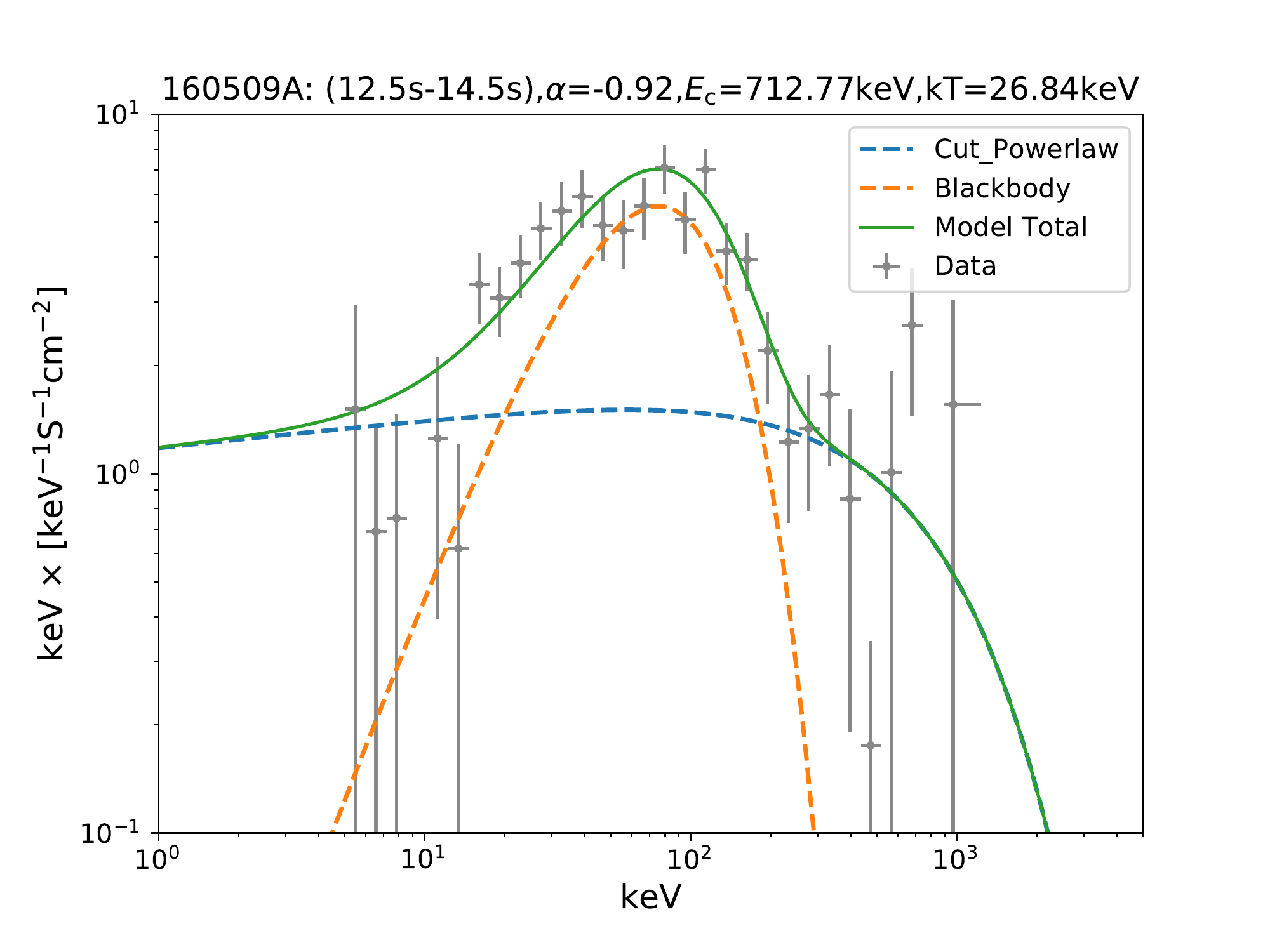}
\includegraphics[angle=0, scale=0.215]{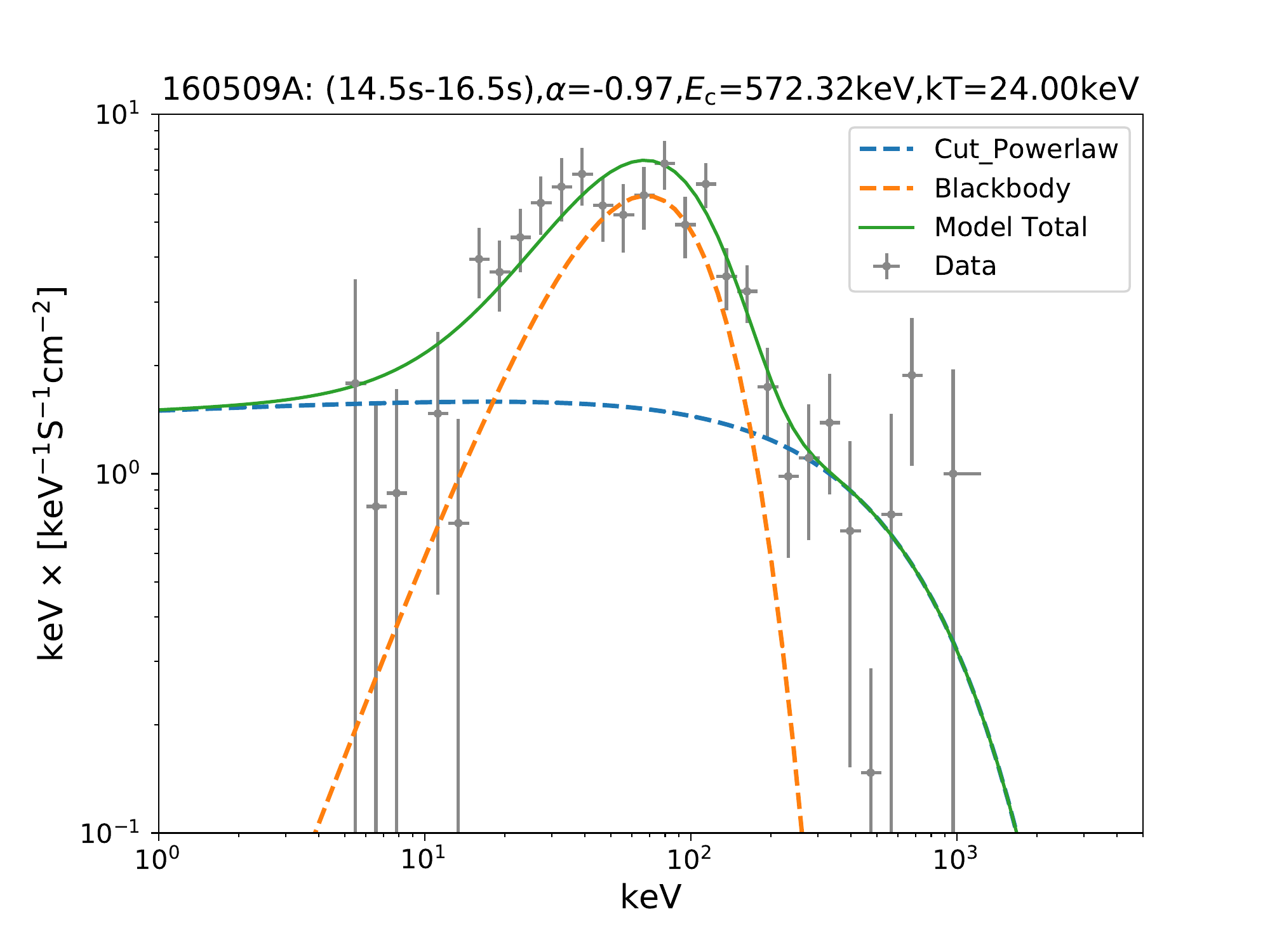}
\includegraphics[angle=0, scale=0.215]{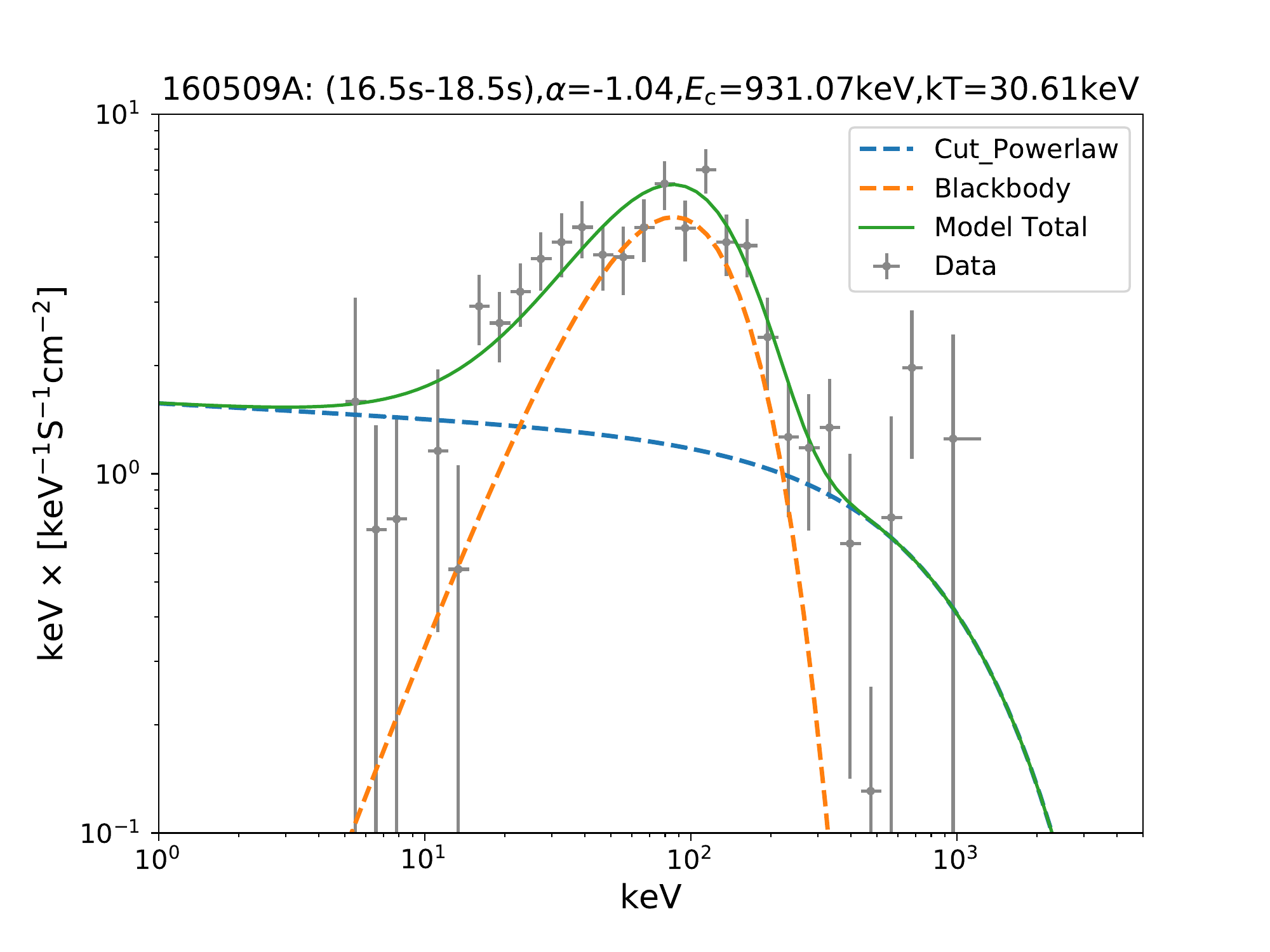}\\
\includegraphics[angle=0, scale=0.105]{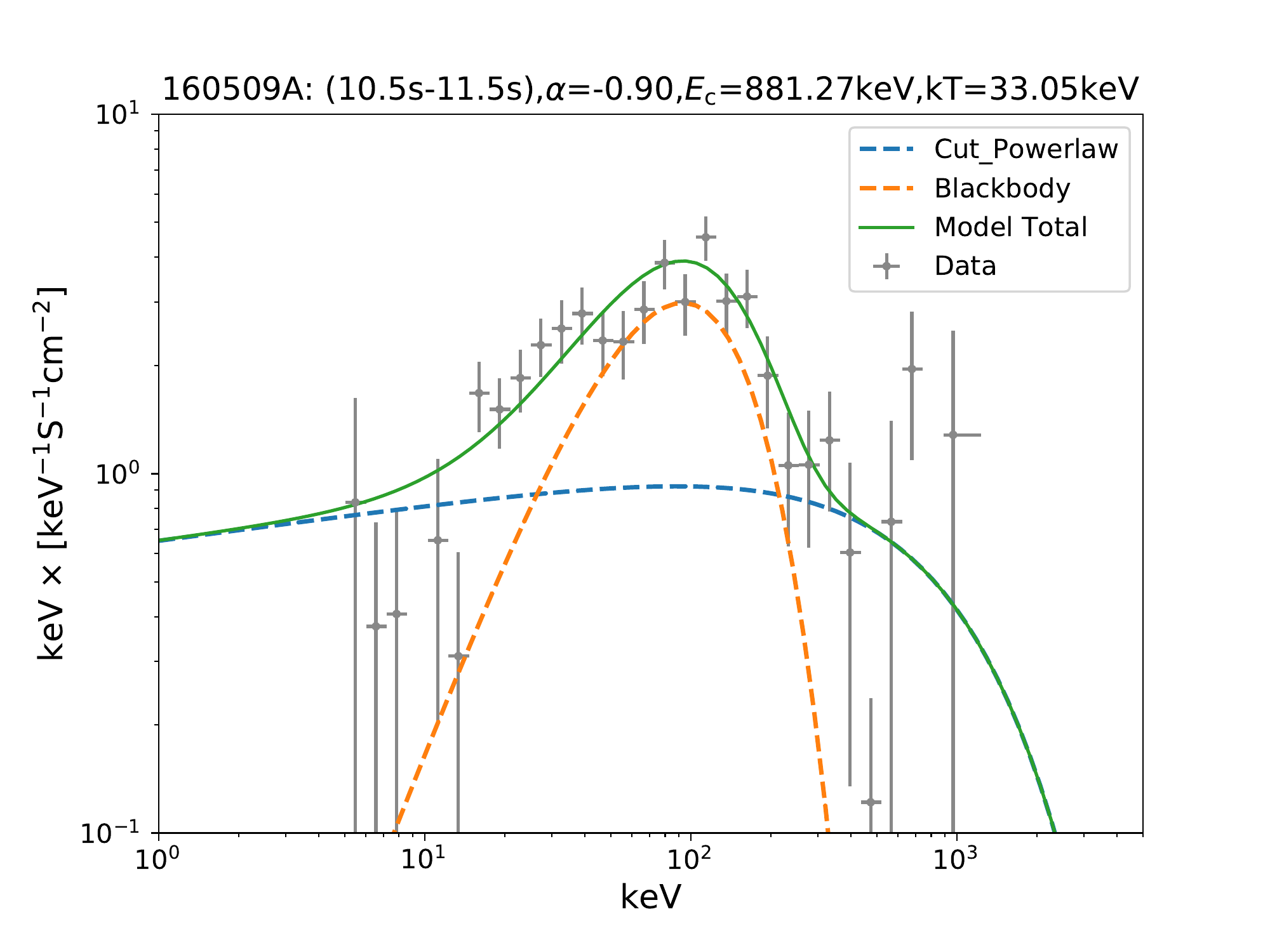}
\includegraphics[angle=0, scale=0.105]{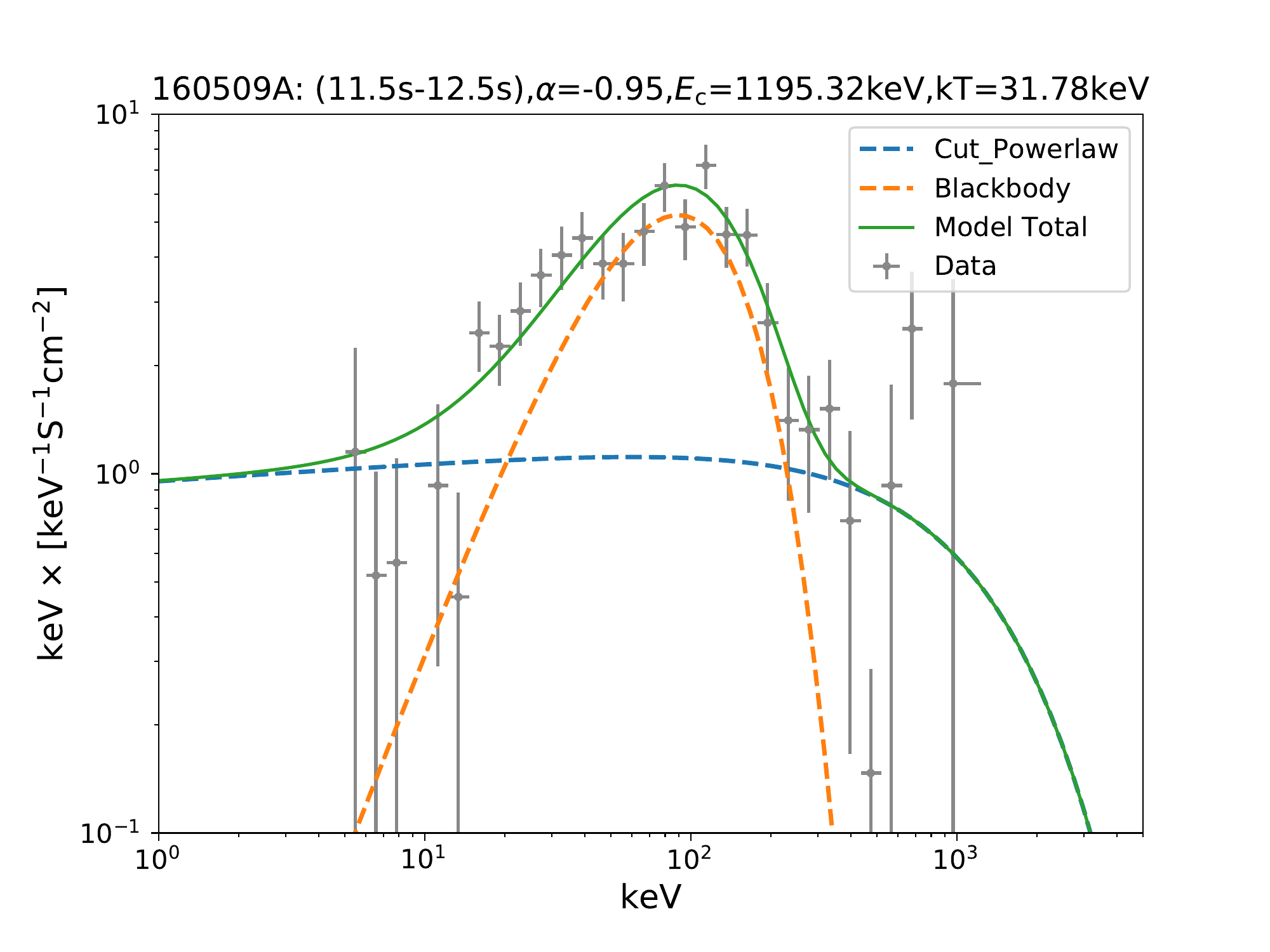}
\includegraphics[angle=0, scale=0.105]{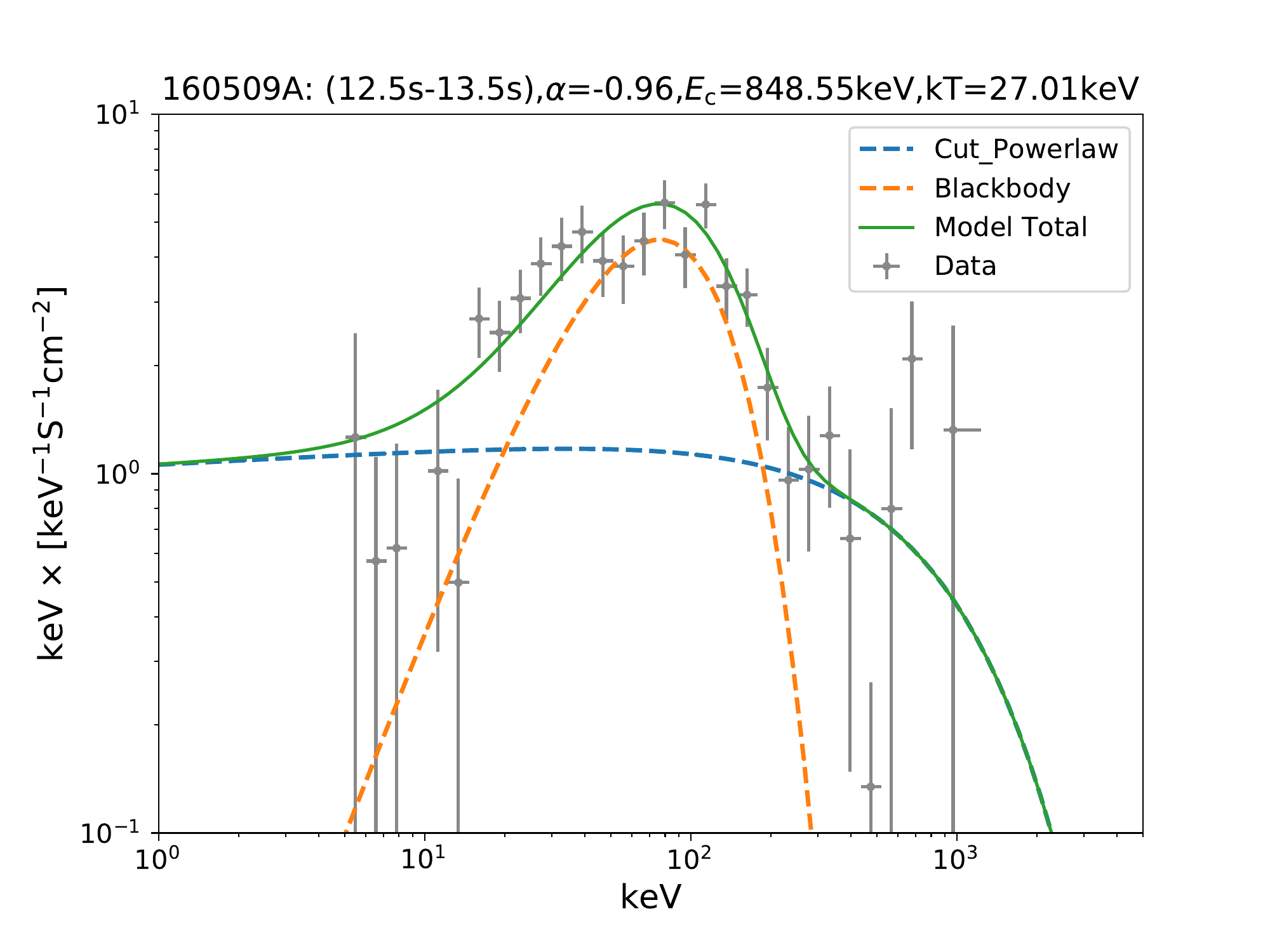}
\includegraphics[angle=0, scale=0.105]{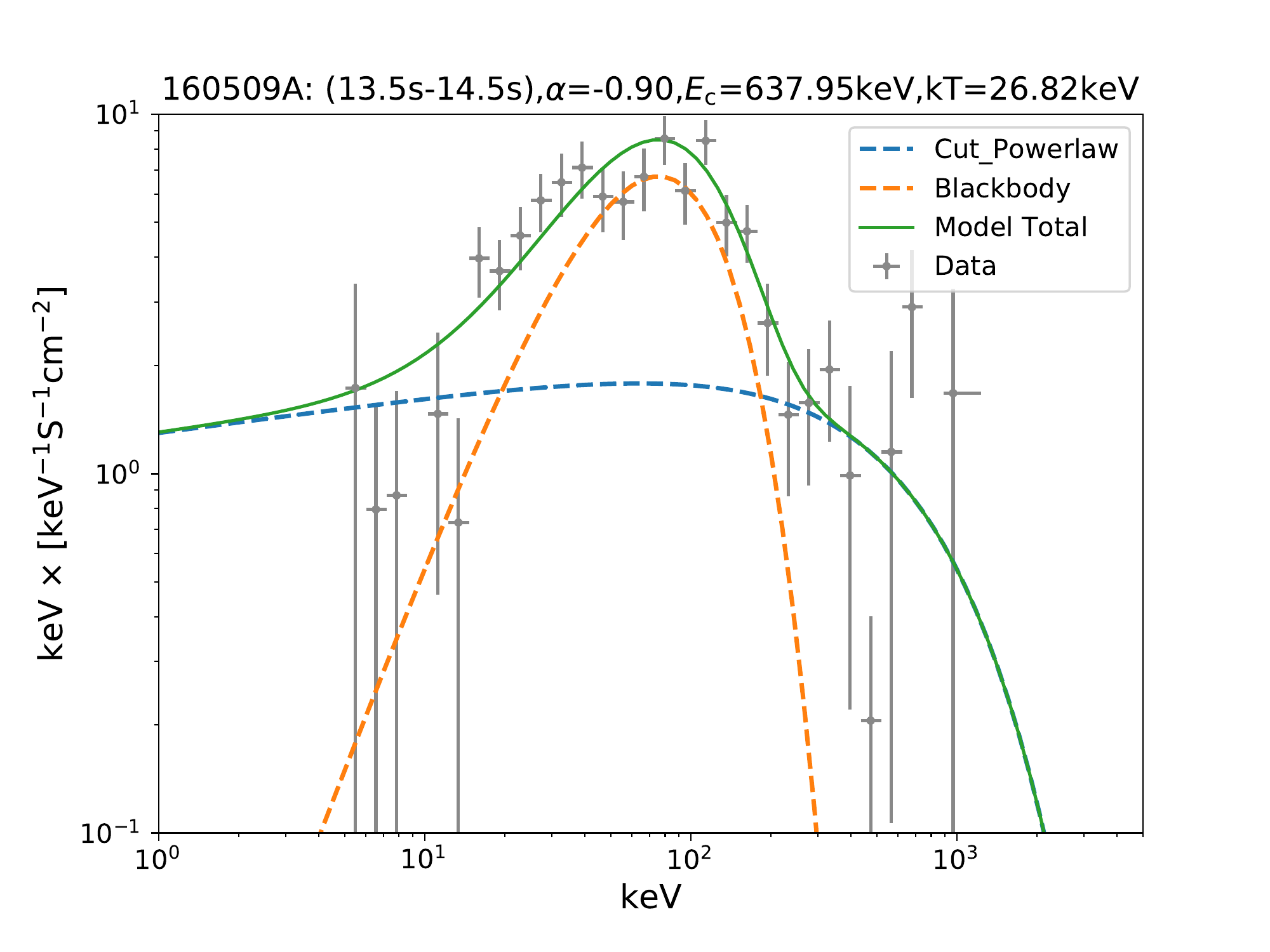}
\includegraphics[angle=0, scale=0.105]{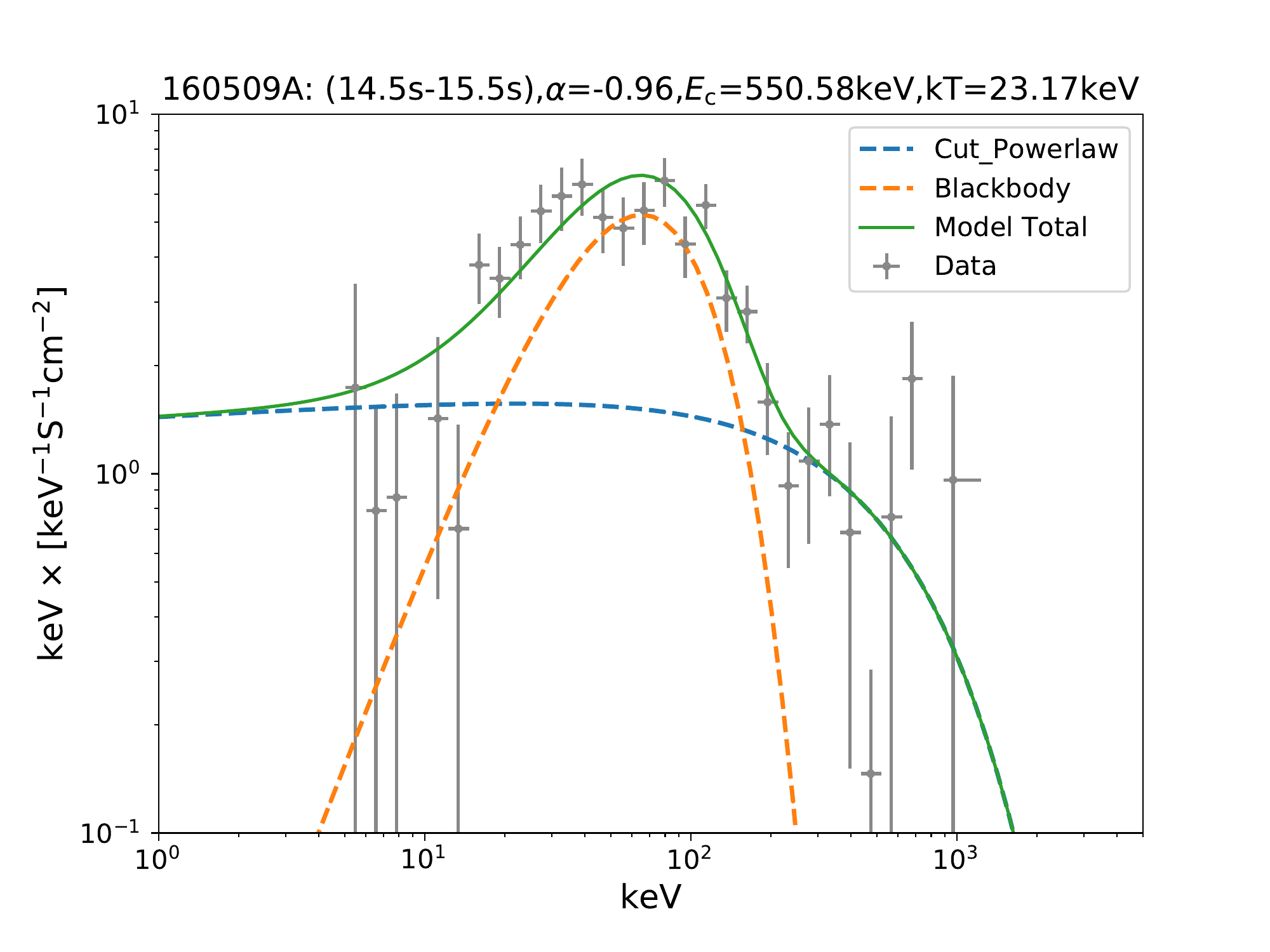}
\includegraphics[angle=0, scale=0.105]{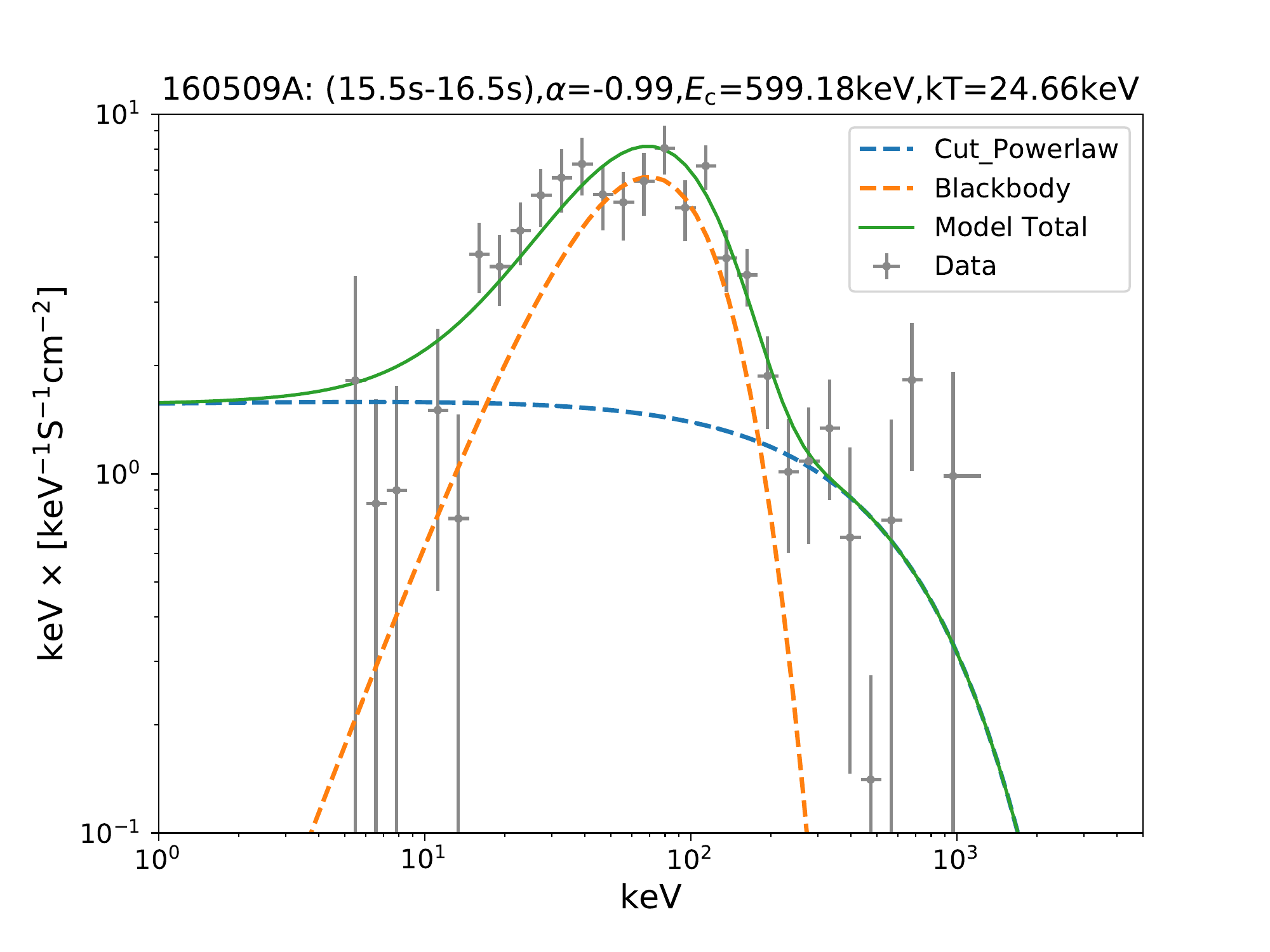}
\includegraphics[angle=0, scale=0.105]{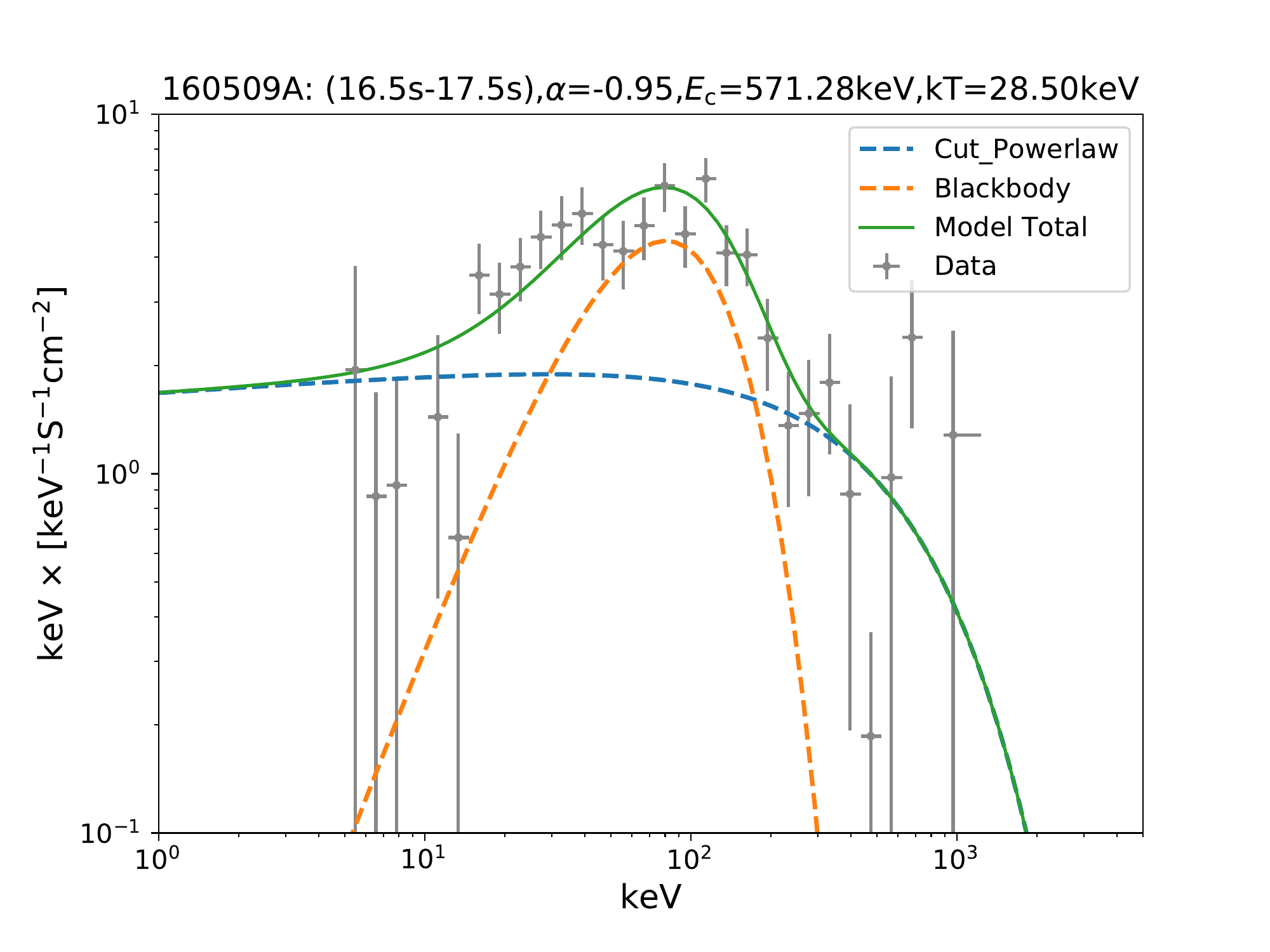}
\includegraphics[angle=0, scale=0.105]{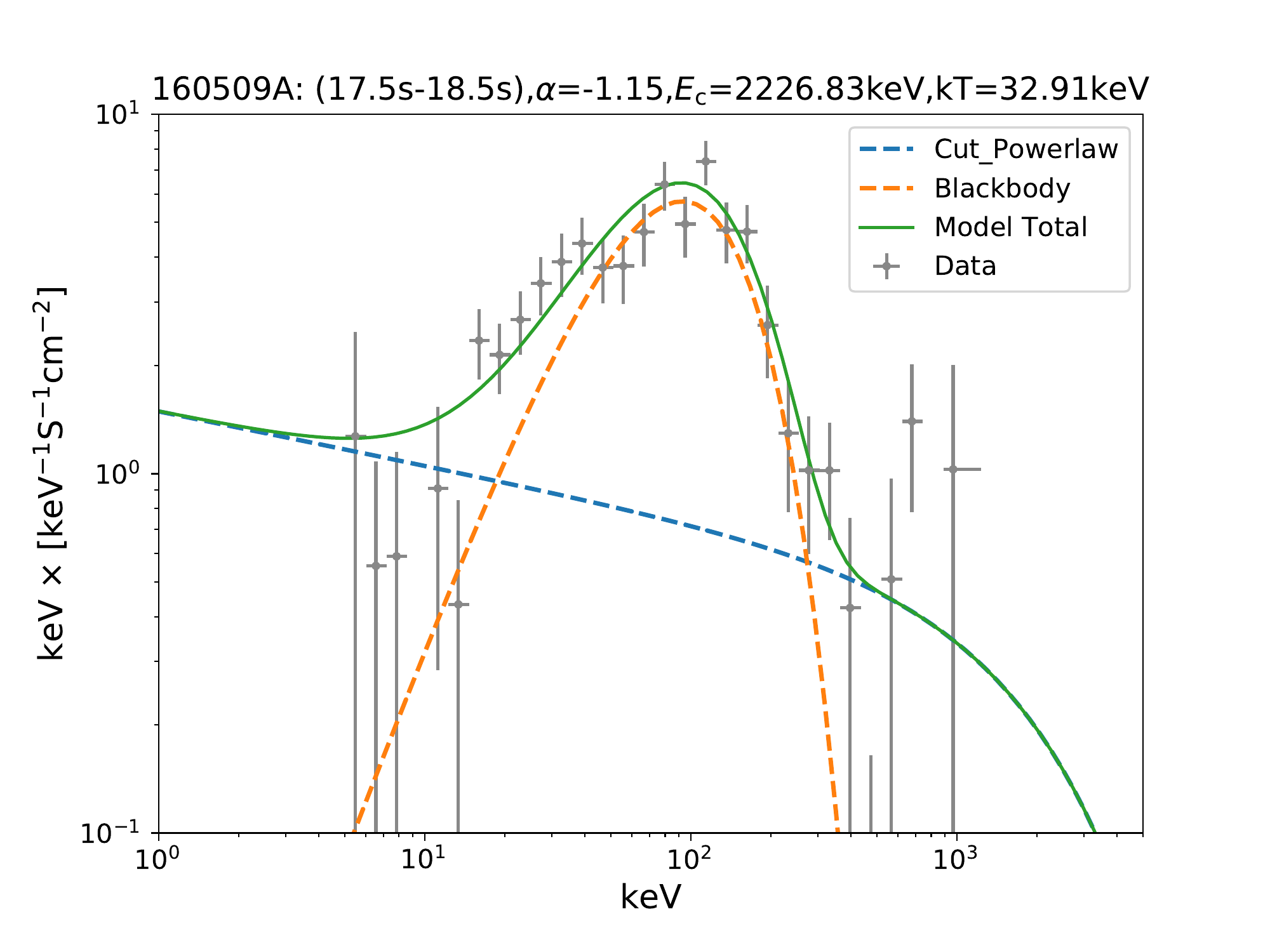}\\
\includegraphics[angle=0, scale=0.05]{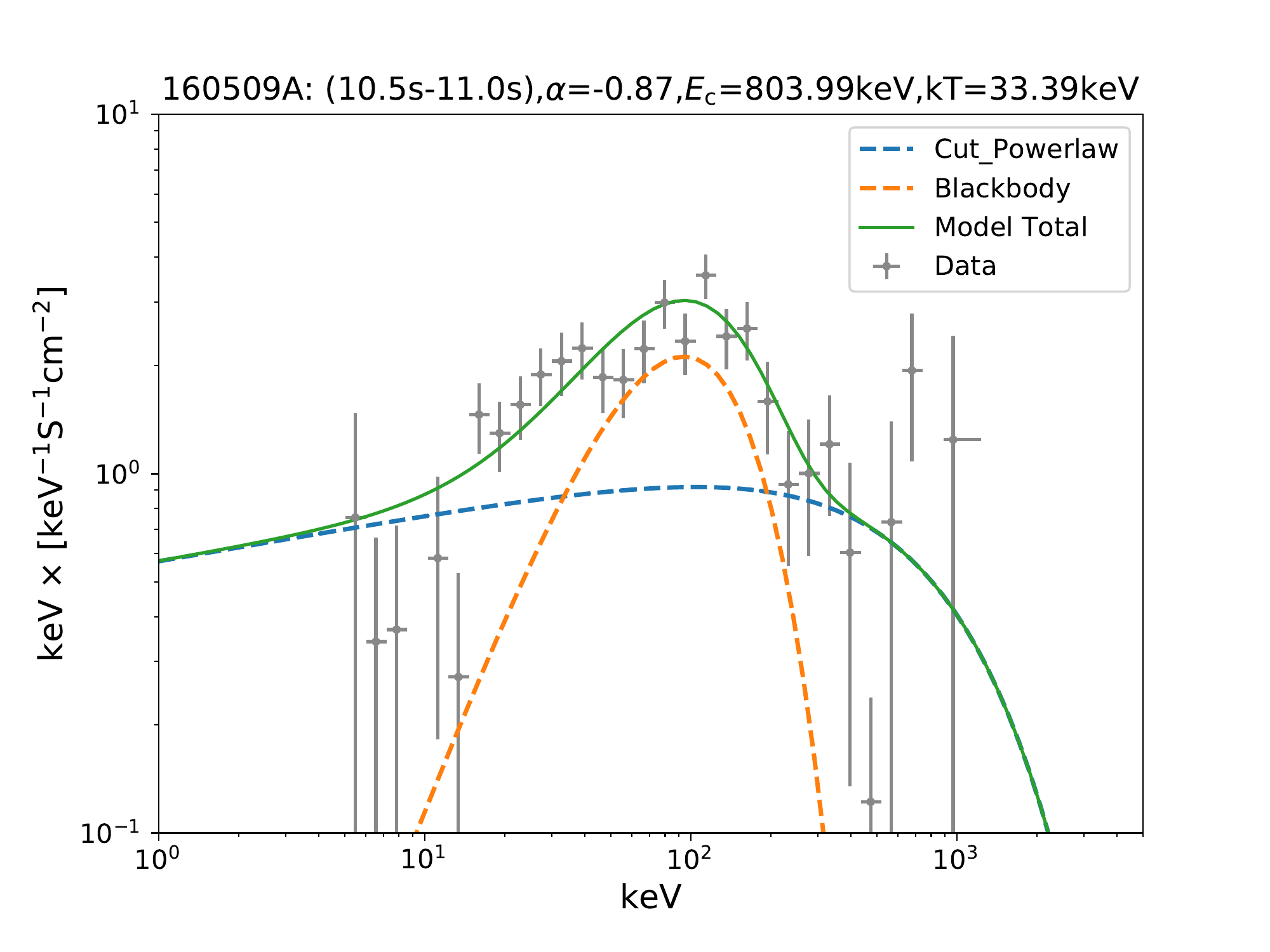}
\includegraphics[angle=0, scale=0.05]{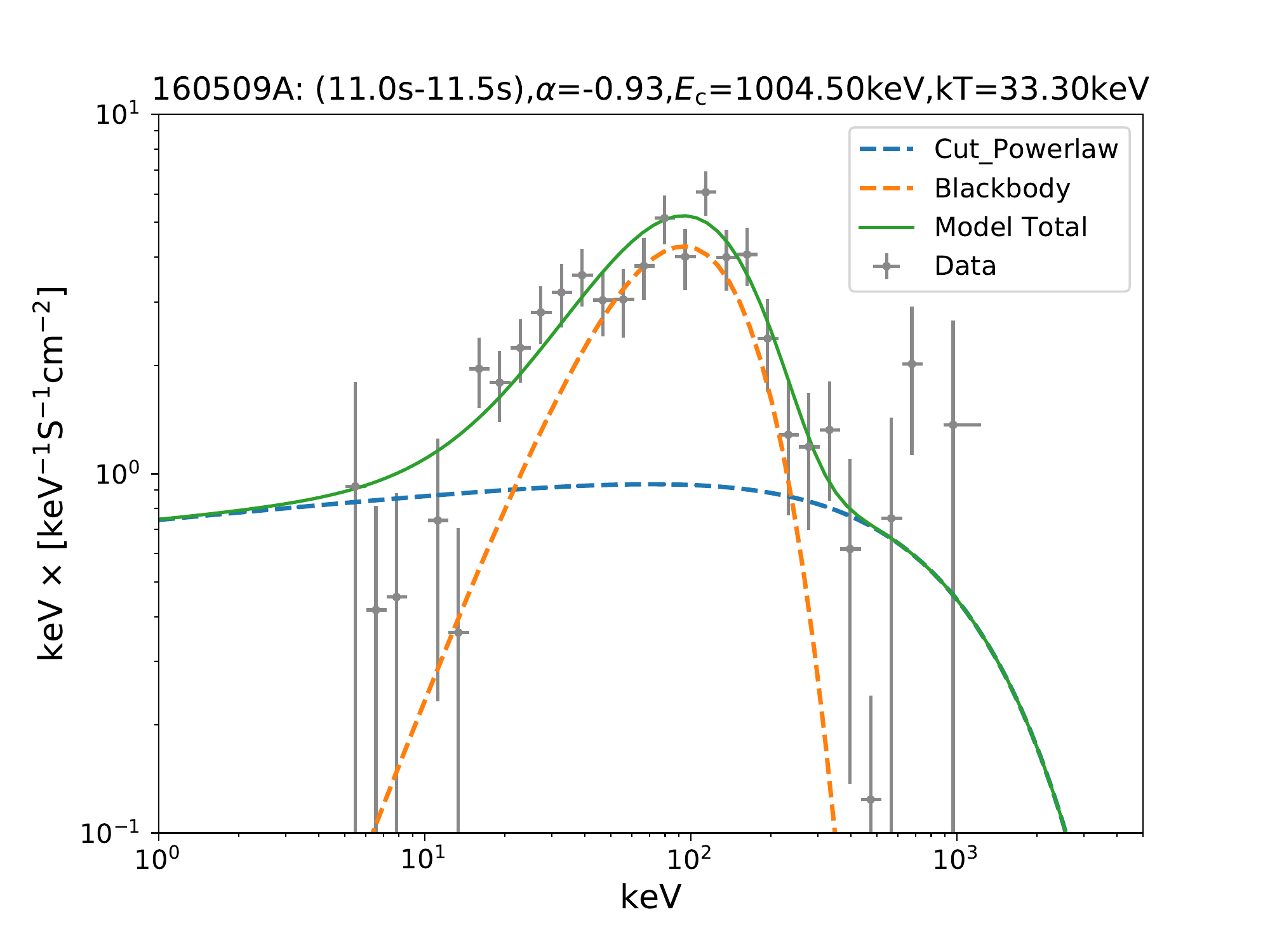}
\includegraphics[angle=0, scale=0.05]{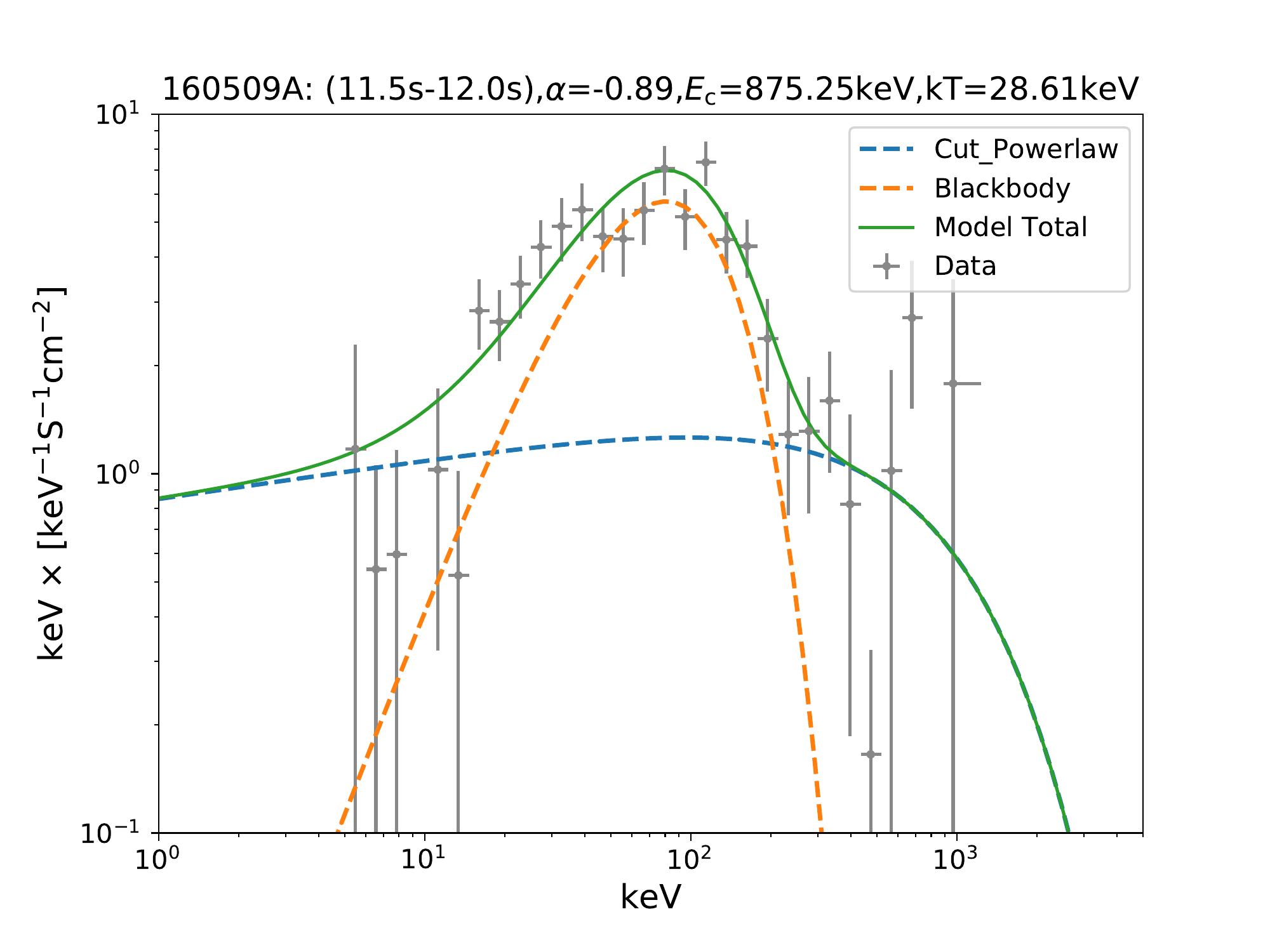}
\includegraphics[angle=0, scale=0.05]{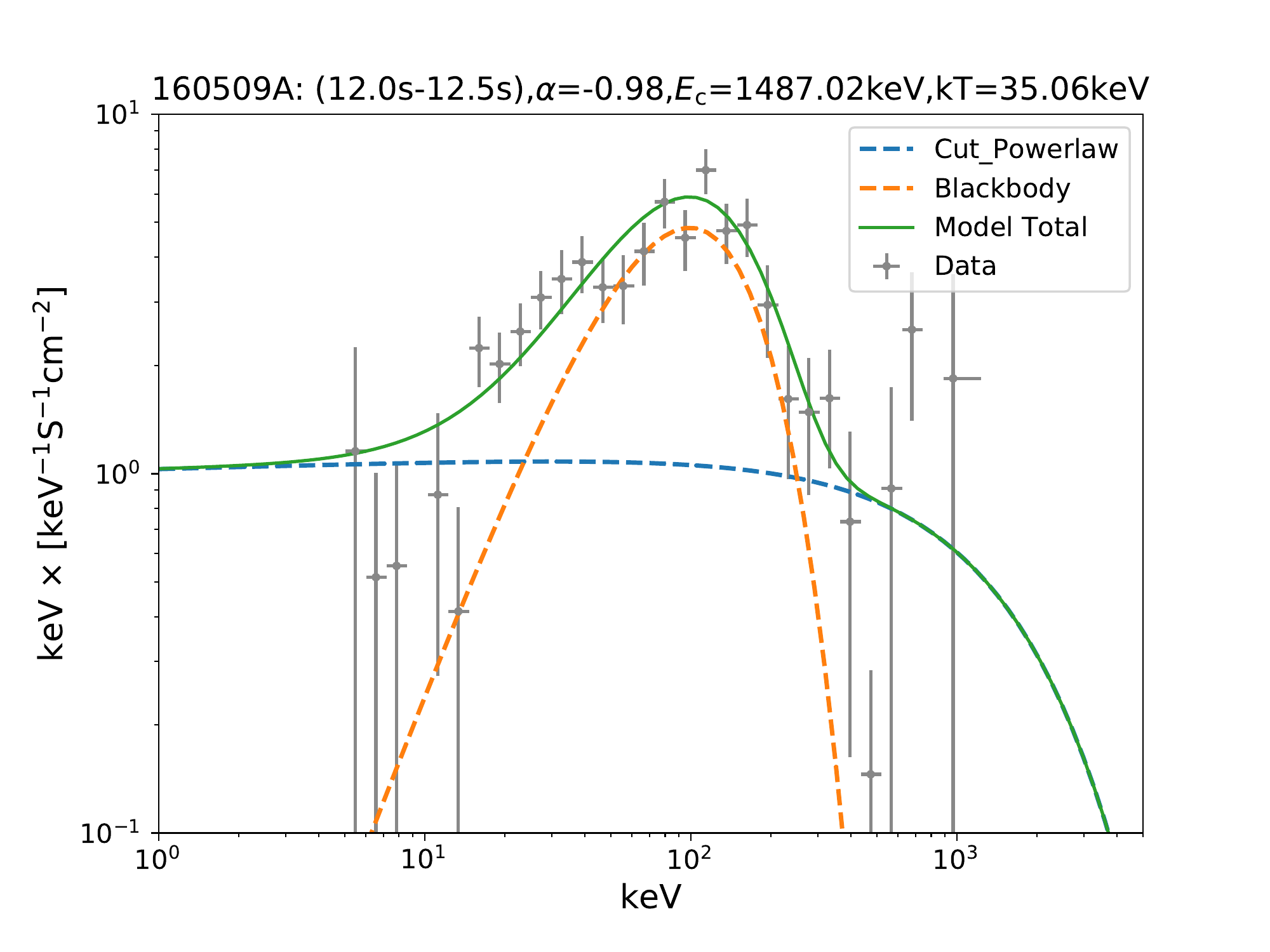}
\includegraphics[angle=0, scale=0.05]{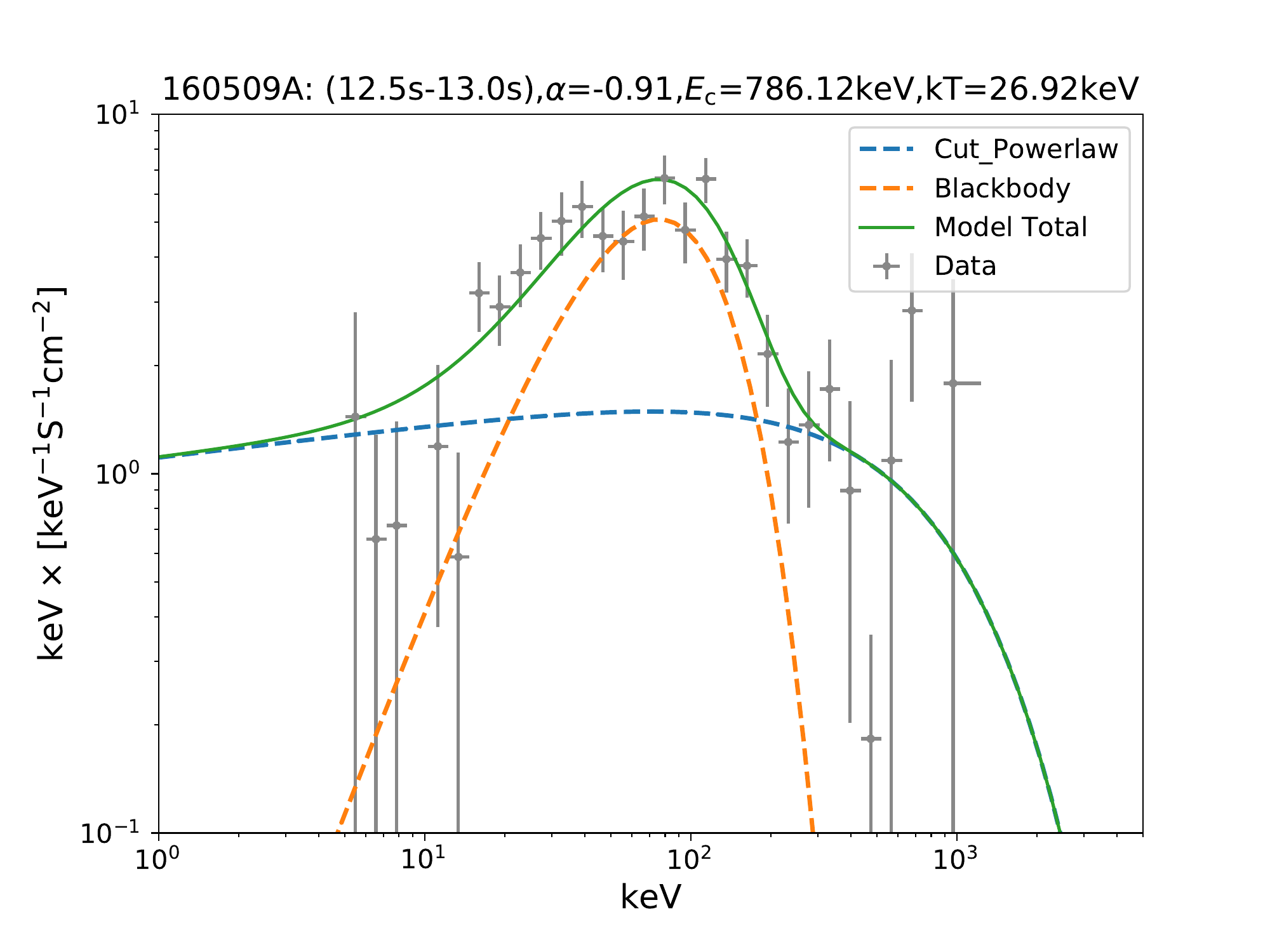}
\includegraphics[angle=0, scale=0.05]{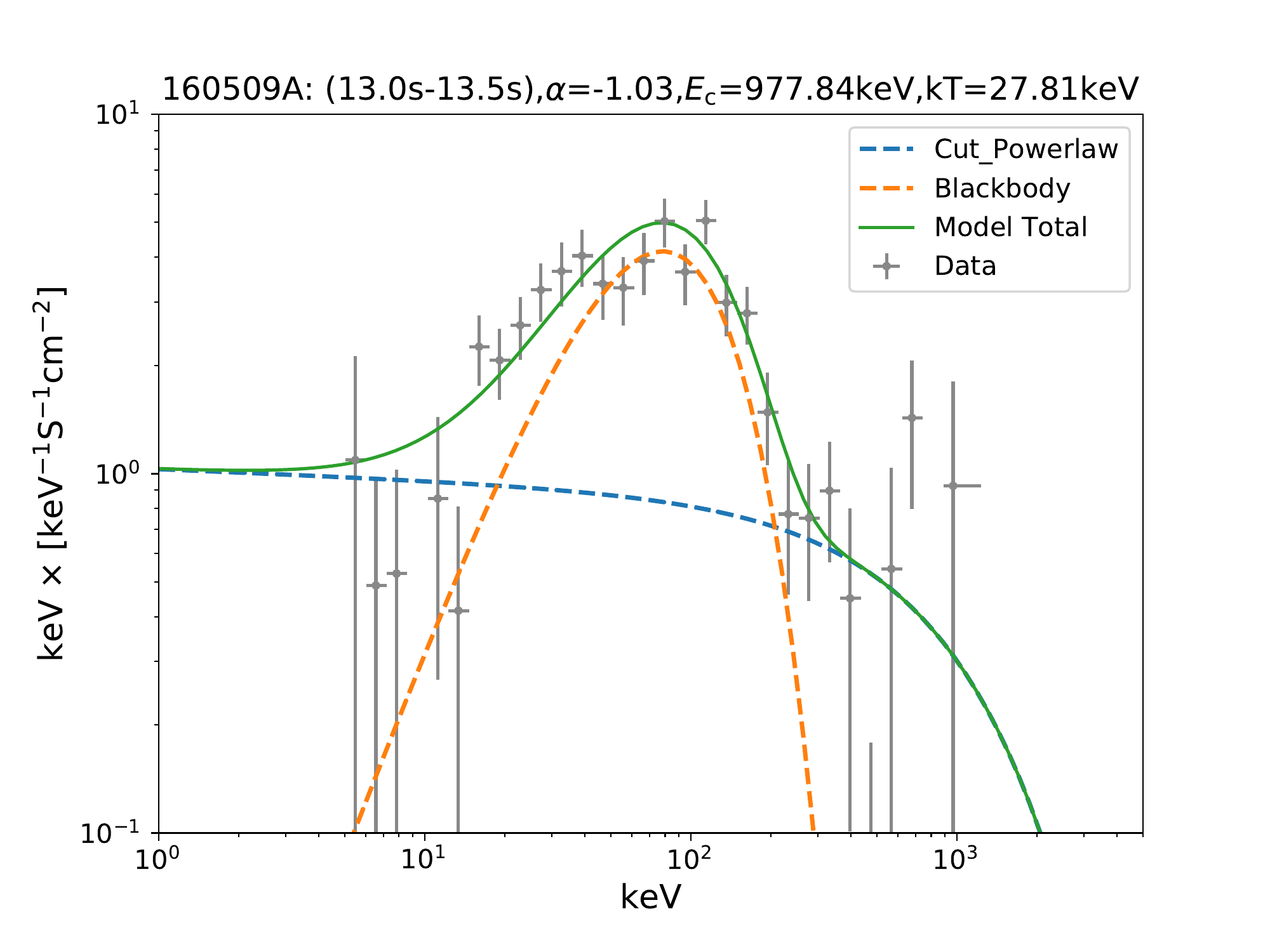}
\includegraphics[angle=0, scale=0.05]{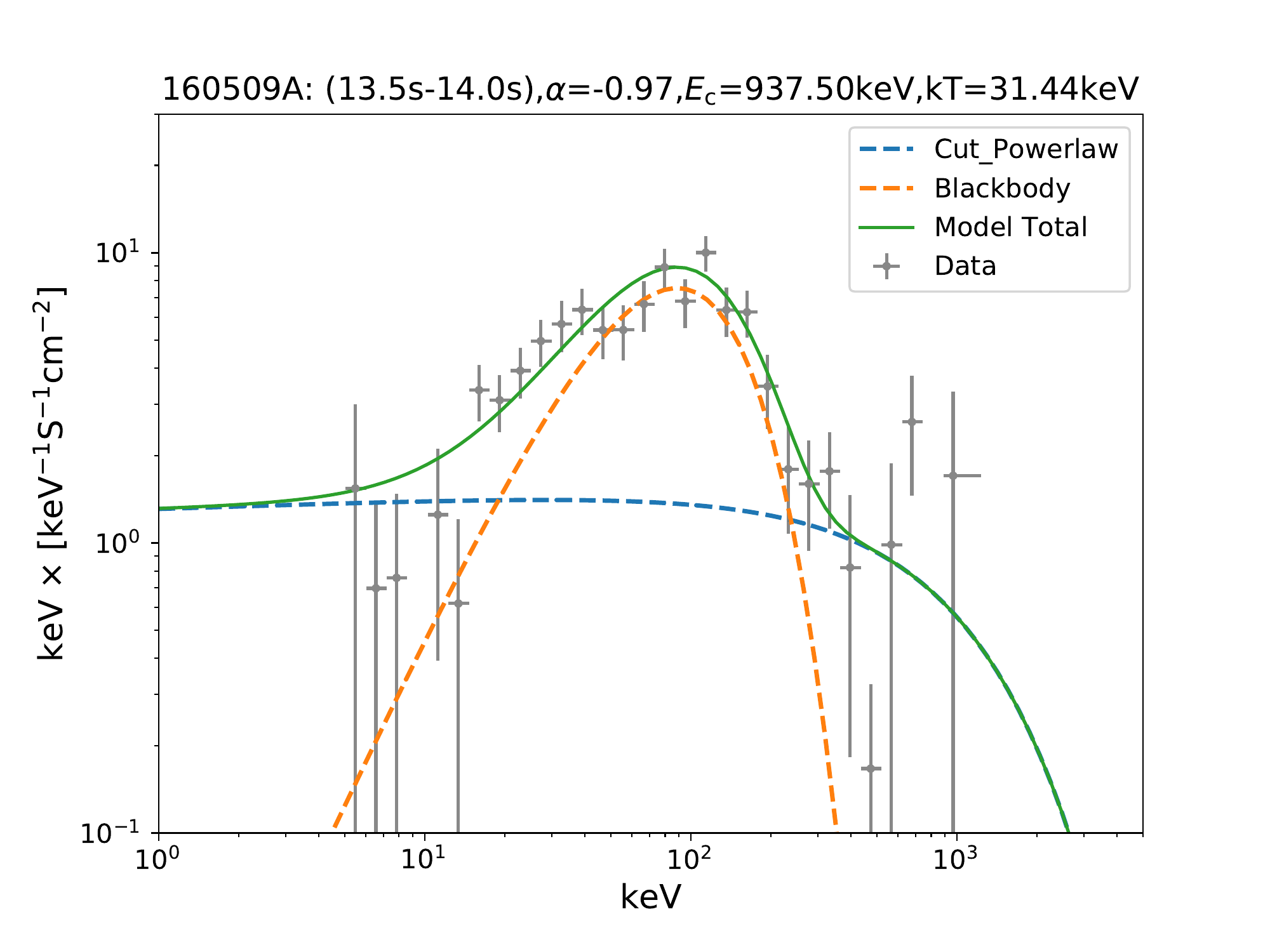}
\includegraphics[angle=0, scale=0.05]{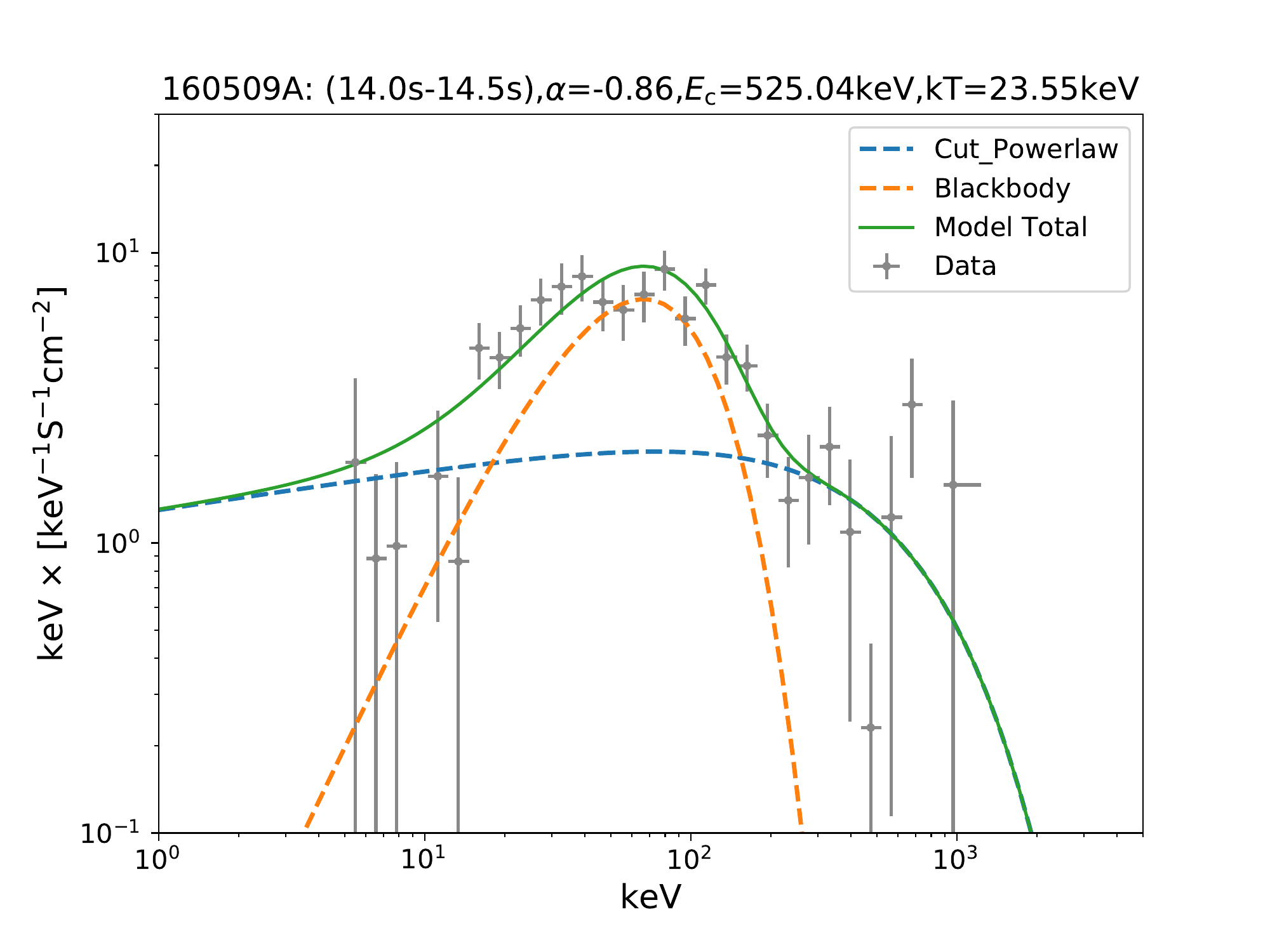}
\includegraphics[angle=0, scale=0.05]{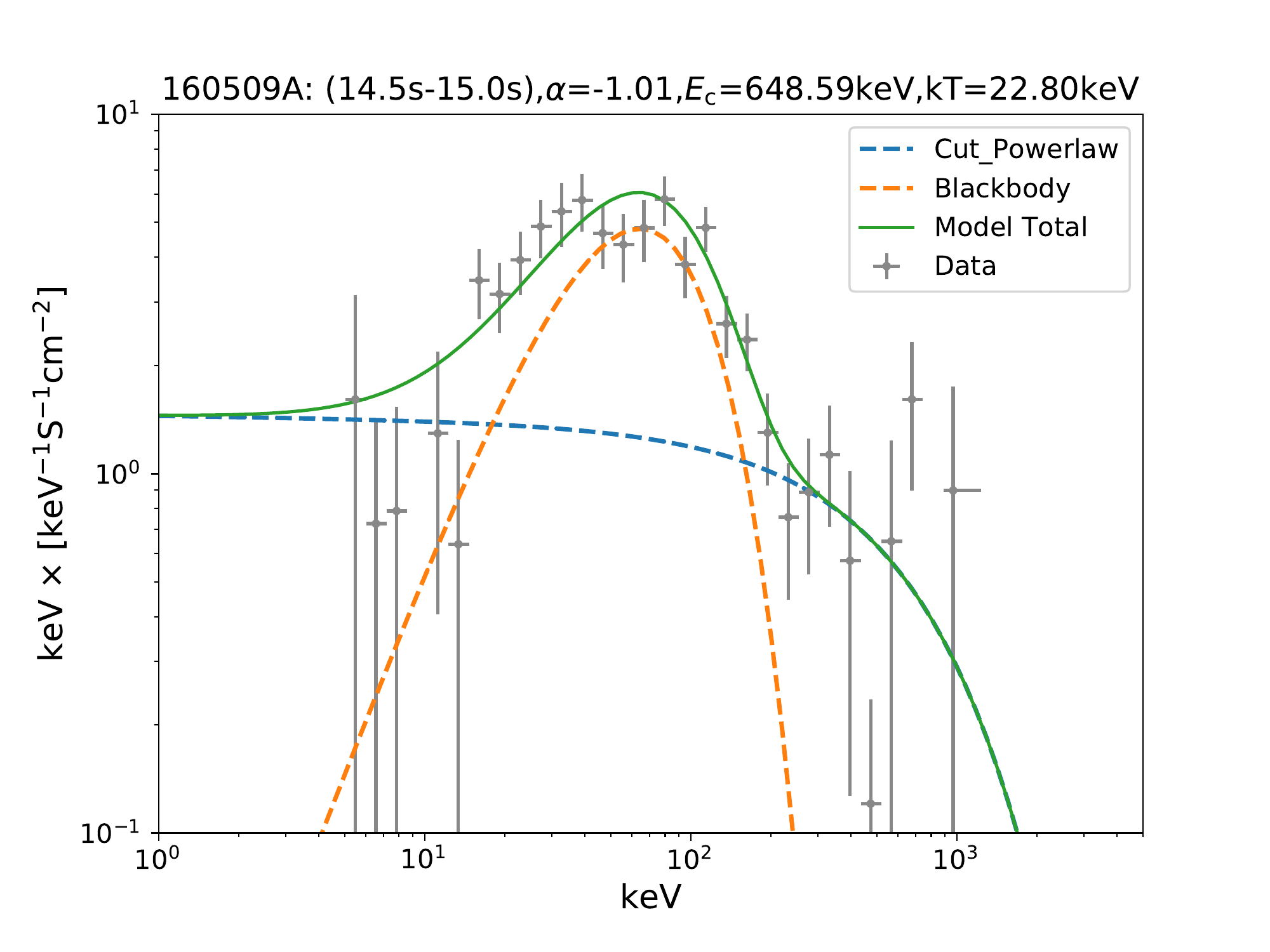}
\includegraphics[angle=0, scale=0.05]{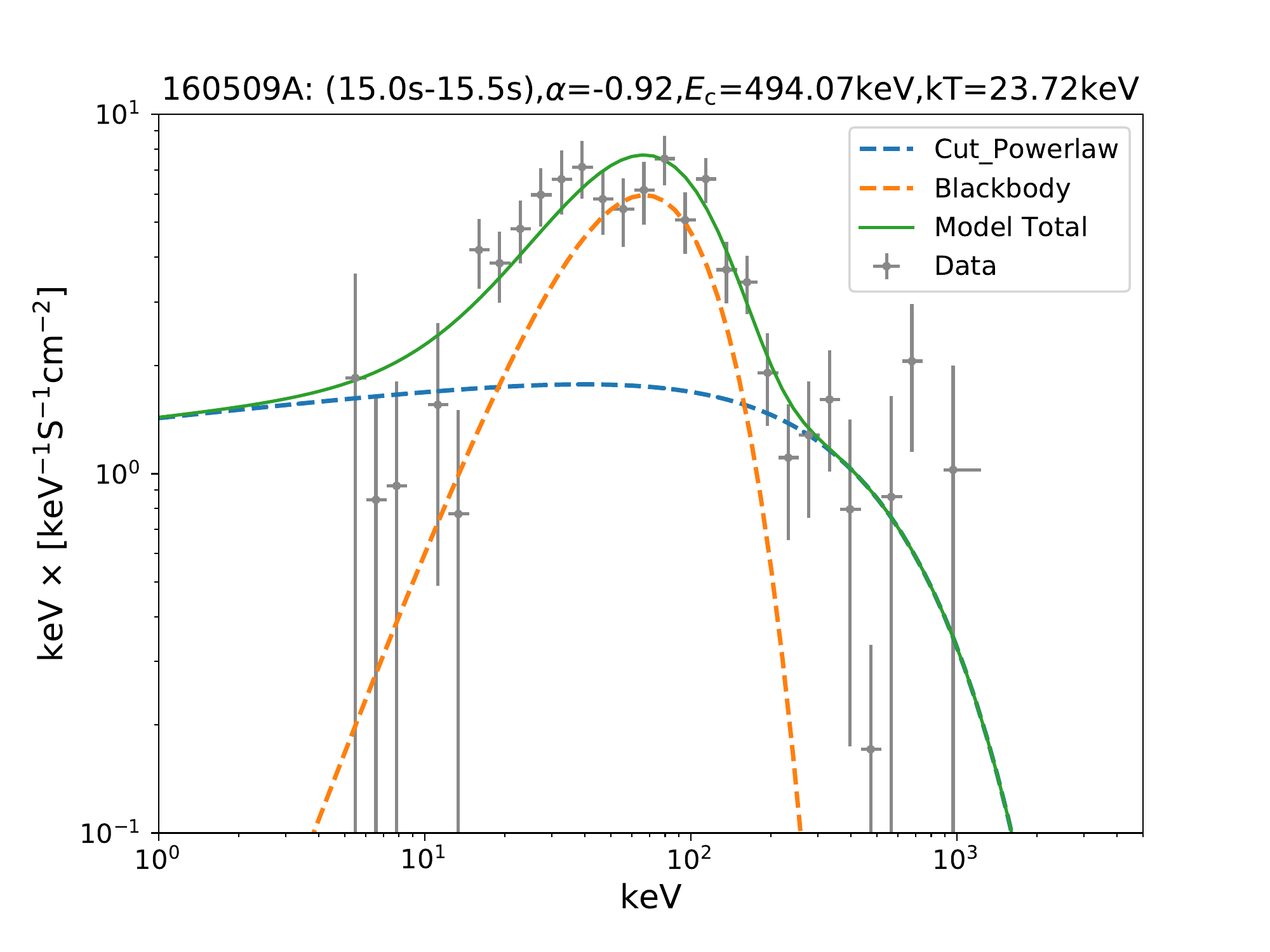}
\includegraphics[angle=0, scale=0.05]{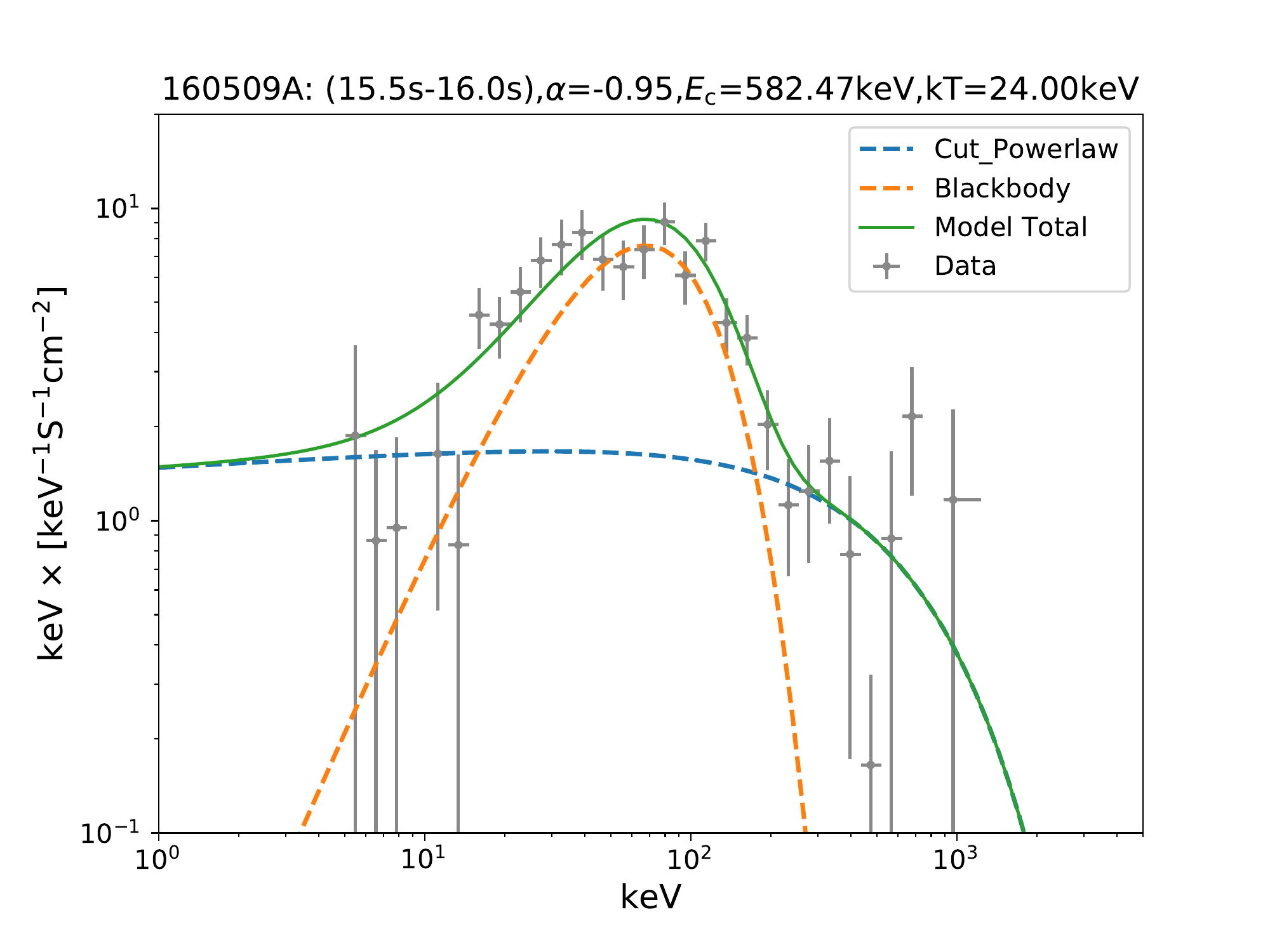}
\includegraphics[angle=0, scale=0.05]{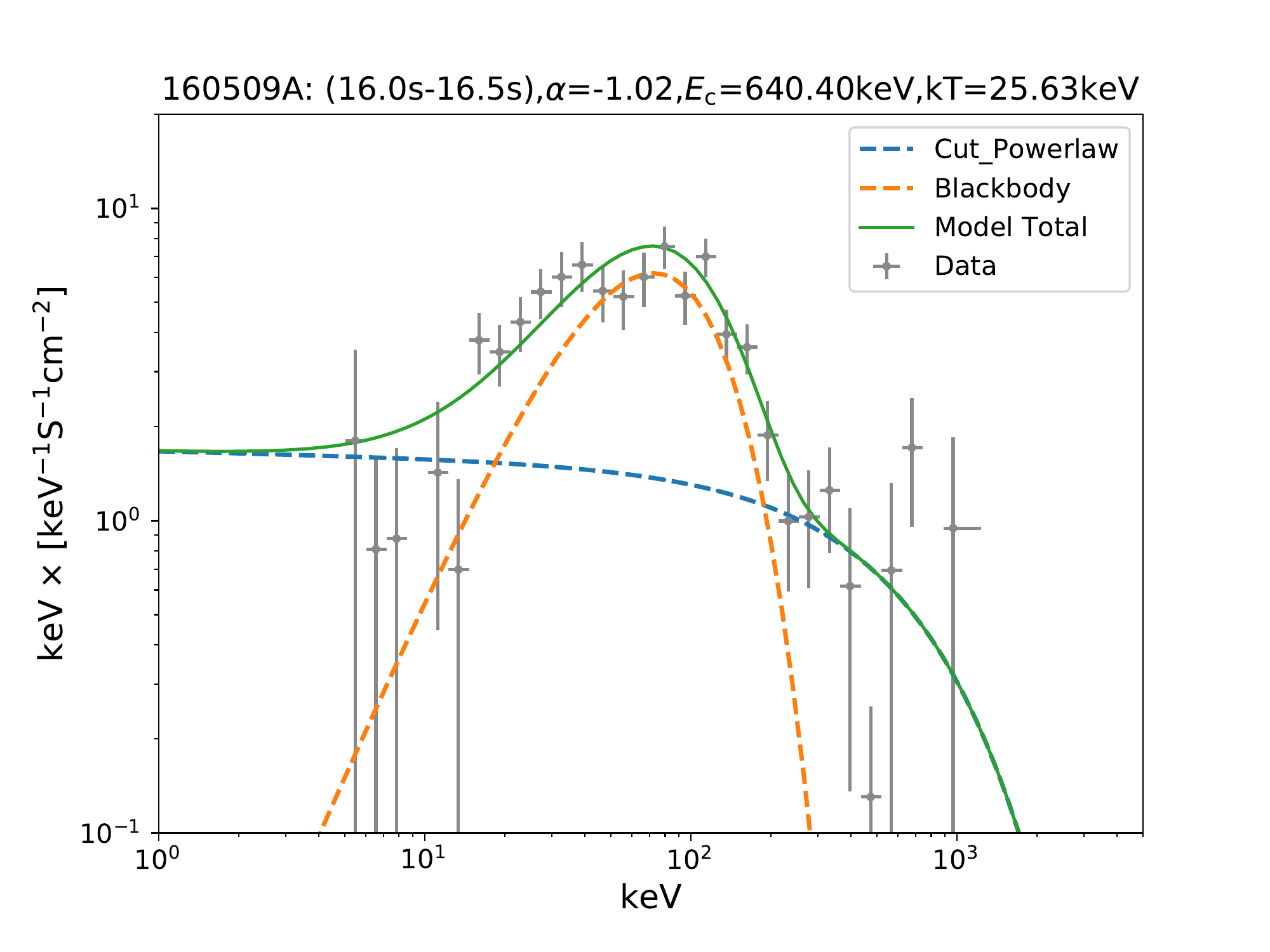}
\includegraphics[angle=0, scale=0.05]{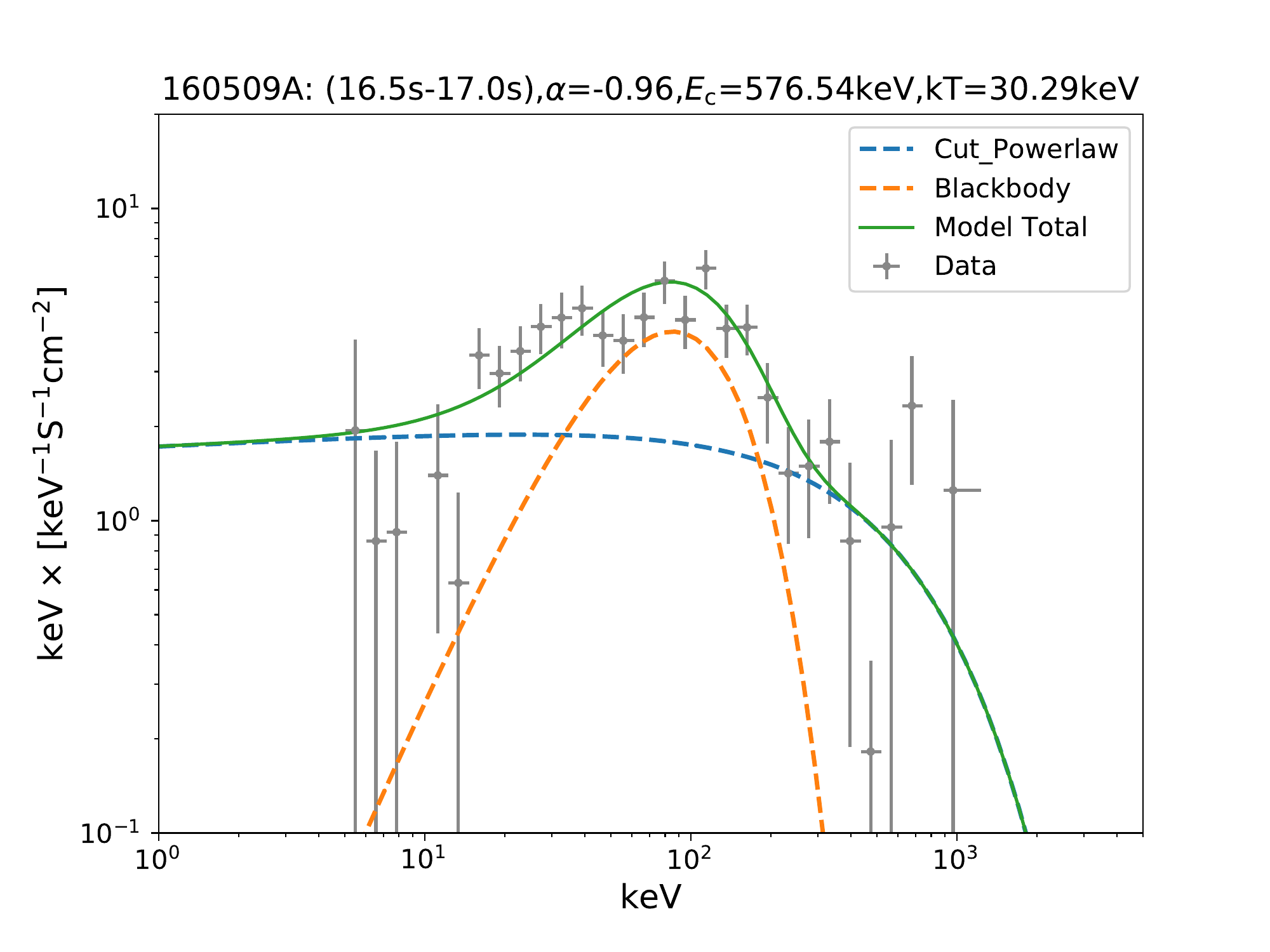}
\includegraphics[angle=0, scale=0.05]{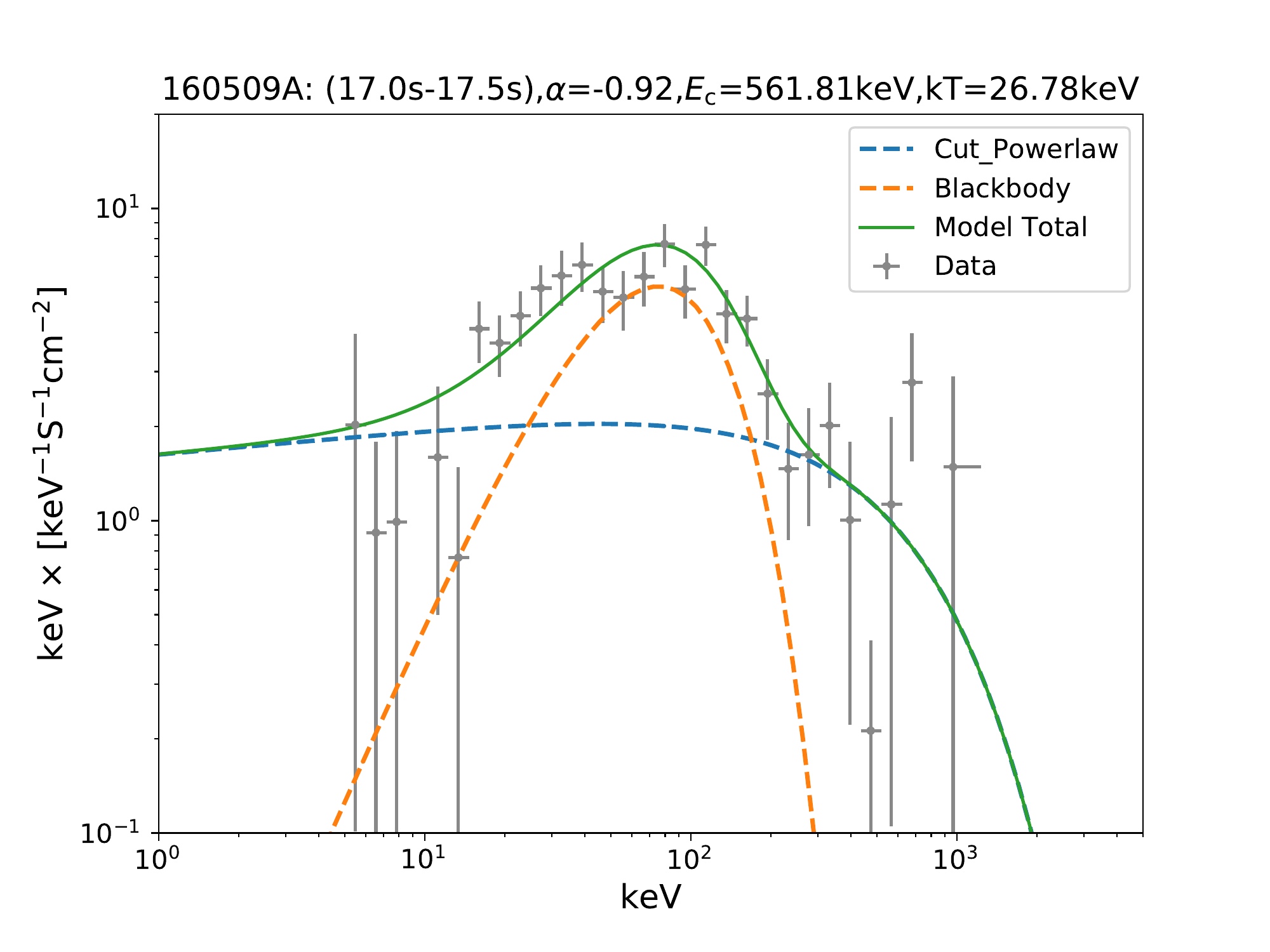}
\includegraphics[angle=0, scale=0.05]{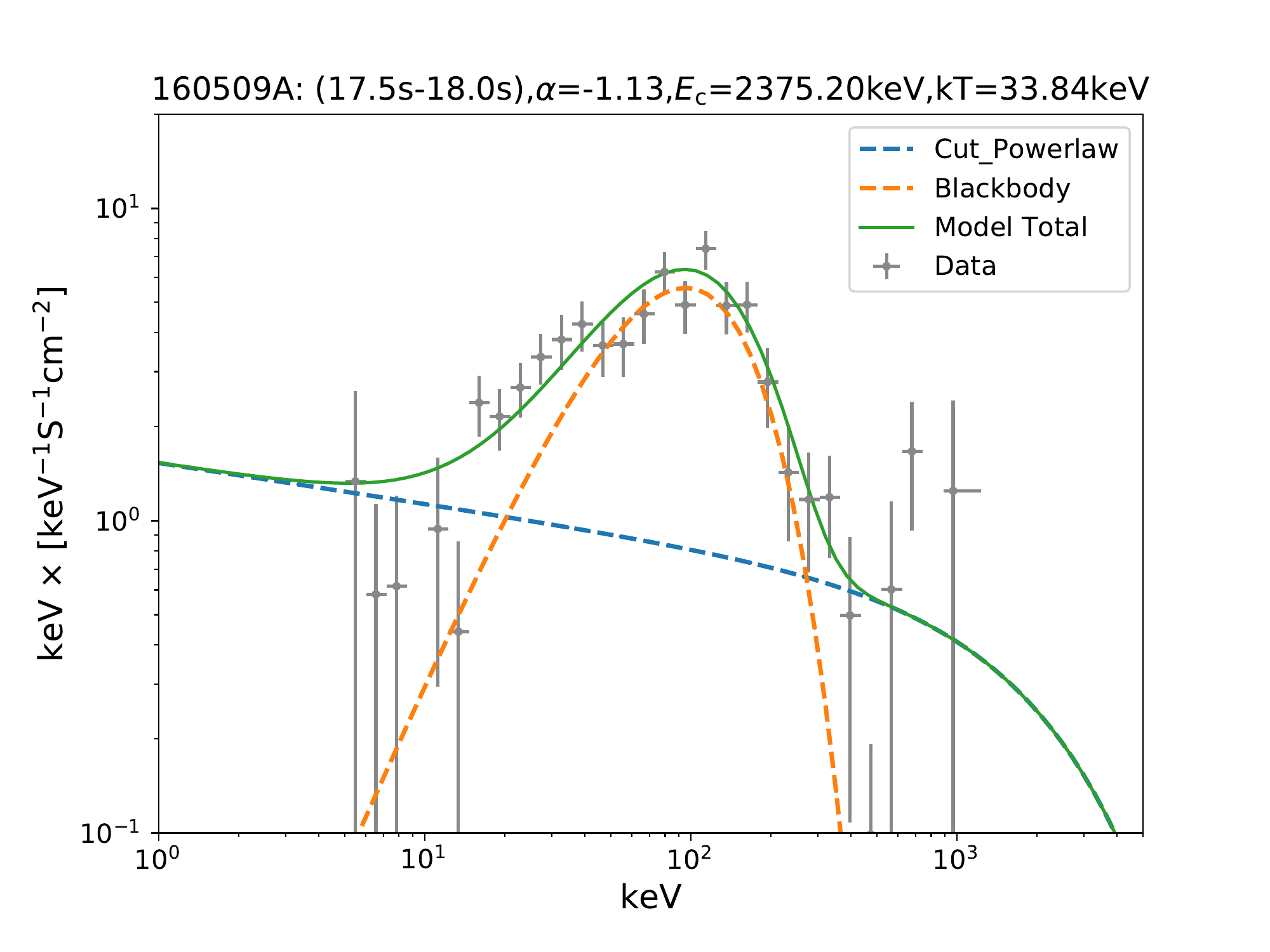}
\includegraphics[angle=0, scale=0.05]{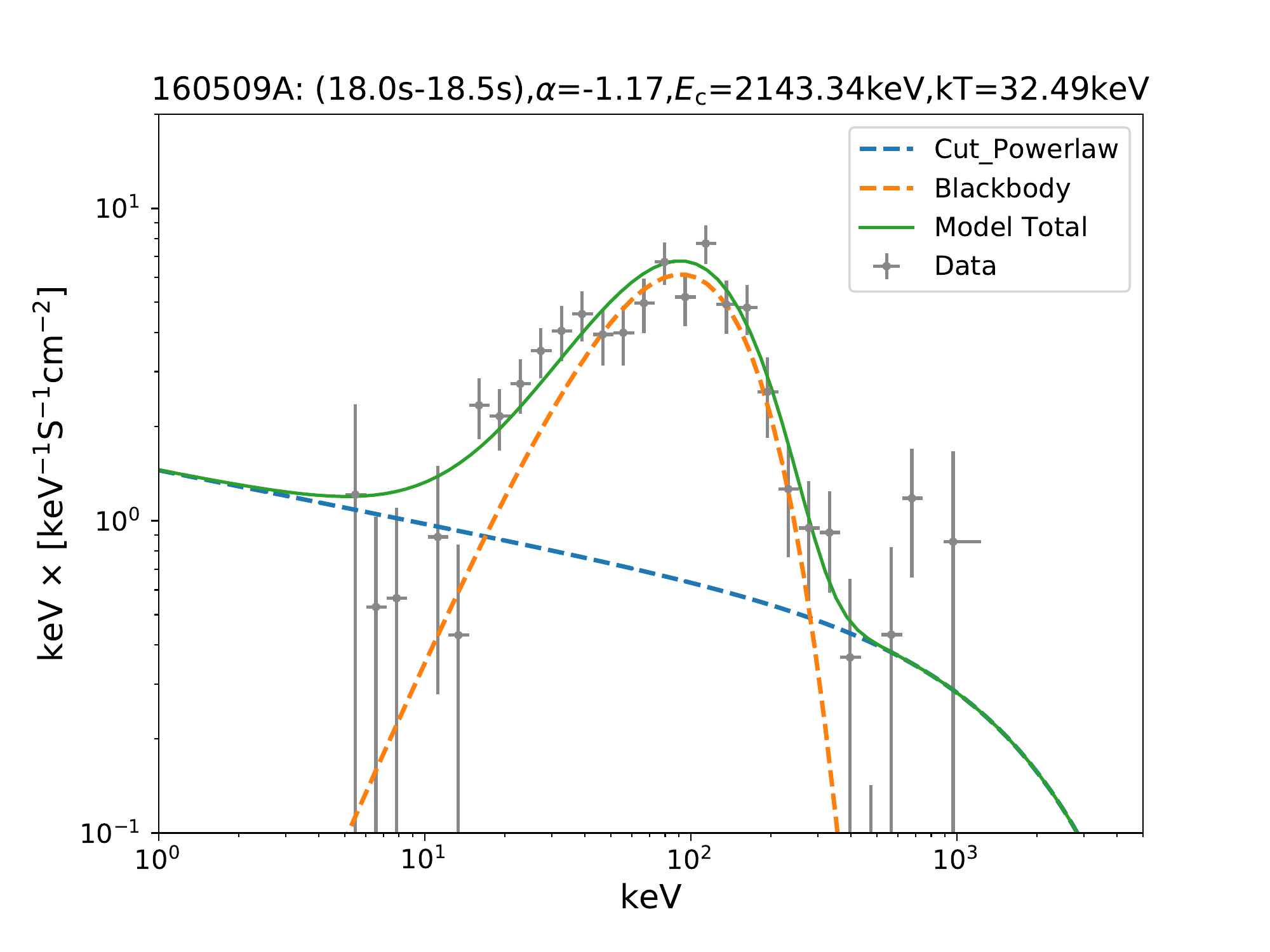}
\caption{Time-resolved spectral analysis of GRB 160509A. All the layers have the same time coverage, from $10.5$~s ($t_{\rm rf} = 4.84$~s) to $18.5$~s ($t_{\rm rf} = 8.53$~s), but with different time divisions: one part (top layer), two equal parts (second layer), four equal parts (third layer), eight equal parts (fourth layer), and sixteen equal parts (bottom layer), respectively. Two dash lines represent CPL (blue) and BB (orange) components, while the solid line represents total model (green). The results of spectral analysis including time duration, temperature and cutoff energy are obtained in the observed frame, as shown in this figure. We have converted to have their corresponding value in the rest-frame, see Table \ref{tab:160509A}, where rest-frame time in column 2, rest-frame cutoff energy in column 5 and rest-frame temperature in column 6.}
\label{fig:spectra_resolved_160509A}
\end{figure*} 

\begin{table*}[ht!]
\small\addtolength{\tabcolsep}{-2pt}
\caption{Results of the time-resolved spectral fits of GRB 160509A (CPL+BB model) from the $t_{\rm rf}=4.84$~s to $t_{\rm rf}=8.53$~s. The definitions of parameters are the same as in table \ref{tab:160625B}.}             
\label{tab:160509A}
\centering                         
\begin{tabular}{ccccccccccc}       
\hline\hline                  
$t_{1}$$\sim$$t_{2}$&$t_{rf,1}$$\sim$$t_{rf,2}$&{$S$}&$\alpha$&$E_{\rm c}$&$kT$&$\Delta$DIC&$F_{\rm BB}$&$F_{\rm tot}$&$F_{\rm ratio}$&$E_{\rm tot}$\\
\hline
(s)&(s)&&&(keV)&(keV)&&(10$^{-6}$)&(10$^{-6}$)&&(erg)\\
Obs&Rest-frame&&&&&&(erg~cm$^{-2}$~s$^{-1}$)&(erg~cm$^{-2}$~s$^{-1}$)\\ 
\hline                        
10.50$\sim$18.50&4.84$\sim$8.53&292.18&-0.96$^{+0.01}_{-0.01}$&745.3$^{+27.6}_{-26.9}$&27.2$^{+0.6}_{-0.6}$&-633.8&0.98$^{+0.13}_{-0.11}$&18.01$^{+1.10}_{-1.00}$&0.05$^{+0.01}_{-0.01}$&5.40e+53\\
\hline
10.50$\sim$14.50&4.84$\sim$6.68&199.13&-0.93$^{+0.02}_{-0.02}$&829.2$^{+47.6}_{-47.0}$&28.7$^{+0.9}_{-0.9}$&-335.4&1.00$^{+0.19}_{-0.15}$&18.50$^{+1.70}_{-1.57}$&0.05$^{+0.01}_{-0.01}$&2.77e+53\\
14.50$\sim$18.50&6.68$\sim$8.53&232.97&-0.99$^{+0.01}_{-0.01}$&679.2$^{+31.9}_{-32.3}$&26.0$^{+0.8}_{-0.8}$&-324.2&0.99$^{+0.19}_{-0.15}$&18.00$^{+1.28}_{-1.27}$&0.05$^{+0.01}_{-0.01}$&2.70e+53\\
\hline
10.50$\sim$12.50&4.84$\sim$5.76&127.55&-0.93$^{+0.02}_{-0.02}$&1054.1$^{+97.5}_{-97.1}$&32.3$^{+1.7}_{-1.7}$&-145.5&0.94$^{+0.30}_{-0.22}$&18.83$^{+2.62}_{-2.40}$&0.05$^{+0.02}_{-0.01}$&1.41e+53\\
12.50$\sim$14.50&5.76$\sim$6.68&161.51&-0.92$^{+0.02}_{-0.02}$&712.8$^{+47.1}_{-47.4}$&26.8$^{+1.1}_{-1.1}$&-205.1&1.08$^{+0.26}_{-0.20}$&18.81$^{+1.96}_{-1.82}$&0.06$^{+0.01}_{-0.01}$&1.41e+53\\
14.50$\sim$16.50&6.68$\sim$7.60&169.80&-0.97$^{+0.02}_{-0.02}$&572.3$^{+32.9}_{-33.2}$&24.0$^{+1.0}_{-1.0}$&-203.4&1.02$^{+0.27}_{-0.20}$&16.14$^{+1.53}_{-1.39}$&0.06$^{+0.02}_{-0.01}$&1.21e+53\\
16.50$\sim$18.50&7.60$\sim$8.53&169.78&-1.04$^{+0.02}_{-0.02}$&931.1$^{+87.8}_{-86.6}$&30.6$^{+1.6}_{-1.6}$&-147.6&1.15$^{+0.31}_{-0.27}$&20.87$^{+2.83}_{-2.24}$&0.06$^{+0.02}_{-0.01}$&1.56e+53\\
\hline
10.50$\sim$11.50&4.84$\sim$5.30&77.67&-0.90$^{+0.04}_{-0.04}$&881.3$^{+135.4}_{-141.4}$&33.0$^{+3.3}_{-3.3}$&-43.8&0.69$^{+0.45}_{-0.27}$&14.27$^{+3.75}_{-2.93}$&0.05$^{+0.03}_{-0.02}$&5.35e+52\\
11.50$\sim$12.50&5.30$\sim$5.76&104.90&-0.95$^{+0.02}_{-0.02}$&1195.3$^{+126.1}_{-124.3}$&31.8$^{+1.9}_{-1.9}$&-117.1&1.21$^{+0.45}_{-0.33}$&23.14$^{+3.53}_{-3.11}$&0.05$^{+0.02}_{-0.02}$&8.67e+52\\
12.50$\sim$13.50&5.76$\sim$6.22&102.77&-0.96$^{+0.03}_{-0.03}$&848.5$^{+91.0}_{-89.5}$&27.0$^{+1.8}_{-1.8}$&-81.4&0.86$^{+0.37}_{-0.25}$&17.63$^{+2.89}_{-2.59}$&0.05$^{+0.02}_{-0.02}$&6.61e+52\\
13.50$\sim$14.50&6.22$\sim$6.68&129.10&-0.90$^{+0.03}_{-0.03}$&638.0$^{+53.1}_{-52.7}$&26.8$^{+1.4}_{-1.4}$&-128.1&1.30$^{+0.38}_{-0.31}$&20.50$^{+3.12}_{-2.65}$&0.06$^{+0.02}_{-0.02}$&7.68e+52\\
14.50$\sim$15.50&6.68$\sim$7.14&117.25&-0.96$^{+0.03}_{-0.03}$&550.6$^{+44.3}_{-44.5}$&23.2$^{+1.5}_{-1.5}$&-85.7&0.86$^{+0.32}_{-0.26}$&15.20$^{+2.37}_{-1.68}$&0.06$^{+0.02}_{-0.02}$&5.69e+52\\
15.50$\sim$16.50&7.14$\sim$7.60&127.21&-0.99$^{+0.03}_{-0.03}$&599.2$^{+52.7}_{-52.5}$&24.7$^{+1.3}_{-1.3}$&-124.5&1.17$^{+0.40}_{-0.28}$&17.14$^{+2.23}_{-2.19}$&0.07$^{+0.03}_{-0.02}$&6.42e+52\\
16.50$\sim$17.50&7.60$\sim$8.06&131.95&-0.95$^{+0.03}_{-0.03}$&571.3$^{+47.4}_{-46.6}$&28.5$^{+2.3}_{-2.4}$&-49.8&0.90$^{+0.46}_{-0.34}$&19.51$^{+2.69}_{-2.19}$&0.05$^{+0.02}_{-0.02}$&7.31e+52\\
17.50$\sim$18.50&8.06$\sim$8.53&112.19&-1.15$^{+0.02}_{-0.02}$&2226.8$^{+325.2}_{-326.2}$&32.9$^{+1.9}_{-1.9}$&-133.0&1.33$^{+0.47}_{-0.34}$&27.25$^{+4.66}_{-3.82}$&0.05$^{+0.02}_{-0.01}$&1.02e+53\\
\hline
10.50$\sim$11.00&4.84$\sim$5.07&48.87&-0.87$^{+0.06}_{-0.06}$&804.0$^{+189.6}_{-191.8}$&33.4$^{+8.6}_{-8.3}$&-23.1&0.33$^{+0.86}_{-0.24}$&12.55$^{+5.59}_{-3.97}$&0.03$^{+0.07}_{-0.02}$&2.35e+52\\
11.00$\sim$11.50&5.07$\sim$5.30&61.64&-0.93$^{+0.05}_{-0.05}$&1004.5$^{+202.6}_{-211.2}$&33.3$^{+3.5}_{-3.5}$&-40.8&1.00$^{+0.67}_{-0.42}$&16.82$^{+6.06}_{-4.34}$&0.06$^{+0.05}_{-0.03}$&3.15e+52\\
11.50$\sim$12.00&5.30$\sim$5.53&74.34&-0.89$^{+0.05}_{-0.05}$&875.2$^{+145.3}_{-147.8}$&28.6$^{+2.4}_{-2.4}$&-64.6&1.13$^{+0.55}_{-0.37}$&19.22$^{+5.94}_{-4.13}$&0.06$^{+0.03}_{-0.02}$&3.60e+52\\
12.00$\sim$12.50&5.53$\sim$5.76&75.45&-0.98$^{+0.03}_{-0.03}$&1487.0$^{+208.8}_{-205.9}$&35.1$^{+3.4}_{-3.4}$&-57.0&1.16$^{+0.74}_{-0.44}$&27.26$^{+6.00}_{-4.46}$&0.04$^{+0.03}_{-0.02}$&5.11e+52\\
12.50$\sim$13.00&5.76$\sim$5.99&81.26&-0.91$^{+0.03}_{-0.03}$&786.1$^{+96.3}_{-96.7}$&26.9$^{+2.3}_{-2.3}$&-49.0&0.94$^{+0.54}_{-0.34}$&20.82$^{+4.08}_{-3.74}$&0.05$^{+0.03}_{-0.02}$&3.90e+52\\
13.00$\sim$13.50&5.99$\sim$6.22&65.10&-1.03$^{+0.05}_{-0.05}$&977.8$^{+201.9}_{-199.4}$&27.8$^{+3.1}_{-3.1}$&-39.8&0.77$^{+0.62}_{-0.34}$&14.55$^{+4.63}_{-3.02}$&0.05$^{+0.05}_{-0.03}$&2.73e+52\\
13.50$\sim$14.00&6.22$\sim$6.45&90.78&-0.97$^{+0.04}_{-0.04}$&937.5$^{+151.2}_{-151.4}$&31.4$^{+2.1}_{-2.1}$&-77.7&1.62$^{+0.68}_{-0.42}$&23.85$^{+6.01}_{-4.95}$&0.07$^{+0.03}_{-0.02}$&4.47e+52\\
14.00$\sim$14.50&6.45$\sim$6.68&93.73&-0.86$^{+0.04}_{-0.04}$&525.0$^{+50.2}_{-49.4}$&23.6$^{+1.9}_{-1.9}$&-65.2&1.13$^{+0.61}_{-0.37}$&18.94$^{+3.46}_{-3.06}$&0.06$^{+0.03}_{-0.02}$&3.55e+52\\
14.50$\sim$15.00&6.68$\sim$6.91&80.00&-1.01$^{+0.04}_{-0.04}$&648.6$^{+79.1}_{-80.5}$&22.8$^{+2.2}_{-2.2}$&-41.5&0.75$^{+0.51}_{-0.30}$&15.08$^{+3.14}_{-2.32}$&0.05$^{+0.04}_{-0.02}$&2.82e+52\\
15.00$\sim$15.50&6.91$\sim$7.14&87.43&-0.92$^{+0.04}_{-0.04}$&494.1$^{+51.9}_{-50.7}$&23.7$^{+1.9}_{-1.9}$&-50.0&0.96$^{+0.54}_{-0.33}$&15.80$^{+3.08}_{-2.56}$&0.06$^{+0.04}_{-0.02}$&2.96e+52\\
15.50$\sim$16.00&7.14$\sim$7.37&91.73&-0.95$^{+0.04}_{-0.04}$&582.5$^{+63.7}_{-65.1}$&24.0$^{+1.6}_{-1.6}$&-80.8&1.30$^{+0.56}_{-0.39}$&17.48$^{+3.38}_{-2.64}$&0.07$^{+0.03}_{-0.02}$&3.27e+52\\
16.00$\sim$16.50&7.37$\sim$7.60&90.06&-1.02$^{+0.04}_{-0.04}$&640.4$^{+91.4}_{-92.2}$&25.6$^{+2.2}_{-2.2}$&-51.2&1.10$^{+0.58}_{-0.41}$&16.62$^{+3.91}_{-2.84}$&0.07$^{+0.04}_{-0.03}$&3.11e+52\\
16.50$\sim$17.00&7.60$\sim$7.83&90.67&-0.96$^{+0.04}_{-0.04}$&576.5$^{+78.4}_{-77.9}$&30.3$^{+5.5}_{-5.2}$&-25.1&0.71$^{+0.87}_{-0.48}$&18.83$^{+4.75}_{-3.42}$&0.04$^{+0.05}_{-0.03}$&3.53e+52\\
17.00$\sim$17.50&7.83$\sim$8.06&97.88&-0.92$^{+0.04}_{-0.04}$&561.8$^{+62.8}_{-63.6}$&26.8$^{+2.8}_{-2.8}$&-40.2&1.03$^{+0.72}_{-0.42}$&20.40$^{+3.97}_{-3.30}$&0.05$^{+0.04}_{-0.02}$&3.82e+52\\
17.50$\sim$18.00&8.06$\sim$8.29&82.94&-1.13$^{+0.03}_{-0.03}$&2375.2$^{+440.5}_{-440.1}$&33.8$^{+3.1}_{-3.2}$&-68.8&1.35$^{+0.83}_{-0.55}$&31.18$^{+7.03}_{-5.27}$&0.04$^{+0.03}_{-0.02}$&5.84e+52\\
18.00$\sim$18.50&8.29$\sim$8.53&77.25&-1.17$^{+0.04}_{-0.04}$&2143.3$^{+521.4}_{-532.9}$&32.5$^{+2.5}_{-2.5}$&-65.0&1.37$^{+0.74}_{-0.45}$&23.06$^{+6.84}_{-4.79}$&0.06$^{+0.04}_{-0.02}$&4.32e+52\\
\hline                                   
\end{tabular}
\end{table*}

\begin{figure*}[h!]
\centering
\includegraphics[width=0.49\hsize,clip]{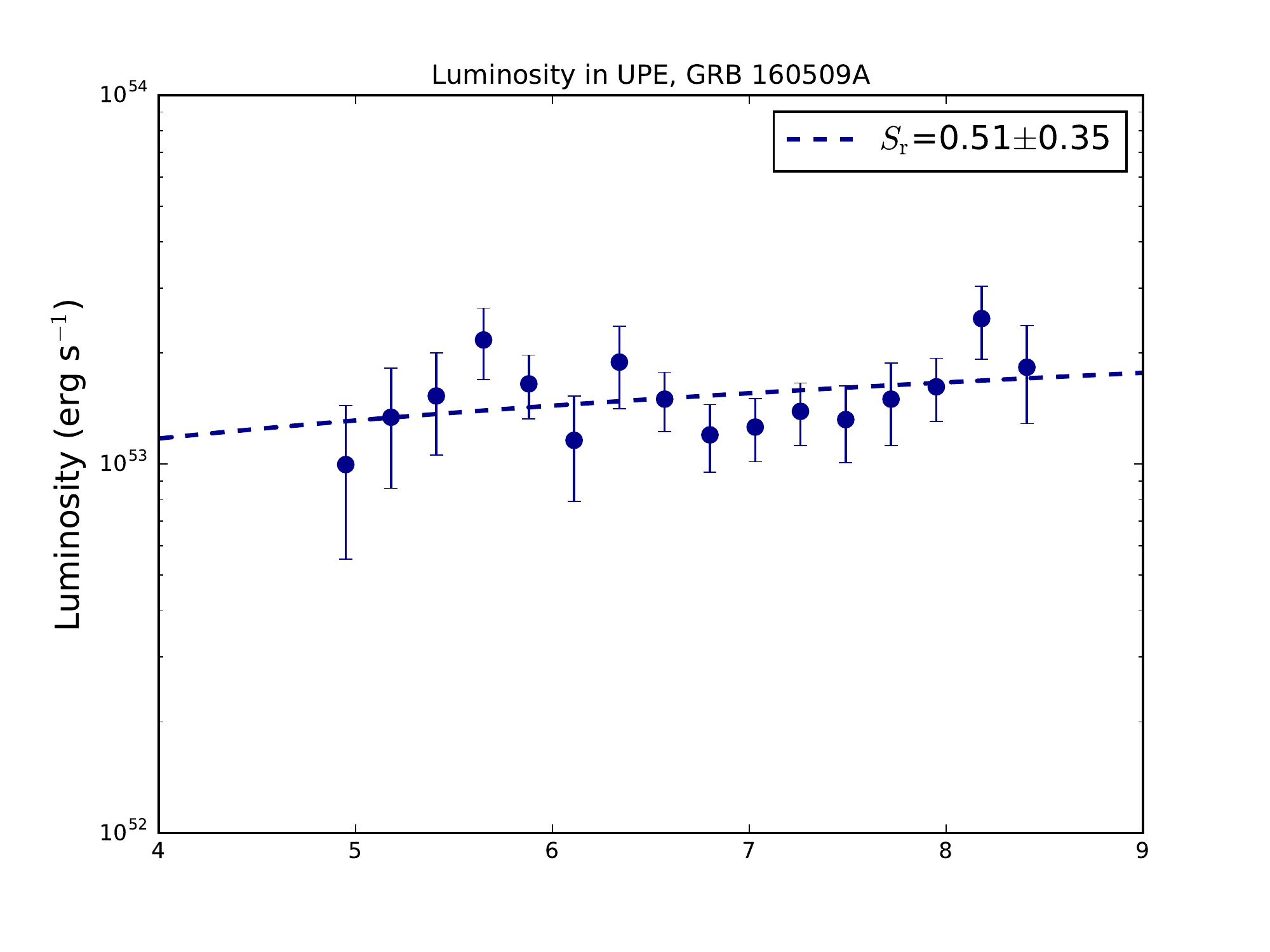}
\includegraphics[width=0.49\hsize,clip]{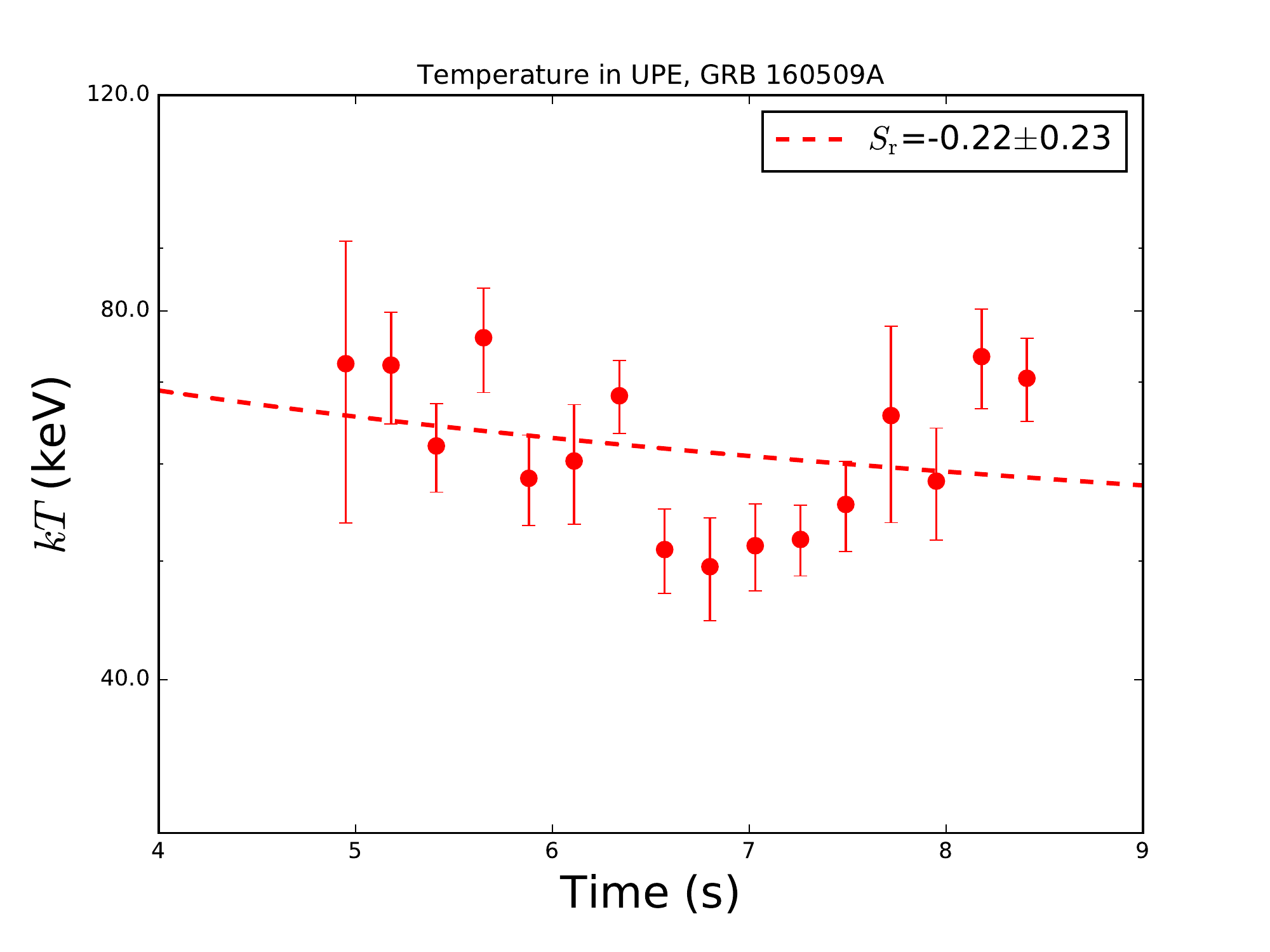}
\caption{\textbf{Left}: GRB 160509A luminosity light-curve of the UPE as derived from the fifth iteration with 16 sub-intervals. The values of the  best fit parameters from Table~\ref{tab:160509A} are used to apply the k-correction and measuring the luminosity as a function of time. The power law index of $0.51 \pm 0.35$ of the luminosity is similar to the one obtained in the GeV emission luminosity {after the UPE phase} with index of $-0.22 \pm 0.23$. For more information about GeV luminosity behavior see \citep{2019arXiv190107505W} and the companion paper (Ruffini, Moradi et al, 2019, submitted). \textbf{Right}: Time evolution of the rest-frame temperature of the UPE as derived from the fifth iteration with $16$ sub-intervals, is reported in Table~\ref{tab:160509A}.}
\label{iterativ160509A}
\end{figure*}

Based on the temporal and spectral analysis, three different physical processes are identified in the keV-MeV energy range (see Fig.~\ref{fig:lc_160509A}, Fig.~\ref{fig:joinedGRB160509A}, and Table~\ref{table:episodes_160509A}):
(1). SN-rise, the time interval ranges from $t_{\rm rf}=0.92$~s to $t_{\rm rf}=1.84$~s.
(2). UPE phase, the time interval ranges from $t_{\rm rf}=4.84$~s to $t_{\rm rf}=8.53$~s.
(3). Cavity, the time interval ranges from $t_{\rm rf}=10.14$~s to $t_{\rm rf}=13.82$~s.

--- \textit{SN-rise}. {Figure~\ref{fig:joinedGRB160509A} (upper left panel) shows the SN-rise and the spectral fitting of the cavity emission during its rest-frame time interval of occurrence. The spectrum of the SN-rise of GRB 160509A are best fitted by CPL+BB model, from  $t_{\rm rf} \simeq 0.92$s to $t_{\rm rf} \simeq 1.84$s. The spectrum contains a BB component of temperature $25.61$~keV and a photon index $\alpha$ of $-1.22$, and $E_{\rm c}=1769.76$~keV.}

--- \textit{UPE phase}. We perform the corresponding time-resolved spectral analysis from which we can see that the self-similarity first discovered in GRB 190114C is confirmed in the case of GRB 160509A. {For the first iteration, we present the best-fit of the spectrum of the UPE entire duration from $t_{\rm rf}=4.84$~s to $t_{\rm rf}=8.53$~s (see Fig.~\ref{fig:spectra_resolved_160509A}, first layer).} 

We then divide the rest-frame time interval in half and perform again the same spectral analysis for the two 1.85 second interval, namely [4.84s-6.68s] and [6.68s-8.53s], obtaining the results shown in Fig.~\ref{fig:spectra_resolved_160509A}. Iteration 3: we then divide each of these half intervals again in half, i.e., $\Delta t_{\rm rf}$= 0.92s corresponding to [4.84s-5.76s], [5.76s-6.68s], [6.68s-7.60s] and [7.60s-8.53s] and redo the previous spectral analysis obtaining the results still in Fig.~\ref{fig:spectra_resolved_160509A}. In a fourth iteration we divide the UPE into 8 sub-intervals of $\Delta t_{\rm rf}$= 0.46s corresponding to the time intervals [4.84s-5.30s], [5.30s-5.76s], [5.76s-6.22s], [6.22s-6.68s], [6.68s-7.14s], [7.14s-7.60s], [7.60s-8.06s] and [8.06s-8.53s], and redo the spectral analysis (see in Fig.~\ref{fig:spectra_resolved_160509A}). In the fifth and final iteration of this process we divide the UPE into 16 sub-intervals of $\Delta t_{\rm rf}$= 0.23s corresponding we perform the spectral analysis and find the self-similar CPL+BB emission in the time intervals [4.84s-5.07s], [5.07s-5.30s], [5.30s-5.53s], [5.53s-5.76s], [5.76s-5.99s], [5.99s-6.22s], [6.22s-6.45s], [6.45s-6.68s], [6.68s-6.91s], [6.91s-7.14s], [7.14s-7.37s], [7.37s-7.60s], [7.60s-7.83s], [7.83s-8.06s], [8.06s-8.29s] and [8.29s-8.53s], and perform the spectral analysis, see Fig.~\ref{fig:spectra_resolved_160509A}. 

--- \textit{Cavity}. Figure~\ref{fig:joinedGRB160509A} (upper right panel) shows the spectral fitting of the cavity emission during the rest-frame time interval of its occurrence, i.e. from $t_{\rm rf}= 10.14$~s to $13.82$~s. The best fit of the spectrum is a CPL model, with a photon index $\alpha$ is $-1.20$ and the cutoff energy is $314$~keV.  

--- \textit{GeV emission}. Figure~\ref{fig:joinedGRB160509A} (lower left panel) shows the luminosity of the GeV emission and the luminosity in the afterglow are given as a function of the rest-frame time.

--- \textit{Afterglow}. Figure~\ref{fig:joinedGRB160509A} (lower right panel) shows the (k-corrected) rest-frame afterglow luminosity ({\it Swift}/XRT data) as a function of rest-frame time, and obtained a best fit parameters with the power-law index of $-1.259\pm 0.025$.

\section{GRB 130427A} \label{sec:GRB130427A}

A very bright burst, GRB 130427A, was announced by {\it Fermi}-GBM at 07:47:06.42 UT on April 27 2013 \citep{2013GCN.14473....1V}.  \textit{Swift}-BAT was triggered $51.1$~s later, \textit{Swift}-UVOT and \textit{Swift}-XRT began to observe at 181 s and 195 s after the trigger \citep{2013GCN.14448....1M}. Its redshift of $z = 0.34$ was detected and confirmed by  Gemini North telescope \citep{2013GCN..14455...1L},  Nordic Optical Telescope (NOT) \citep{2013GCN..14478...1X} and VLT/X-shooter \citep{2013GCN..14491...1F}.  The isotropic energy is $1.4 \times 10^{54}$~erg, as details in \citet{2013GCN..14455...1L, 2013GCN.14473....1V, 2013GCN..14478...1X, 2013GCN..14491...1F, 2015ApJ...798...10R}, GRB 130427A records a well observed fluence in the optical, X-ray, gamma-ray and GeV bands. 

\begin{figure*}[ht!]
\centering
\includegraphics[width=0.96\hsize,clip]{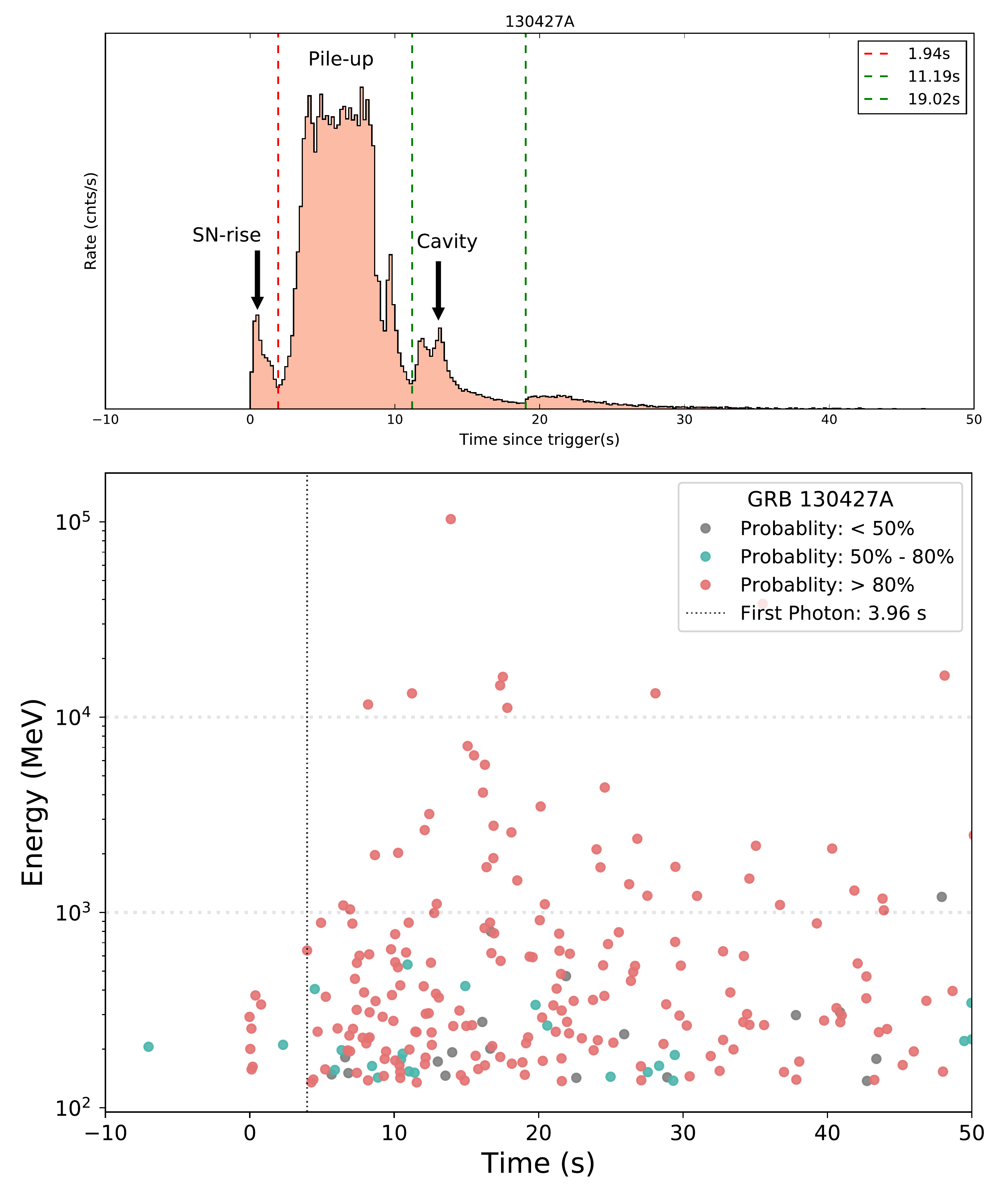}
\caption{\textbf{Upper panel}: Time structure of the prompt emission phase of GRB 130427A presented in the rest-frame. The \textit{Fermi}-GBM observation is strongly piled up due to the high flunece, hence the detection of each episode, especially the starting time of UPE phase can not be determined accurately. \textbf{Lower panel}:  the energy and time of each {\it Fermi}-LAT photon in the rest-frame, the first GeV energy photon occurs at {$t_{rf}=3.96$~s}. The onset of the GeV radiation coincides with the onset of the UPE. The detailed information for each episode {(SN-rise, cavity, GeV, and afterglow)}, see section \ref{sec:GRB130427A} and Table \ref{table:episodes_130427A}, which include the starting time, the duration, the isotropic energy, and the preferred model.}
\label{fig:lc_130427A}
\end{figure*}

\begin{table*}[ht!]
\caption{The Episodes of GRB 130427A, the definitions of  parameters are the same as in Table~\ref{table:episodes_160625B}.}             
\label{table:episodes_130427A}
\centering                         
\begin{tabular}{cccccc}       
\hline\hline                  
Episode &Starting Time&Ending Time&Energy&Spectrum&Reference\\   &Rest-frame&Rest-frame\\  
\hline                        
SN-rise&$  0 $~s&$ 0.49 $~s&$ 6.5 \times10^{51}$~erg&CPL+BB&New in this paper\\
UPE&$ 1.94$~s&$ 11.19$~s&$\sim 1.4\times10^{54}$~erg&CPL+BB&New in this paper\\
Cavity&$ 11.19$~s&$ 19.03$~s&$ 1.97\times 10^{52}$~erg&CPL&New in this paper\\
GeV&$ 3.96$~s&  $> 2 \times 10^4$~s&$ 5.69 \times 10^{52}$~erg&PL&\cite{2015ApJ...798...10R}\\
Afterglow&$ 107$~s&$ >10$ days&$2.65 \times 10^{52}$~erg&PL&\cite{2015ApJ...798...10R}\\
\hline                                   
\end{tabular}
\end{table*}

\begin{figure*}[ht!]
\centering
\includegraphics[width=0.49\hsize,clip]{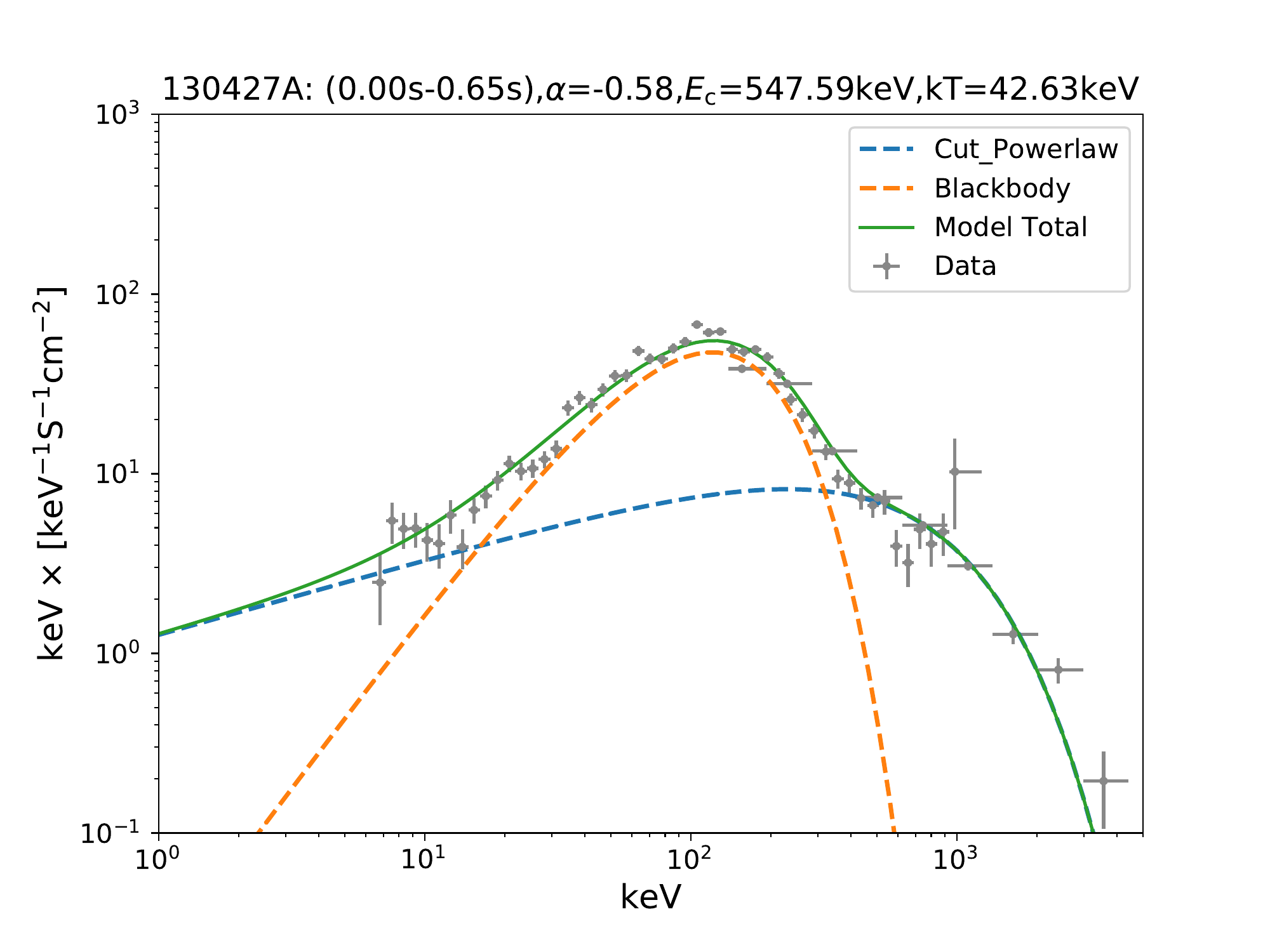}
\includegraphics[width=0.49\hsize,clip]{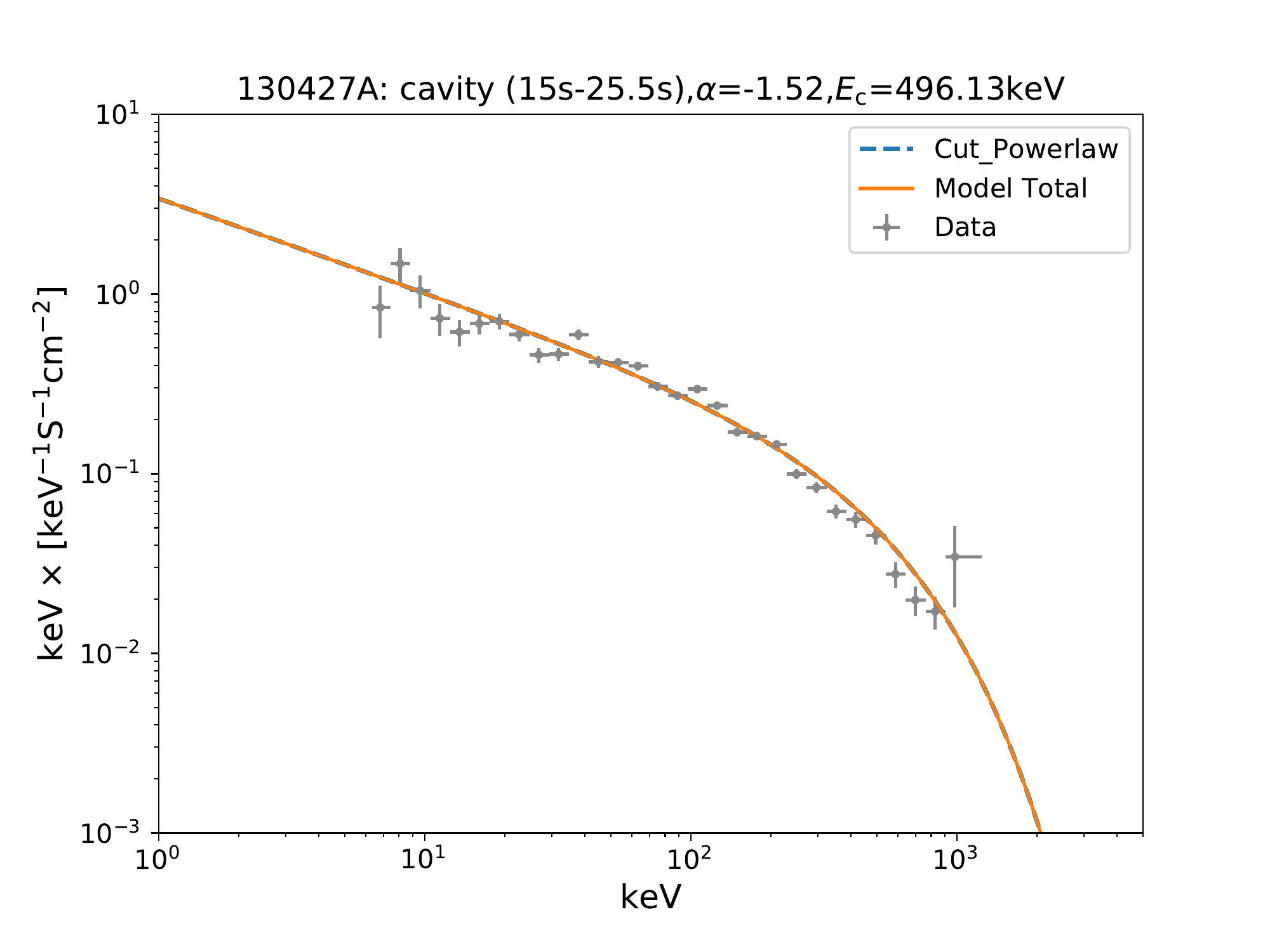}\\
\includegraphics[width=0.49\hsize,clip]{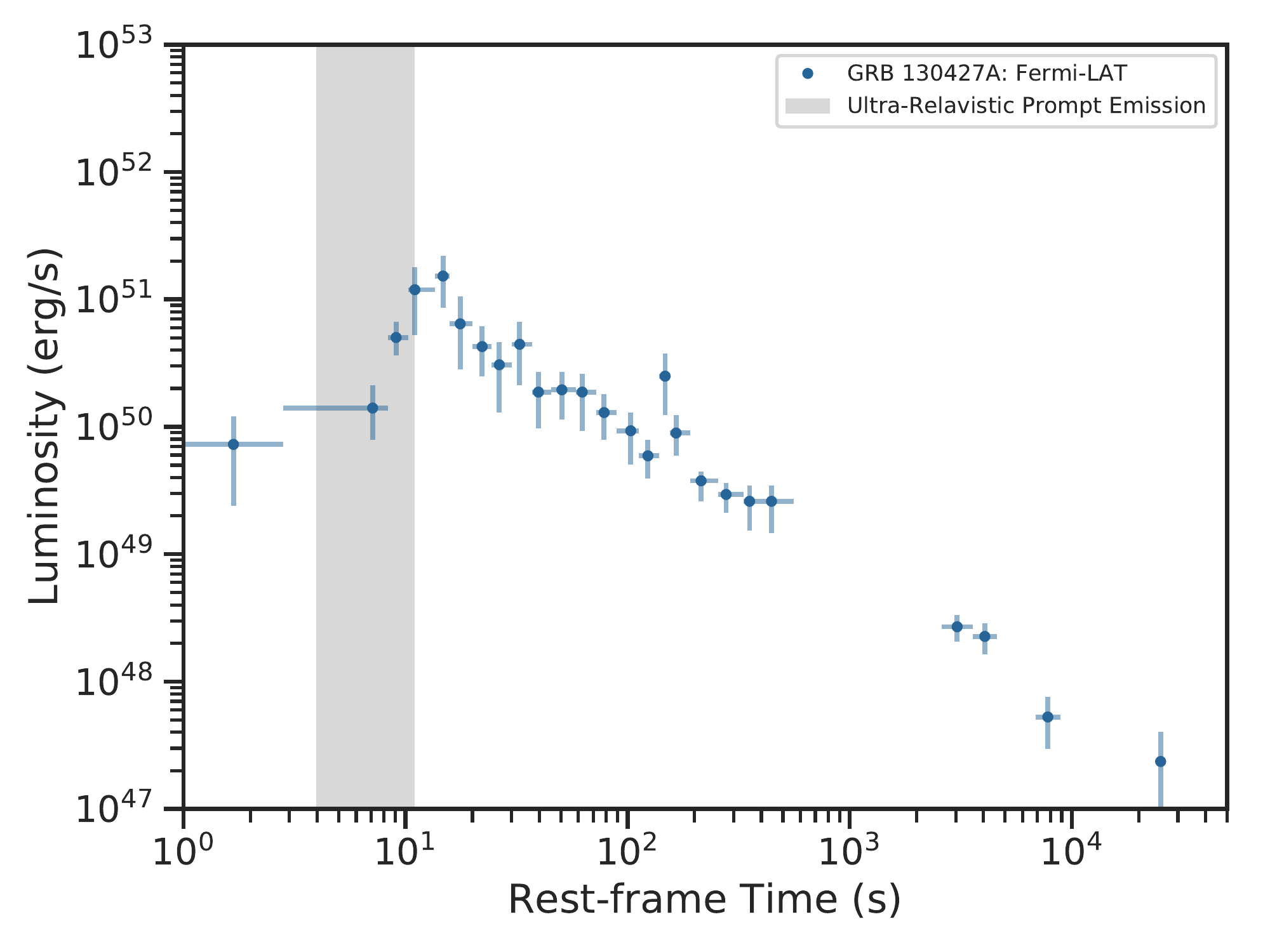}
\includegraphics[width=0.49\hsize,clip]{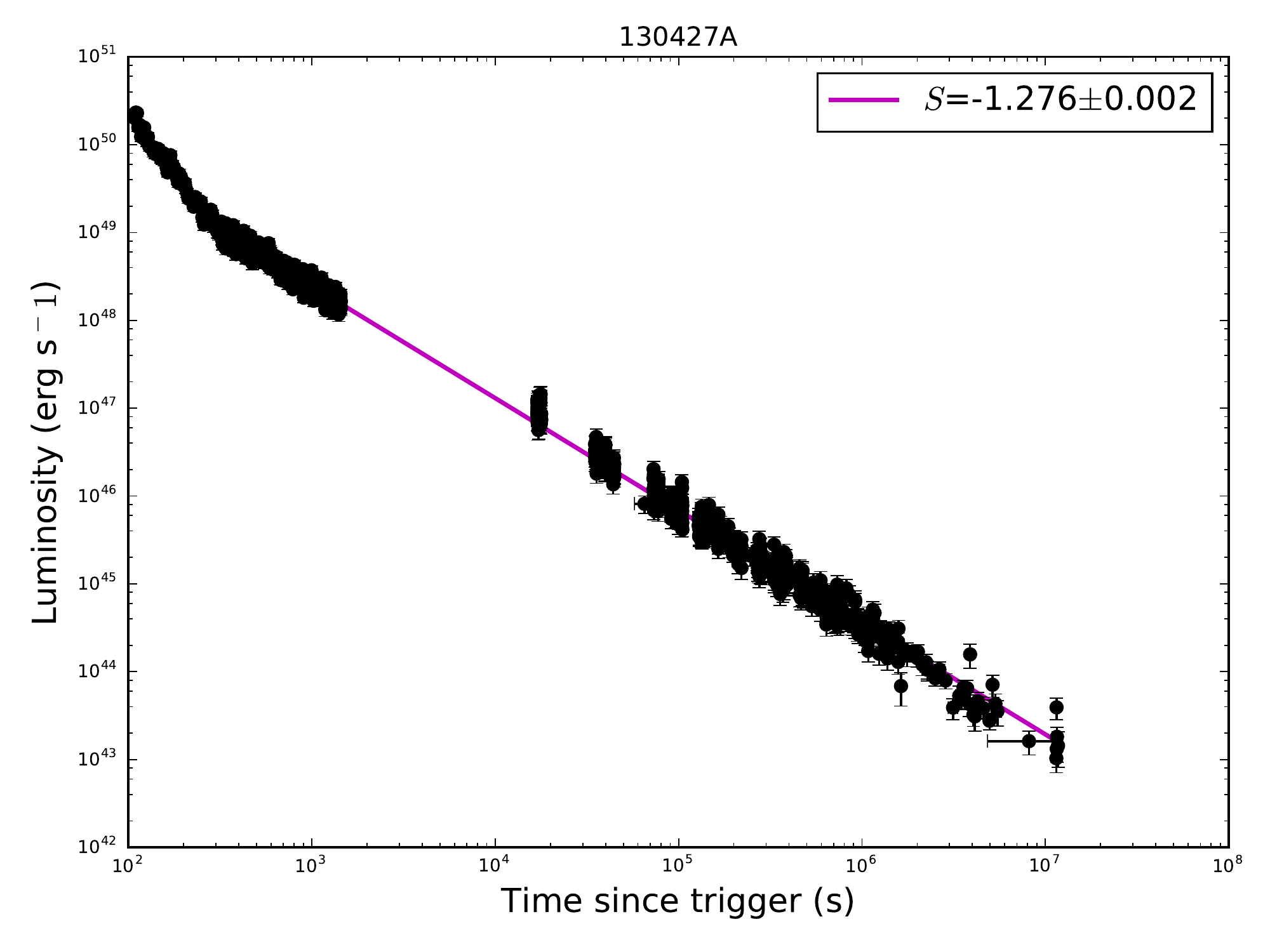}
\caption{SN-rise, Cavity, GeV, and Afterglow of GRB 130427A; see also Table \ref{table:episodes_130427A} that includes, for each Episode, the starting time, the duration, the isotropic energy, and the model that best fits the spectrum. \textbf{Upper left:} SN-rise spectrum, well fitted by a CPL+BB model, from $0$ to $0.65$s ($t_{\rm rf} \simeq 0.49$s); the spectral index $\alpha$ is -0.58, cutoff energy $E_{\rm c}$ is $547.59$ keV, and the BB temperature is $42.63$~keV in the observer's frame. 
\textbf{Upper right:} Featureless spectrum of the cavity emission, from $\approx 15$~s ($t_{\rm rf} = 11.19$~s) to $\approx 25.5$~s ($t_{\rm rf} = 19.03$~s) fitted by a CPL, with the photon index $\alpha$ is $-1.52$ and the cutoff energy is at $496.13$~keV in the observer's frame. 
\textbf{Lower left:} {\it Fermi}-LAT rest-frame light-curve in the $100$~MeV to $100$~GeV energy range. The UPE region is marked with the grey shadow. \textbf{Lower right:} k-corrected X-ray afterglow luminosity observed by {\it Swift}-XRT in the $0.3$--$10$~keV energy range, as a function of the rest-frame time. It is best fitted by a power-law with index $1.276\pm0.002$.}
\label{fig:joinedGRB130427A}
\end{figure*}

The {\it Fermi}-GBM count rate of GRB 130427A is shown in Fig.~\ref{fig:lc_130427A}.  During the UPE phase the event count rate of n9 and n10 of {\it Fermi}-GBM surpasses $\sim 8\times 10^4$ counts per second in the prompt radiation between rest-frame times $T_0 + 3.4$~s and $T_0 + 8.6$~s. The GRB is there affected by pile-up, which significantly deforms the spectrum; details in \citet{2014Sci...343...42A, 2015ApJ...798...10R}, only the data between $t_{\rm rf}=0.0$ and $t_{\rm rf}=$ $1.49$~s can be used for a spectral analysis in the prompt phase. As shown in Fig.~\ref{fig:lc_130427A} (see also Table~\ref{table:episodes_130427A}) clearly identified parts are:

--- \textit{SN-rise}. Figure~\ref{fig:joinedGRB130427A} (upper left panel) shows the clear identification of the SN-rise, as also reported in Fig.~\ref{fig:lc_130427A}. The spectrum of the SN-rise of GRB 130427A are best fitted by CPL+BB model, from $0.0$s ($t_{\rm rf} \simeq 0.0$s) to 0.65 s ($t_{\rm rf} \simeq 0.49$s). The spectrum contains a BB component of temperature $42.63$~keV and a photon index $\alpha=-0.58$, and $E_{\rm c}=547.59$~keV. 

--- \textit{Cavity}. {Figure~\ref{fig:joinedGRB130427A} (upper right panel) shows the featureless spectrum of the cavity emission of GRB 130427A, fitted by a CPL model, from $\approx 15$~s ($t_{\rm rf} = 11.19$~s) to $\approx 25.5$~s ($t_{\rm rf} = 19.03$~s) fitted by a CPL, with the photon index $\alpha=-1.52$ and the cutoff energy is at $496.13$~keV.}

--- \textit{GeV emission}. Figure~\ref{fig:joinedGRB130427A} (lower left panel) shows the rest-frame luminosity of the GeV emission as a function of the rest-frame time.

--- \textit{Afterglow}. Figure~\ref{fig:joinedGRB130427A} (lower right panel) shows the (k-corrected) luminosity of the afterglow ({\it Swift}/XRT data) as a function of the rest-frame time, and obtained as best-fit a power-law index of $-1.276\pm0.002$.

\section{BdHN II: GRB 180728A} \label{sec:GRB180728A}

GRB 180728A triggered \textit{Swift}-BAT at 17:29:00 UT, on 2018 July 28 \citep{GCN23046}. Due  to  the Earth  limb, \textit{Swift}-XRT   began the observation  $1730.8$ s  after the trigger \citep{GCN23049}. \textit{Fermi} was triggered at 17:29:02.28 UT, no GeV  photon  was  detected though the initial \textit{Fermi}-LAT bore-sight angle was only 35 degrees \citep{GCN23053}. This burst occurred at a close distance of redshift $z = 0.117$ detected by VLT/X-shooter\citep{GCN23055}. On July 28, we made a prediction of the SN appearance in $\sim 15$ days \citep{GCN23066,2019ApJ...874...39W}, and indeed the SN optical peak was confirmed then \citep{GCN23142,GCN23181}.
 
\begin{figure*}
\centering
\includegraphics[width=0.6\hsize,clip]{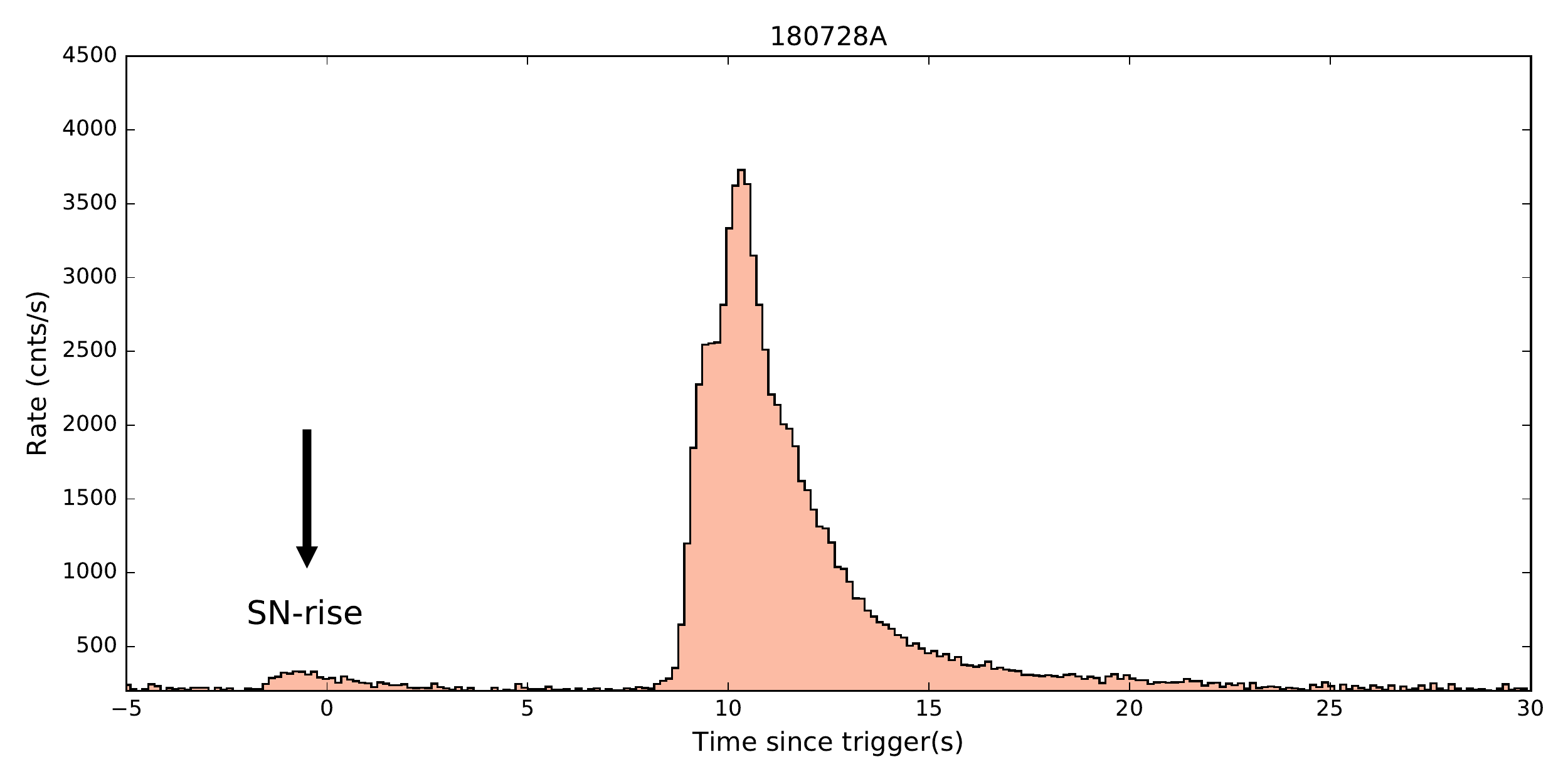}
\caption{We identify the SN-rise from the CO$_{\rm core}$ of a BdHN II in GRB 180728A \citep{2019ApJ...874...39W}. This GRB is composed of two spikes. The first spike, the precursor, shows a power-law spectrum with a power-law index $-2.31\pm 0.08$ in its $2.75$~s duration. The averaged luminosity is $3.24^{+0.78}_{-0.55} \times 10^{49}$~erg~s$^{-1}$, and the integrated energy gives $7.98^{+1.92}_{-1.34}\times 10^{49}$~erg in the energy range from $1$~keV to $10$~MeV. This energy emitted is in agreement the conversion of the SN-rise kinetic energy into electromagnetic emission. We consider second pulse (prompt emission without self similarity) as due to the hypercritical accretion of the SN ejecta onto the companion NS, starting from $8.72$~s and lasts $13.82$~s. This pulse contains $2.73\times 10^{51}$~erg isotropic energy. The best-fit is a CPL+BB model of temperature $\approx 7$~keV in the observer frame. The BB component is interpreted as a matter outflow driven by the Rayleigh-Taylor convective instability developed in the accretion process. From the time of observing the SN-rise to the starting time of the hypercritical accretion, $\Delta t\approx 10$~s, it has been inferred a binary separation of $\approx 3 \times 10^{10}$~cm. From the binary separation, by angular momentum conservation, it has been inferred that the spin of the $\nu$NS left from the collapse of the CO$_{\rm core}$, is $\approx 2.5$~ms \citep{2019ApJ...874...39W}. This $\nu$NS powers the afterglow by dissipating its rotational energy.}
\label{fig:180728A}
\end{figure*} 

\begin{table*}
\caption{The Episodes of GRB 180728A, the definitions of  parameters are the same as in table \ref{table:episodes_160625B} except this GRB as a BdHN II has no GeV emission. Prompt emission without self similarity.}             
\label{table:episodes_180728A}
\centering                         
\begin{tabular}{cccccc}       
\hline\hline                  
Episode &Starting Time &Ending Time&Energy&Spectrum&Reference\\   &Rest-frame&Rest-frame\\  
\hline                        
SN-rise&$  0 $~s&$ 2.46 $~s&$7.98\times10^{49}$~erg&PL&\cite{2019ApJ...874...39W}\\
Prompt emission&$ 7.81$~s&$ 11.82$~s&$2.73\times10^{51}$~erg&CPL+BB&\cite{2019ApJ...874...39W}\\
Afterglow&$ 1556$~s&$ >10$ days&$ 5.81\times 10^{50}$~erg&PL&\cite{2019ApJ...874...39W}\\
\hline                                   
\end{tabular}
\end{table*}

This GRB is composed of two pulses, see Fig.~\ref{fig:180728A} and Table~\ref{table:episodes_180728A}:

--- \textit{First pulse as SN-rise}. The first spike, the precursor, shows a power-law spectrum with a power-law index $-2.31\pm 0.08$ in its $2.75$~s duration. The averaged luminosity is $3.24^{+0.78}_{-0.55} \times 10^{49}$~erg~s$^{-1}$, and the integrated energy gives $7.98^{+1.92}_{-1.34}\times 10^{49}$~erg in the energy range from $1$~keV to $10$~MeV. This energy emitted is in agreement the conversion of the SN-rise kinetic energy into electromagnetic emission.

--- \textit{Second pulse as the hypercritical accretion of the SN ejecta onto the companion NS}.
This pulse starting from $8.72$~s and lasts $13.82$~s which contains $2.73\times 10^{51}$~erg isotropic energy. The best-fit is which is  a CPL+BB model of temperature $\approx 7$~keV in the observer frame is shown in Fig.~\ref{fig:180728A}. The BB component is interpreted as a matter outflow driven by the Rayleigh-Taylor convective instability developed in the accretion process  \citep[see e.g.][]{2012A&A...548L...5I}. From the time of observing the SN-rise to the starting time of the hypercritical accretion, $\Delta t\approx 10$~s, it has been inferred a binary separation of $\approx 3 \times 10^{10}$~cm. The binary separation determines, by angular momentum conservation, the spin period of $\approx 2.5$~ms of the $\nu$NS left from the collapse of CO$_{\rm core}$. This $\nu$NS powers the afterglow by dissipating its rotational energy \citep{2019ApJ...874...39W}.

\section{Discussion}\label{sec:discussion}

\begin{table*}
\small\addtolength{\tabcolsep}{-2pt}
\caption{The properties of the SN-rise in BdHN I: GRB 190114C, GRB 130427A, GRB 160509A, and GRB 160625B; and the properties of the SN-rise in BdHN II: GRB 180728A.}             
\label{tab:shockbreakout}
\centering                         
\begin{tabular}{ccccccccc}       
\hline\hline    
GRB&$t_{1}$$\sim$$t_{2}$ &Duration&Flux&$E_{\rm sh}$&$E_{\rm iso}$&Temperature&redshift&Reference\\
&(s)&(s)&(erg cm$^{-2}$ s$^{-1}$)&(10$^{52}$ erg)&(erg)&(keV)\\
&(Observation)&(Rest)&&(SN-rise)&(Total)&(Rest)&&(For SN-rise)\\
\hline                        
190114C&1.12$\sim$1.68&0.39&1.06$^{+0.20}_{-0.20}$(10$^{-4}$)&2.82$^{+0.13}_{-0.13}$&(2.48$\pm$0.20)$\times$10$^{53}$&27.4$^{+45.4}_{-25.6}$&0.424&\cite{GCN23983}\\
\hline
130427A&0.0$\sim$0.65&0.49&2.14$^{+0.28}_{-0.26}$(10$^{-5}$)&0.65$^{+0.17}_{-0.17}$&$\sim$1.40$\times$10$^{54}$&44.9$^{+1.5}_{-1.5}$&0.3399&\cite{2013ApJ...776...98X}\\
160509A&2.0$\sim$4.0&0.92&1.82$^{+1.23}_{-0.76}$(10$^{-6}$)&1.47$^{+0.6}_{-0.6}$&$\sim$1.06$\times$10$^{54}$&25.6$^{+4.8}_{-4.7}$&1.17&\cite{2017ApJ...844L...7T}\\
160625B&0$\sim$2.0&0.83&6.8$^{+1.6}_{-1.6}$(10$^{-7}$)&1.09$^{+0.2}_{-0.2}$&$\sim$3.00$\times$10$^{54}$&36.8$^{+1.9}_{-1.9}$&1.406&This paper\\
\hline
180728A&$-1.57\sim1.18$&$0.83$&4.82$^{+1.16}_{-0.82}(10^{-8}$)&7.98$^{+1.92}_{-1.34} \times 10^{49}$&$2.76^{+0.11}_{-0.10}\times$10$^{51}$& - &0.117& \citet{2018GCN.23142....1I}\\
\hline                                   
\end{tabular}
\end{table*}

In Table~\ref{tab:shockbreakout}, we compare and contrast the duration, the fluxes, the energy, the temperature of the BB component associated with the SN-rise of the above BdHNe I and II, we give as well, for each GRB, the corresponding redshift and $E_{\rm iso}$.

In the case of BdHN I, all of them have a similar SN-rise duration of nearly a second, consistent with the radius of the CO$_{\rm core}$ of $10^{10}$~cm, and energies of the order of $10^{52}$~erg. These energies are much larger than the one we have here found in in the SN-rise of BdHNe II, $\sim 10^{50}$~erg, comparable to the one of isolated SNe \citep[see, e.g.,][]{1982ApJ...253..785A,1990RvMP...62..801B,2017hsn..book..967W}.

\subsection{The SN-rise energetics of BdHN I}\label{sec:discussion_energy}

The larger energies of the SN-rise associated with BdHNe I here discovered can also be ascribed to a more energetic, rapidly rotating CO$_{\rm core}$. This can be the result of the binary nature of the progenitor with a short orbital period of the order of $4$--$5$~min in which angular momentum transfer by tidal effects during the previous evolutionary stages has been at work very efficiently. 

Let us estimate the rotational energy of the CO$_{\rm core}$ assuming that the binary is tidally locked. In this case the CO$_{\rm core}$ rotation period, $P_{\rm CO}$, equals the binary orbital period, $P_{\rm orb}$ \citep[see, e.g.,][]{2002MNRAS.329..897H}, i.e.
\begin{equation}
    P_{\rm CO} = P_{\rm orb} = 2 \pi \sqrt{\frac{a^3_{\rm orb}}{G M_{\rm tot}}},
\label{eq:CO_period}
\end{equation}
which is related to the binary separation $a_{\rm orb}$ and the total mass of the system $M_{\rm tot}$ and $G$ is the gravitational constant. 
Let us adopt a typical progenitor of a BdHN from \citet{2019ApJ...871...14B}: a CO$_{\rm core}$ obtained from the evolution of a $30~M_\odot$ zero-age main-sequence (ZAMS) progenitor star, which have a total mass of $M_{\rm CO}=8.9~M_\odot$ and radius $R_{\rm CO}=7.83\times 10^9$~cm, forming a binary with a NS companion of $M_{\rm NS}=2~M_\odot$. As for the orbital period/separation, we constrain our systems by the condition that there is no Roche-lobe overflow at the moment of the supernova explosion of the CO$_{\rm core}$. The Roche lobe radius of the CO$_{\rm core}$ can be estimated as \citep{1983ApJ...268..368E}
\begin{equation}
    \frac{R_{\rm RL}}{a_{\rm orb}} = \frac{0.49 q^{2/3}}{0.6 q^{2/3} + \ln(1 + q^{1/3})},
\end{equation}
where $q=M_{\rm CO}/M_{\rm NS}$. Therefore, the minimum orbital period of the binary, $a_{\rm orb,min}$, is obtained when $R_{\rm CO} = R_{\rm RL}$. For the above parameters, $a_{\rm orb,min}\approx 1.53\times 10^{10}$~cm and correspondingly the minimum orbital period is $P_{\rm orb,min}\approx 5.23$~min.

The rotational energy for a CO$_{\rm core}$ is
\begin{equation}
    E_{\rm rot,CO}=\frac{1}{2} I_{\rm CO} \omega_{\rm CO}^2 = \frac{1}{2} I_{\rm CO}\left(\frac{2 \pi}{P_{\rm CO}}\right)^2,
\label{eq:rot_energy}
\end{equation}
where $I_{\rm CO}$ is the CO$_{\rm core}$ moment of inertia. So, adopting $P_{\rm CO}=P_{\rm orb,min}$ ($\omega_{\rm CO}\approx 0.03$~rad~s$^{-1}$) and $I_{\rm CO}\approx (2/5)M_{\rm CO} R_{\rm CO}^2$, we obtain $E_{\rm rot,CO}\approx 8.7\times 10^{49}$~erg. This is of course lower than the gravitational binding energy $|W|\approx (3/5)G M_{\rm CO}^2/R_{\rm CO} \approx 1.6\times 10^{51}$~erg and lower than the internal thermal energy as from the virial theorem. If we adopt the CO$_{\rm core}$ from the $25~M_\odot$ ZAMS progenitor \citep[see Table~1 in][]{2019ApJ...871...14B}, characterized by $M_{\rm CO}=6.85~M_\odot$ and $R_{\rm CO}=5.86\times 10^9$~cm, and for the corresponding minimum orbital period $P_{\rm orb,min}\approx 4$~min ($\omega_{\rm CO}\approx 0.02$~rad~s$^{-1}$), we obtain $E_{\rm rot,CO}\approx 6.3\times 10^{49}$~erg.

Therefore, a much more energetic SN-rise can be the result of an exploding CO$_{\rm core}$ which rotates at a much higher rotation rate with respect to the one set by tidal synchronization. In the above two examples, the rotational to gravitational energy ratio is $E_{\rm rot}/|W|\approx 0.05$. However, from the stability point of view, it is known from the theory of Newtonian ellipsoids that secular axisymmetric instability sets in at $E_{\rm rot}/|W|\approx 0.14$ and dynamical instability at $E_{\rm rot}/|W|\approx 0.25$ \citep{1969efe..book.....C}. 

Indeed, three-dimensional simulations of SN explosions confirm these stability limits and so explore SN explosions from pre-SN cores with high rotation rates of the order of $1$~rad~s$^{-1}$ \citep[see, e.g.,][]{2014ApJ...793...45N,2018MNRAS.474.2419G,2019ApJ...872..155F}. These angular velocities are a factor $30$--$50$ faster than the ones we have considered above. This implies that the rotational energy of the pre-SN core can be up to a factor $10^3$ higher, namely $E_{\rm rot}\sim {\rm few}\times 10^{52}$~erg.

\subsection{The SN-rise energetics of BdHN II}

In the case of BdHN II, the SN-rise has been shown to have a much smaller energy, $10^{49}$--$10^{50}$~erg. A similar case in the literature is represented by SN 2006aj, associated with GRB 060218 \citep{2006Natur.442.1008C,2006ApJ...643L..99M,2006Natur.442.1011P,2006AA...454..503S,2006A&A...457..857F}. The GRB 060218/SN 2006aj association was indeed interpreted in \citet{2016ApJ...833..107B} as a BdHN II (at that time called ``X-ray flash''). As we have mentioned, the energetics of these SN-rises are closer to the typical ones encountered in isolated SNe \citep[see, e.g.,][]{1982ApJ...253..785A,1990RvMP...62..801B,2017hsn..book..967W}. This is consistent with the longer orbital periods of BdHNe II \citep{2016ApJ...833..107B} since, being farther apart, in the prior evolutionary stages binary interactions have been less effective in transferring angular momentum to the CO$_{\rm core}$. This explains why the SNe associated with BdHN II, even if they occur in a binary, are more similar to isolated SNe. 

As a final remark, we recall that the occurrence of the SN is deduced from direct optical observations for GRB sources at $z<1$, and for all cases the SN occurrence is also inferred, indirectly, by the observation of the afterglows. Indeed, the afterglow originates from the feedback of the emission of the $\nu$NS, originated in the SN event, into the expanding SN ejecta, given the proof of the SN occurrence \citep[see][for details]{2018ApJ...869..101R,2019ApJ...874...39W,2019arXiv190511339R}.

\section{Conclusions} \label{sec:conclusion}

In a companion paper \citep{2019arXiv190404162R}, we have introduced a novel time-resolved spectral analysis technique, adopting ever decreasing time steps, in the analysis of GRB 190114C. This has led to the discovery of the three Episodes and the self-similarity and power-laws in BdHN I.

In this paper, we have made a major effort of applying, such a time-resolved spectral analysis to BdHNe I: GRB 130427A, GRB 160509A, and GRB 160625B; see sections~\ref{sec:GRB160625B}--\ref{sec:GRB130427A}. We have proved that indeed, all the results obtained in GRB 190114C, far from being an exception, do characterize the physics of BdHNe I.

These results open new perspective of research:

1) to study the new physical process characterising each single Episode of a BdHN in the context of previously unexplored regimes: e.g. the analysis of the SN not following the traditional description as an isolated system and identifying their properties within a BdHN I and alternatively, in a BdHN II.  

2) to insert the BdHN evolution in the framework a population synthesis analysis. 

3) to address the new physical process underlying the existence of the observed self-similarities and power-laws revealing a discrete sequence of quantized events with quanta of $10^{37}$~erg on new timescales of $10^{-14}$~s; see the companion papers \citep{2018arXiv181200354R,2019arXiv190708066R}, Ruffini, Moradi, et al. (2019, submitted) and explore the vast new directions open to the identification of the fundamental new laws of our Universe. 

\begin{acknowledgements}
Some important points of this work were performed during the Remo Ruffini and Liang Li's visit to University of Science and Technology of China, Hefei, member of ICRA.
\end{acknowledgements}

\end{document}